\documentclass[review,12pt]{elsarticle}
\usepackage{amssymb}
\usepackage{amsthm}
\usepackage{amsmath}
\usepackage{mathrsfs}
\usepackage{graphicx}
\usepackage{epstopdf}
\usepackage{float}
\usepackage{caption}
\usepackage{subcaption}
\usepackage{bm}
\usepackage{bbm}
\usepackage{booktabs}
\usepackage{mathrsfs}
\usepackage{multirow}
\usepackage{parskip}
\usepackage{amsmath}
\usepackage{fancyhdr}
\usepackage{mathrsfs}
\usepackage{hyperref}
\usepackage{cleveref}
\usepackage{soul}
\usepackage{color,soul} %to highlight text
\usepackage{color} %To highlight references in the text
\usepackage{siunitx}
\usepackage[margin=2.5cm]{geometry}
\usepackage[table]{xcolor}
\biboptions{sort&compress}
\graphicspath{ {./images/} }

\journal{Elsevier}

\makeatletter
\def\@author#1{\g@addto@macro\elsauthors{\normalsize%
    \def\baselinestretch{1}%
    \upshape\authorsep#1\unskip\textsuperscript{%
      \ifx\@fnmark\@empty\else\unskip\sep\@fnmark\let\sep=,\fi
      \ifx\@corref\@empty\else\unskip\sep\@corref\let\sep=,\fi
      }%
    \def\authorsep{\unskip,\space}%
    \global\let\@fnmark\@empty
    \global\let\@corref\@empty  %% Added
    \global\let\sep\@empty}%
    \@eadauthor={#1}
}
\makeatother

\begin{document}

\begin{frontmatter}



\title{Multi-objective Bayesian Optimisation of Spinodoid Cellular Structures for Crush Energy Absorption}


% The author list is pending, we will decide in the later stage on the contribution from each author

\author{Hirak Kansara\fnref{QMUL}}
\author{Siamak F. Khosroshahi \fnref{QMUL}}
\author{Leo Guo\fnref{TU}}
% \author{Lucas Meza \fnref{UoW}}
% \author{John Dear \fnref{IC}} 
\author{Miguel A. Bessa \fnref{BU}}
%\author{Siddhant Kumar\fnref{TU}}

\author{Wei Tan\corref{cor1}\fnref{QMUL}} \ead{wei.tan@qmul.ac.uk}

\address[QMUL]{School of Engineering and Materials Science, Queen Mary University of London, Mile End Road, London, E1 4NS, UK}

\address[TU]{Department of Materials Science and Engineering, Delft University of Technology, 2628 CD Delft}

% \address[UoW]{University of Washington, Mechanical Engineering Department, Stevens Way, Box 352600 Seattle, WA 98195}

% \address[IC]{Department of Mechanical Engineering, Imperial College London, London SW7 2AZ, United Kingdom}

\address[BU]{School of Engineering, Brown University, Providence, Rhode Island, USA}
\cortext[cor1]{Corresponding author.}

\begin{abstract}

In the pursuit of designing safer and more efficient energy-absorbing structures, engineers must tackle the challenge of improving crush performance while balancing multiple conflicting objectives, such as maximising energy absorption and minimising peak impact forces. Accurately simulating real-world conditions necessitates the use of complex material models to replicate the non-linear behaviour of materials under impact, which comes at a significant computational cost. This study addresses these challenges by introducing a multi-objective Bayesian optimisation framework specifically developed to optimise spinodoid structures for crush energy absorption. Spinodoid structures, characterised by their scalable, non-periodic topologies and efficient stress distribution, offer a promising direction for advanced structural design. However, optimising design parameters to enhance crush performance is far from straightforward, particularly under realistic conditions. Conventional optimisation methods, although effective, often require a large number of costly simulations to identify suitable solutions, making the process both time-consuming and resource intensive. In this context, multi-objective Bayesian optimisation provides a clear advantage by intelligently navigating the design space, learning from each evaluation to reduce the number of simulations required, and efficiently addressing the complexities of non-linear material behaviour. By integrating finite element analysis with Bayesian optimisation, the framework developed in this study tackles the dual challenge of improving energy absorption and reducing peak force, particularly in scenarios where plastic deformation plays a critical role. The use of scalarisation and hypervolume-based techniques enables the identification of Pareto-optimal solutions, balancing these conflicting objectives while accounting for the intricacies of plastic material behaviour. Furthermore, the approach ensures that the optimised designs avoid problematic densification, which is essential for maintaining structural integrity during impact. The results not only demonstrate the framework's ability to outperform the NSGA-II algorithm but also highlight its potential for wider applications in structural and material optimisation. The framework's adaptability to various design requirements underscores its capability to address complex, multi-objective optimisation challenges associated with real-world conditions.

\end{abstract}

%  \begin{figure}[ht]
%     \centering
%     \includegraphics[width=0.8\textwidth]{Images/graphical_abstract.pdf}
% \end{figure}

\begin{keyword}

Multi-objective \sep Bayesian optimisation \sep Cellular structures \sep Energy absorption %\sep Finite Element Analysis



\end{keyword}

\end{frontmatter}

%%%% CS \& T Rules
%CS\&T guide for authors: Submissions must be accompanied by a letter stating 1) the significance of the paper for the research community and 2) what it contains that is most important, new or original. The new guideline of manuscript length is a maximum of 22 pages including figures and tables. The text should be 12–point with double spacing and there should not be more than two figures per page and no more than two tables per page, depending on their sizes. Please note 12–pt font size and double spacing should be used throughout the whole manuscript (inclusive of text, references and tables), no schemes are allowed but integrated within the text

%\noindent \textbf{Graphical abstract}
%\begin{figure}[h]
%    \centering
%    \includegraphics[width=1.0\textwidth]{images/Figure0.pdf}
%    \label{Graphical Abstract}
%\end{figure}


\section{Introduction}
\label{Introduction}

Energy-absorbing structures, like crumple zones, play a vital role in dynamic events such as vehicle collisions, converting kinetic energy into plastic deformation energy. The primary goal of these structures is to protect occupants by minimising the impact forces they experience, which is defined as the crashworthiness of a structure. Ensuring the protection of occupants during crashes is an essential part of the design and manufacturing process of vehicles. Obtaining the correct combination of materials and structural design is crucial for achieving an efficient energy-absorbing structure. The type of materials used to fabricate such devices dictates their performance during impact, i.e., their energy-absorbing efficiency. The use of metals such as steel and aluminium has continued to gain popularity, accelerated by improvements in metal 3D printing technology, enabling the production of complex prototypes \cite{zhang20203d, wu2023additively, harris2021impact, alkhatib2019deformation}. 

In recent years, there has been a significant surge in the utilisation of composite materials for enhancing crashworthiness. This increased adoption can be attributed to the numerous advantages they offer over traditional metallic materials, including superior strength, a high strength-to-weight ratio and corrosion resistance. Notably, the most significant benefit of replacing metallic components with composite structures in vehicles is their lightweight nature, which contributes to reducing carbon dioxide emissions through decreased fuel consumption \cite{fontaras2017fuel}. 

However, solely substituting the material used for energy-absorbing structures may not lead to sufficient weight reduction. Consequently, the focus has shifted to optimising the mechanical properties of structures, leading to the development of superior mechanical metamaterials \cite{mohsenizadeh2018additively, ai2018three, li2017harnessing, yu2019investigation, portela2020extreme, bowen2022hierarchically}. These architected materials can be classified as a subclass of cellular solids, characterised by porous structures with interconnected networks occurring periodically or at irregular intervals \cite{gibson2003cellular}. Naturally occurring cellular structures such as bone, wood, bee honeycombs, and sea sponges exhibit high stiffness-to-weight ratios \cite{launey2010mechanistic, jakob2022strength, mirkhalaf2014overcoming, woesz2006micromechanical}. With advancements in fabrication techniques, prototyping of beam- and plate-based structures has become seamless. However, these structures often face a major drawback -- high-stress concentrations at the junctions, leading to premature failure and poor recoverability \cite{latture2018effects, portela2018impact}.

In contrast, shell-based structures offer a solution to alleviate stress concentrations by utilising doubly curved surfaces \cite{han2015new}. Employing intersection-free smooth surfaces effectively avoids the occurrence of sharp junctions, a feature prominently observed in triply periodic minimal surface (TPMS) architectures such as the Schwarz P surface, the Schwarz D surface \cite{hyde1996language, al2019multifunctional, schwarz1972gesammelte}, and the gyroid surface \cite{schoen1970infinite}. Moreover, the double curvature orientation of these smooth surfaces enables TPMS structures to exhibit high stiffness and has been shown to reduce stress localisation \cite{hyde1996language}. Nonetheless, it is crucial to note that these structures are sensitive to imperfections during fabrication, and any break in symmetry would result in a loss of stiffness and premature failure \cite{hutchinson2018imperfections}. This limitation significantly impacts their manufacturability and scalability.

By leveraging self-assembly techniques, potentially complemented by the integration of additive manufacturing on a larger scale, there exists a significant opportunity to greatly enhance scalability. Spinodal metamaterials exemplify structures formed through such self-assembly processes. These metamaterials undergo spinodal decomposition -- a diffusion-driven phase transformation \cite{cahn1961spinodal}. This process involves the spontaneous separation of solid or liquid solutions into two phases, a phenomenon evident in various materials such as metal foams \cite{miller1995spinodal, hodge2007scaling}, microemulsions \cite{bates1997polymeric}, and polymer blends \cite{de1980dynamics, binder1983collective}. Furthermore, the nature of spinodal topologies makes them resilient to imperfections and symmetry-breaking defects \cite{portela2020extreme, hsieh2019mechanical} Additionally, they demonstrate high energy absorption (EA), a desirable trait for crashworthiness applications \cite{guell2019ultrahigh, zhang2021mechanical}. Such topologies also ensure scalability when paired with additive manufacturing. Building on these advantages, recent improvements in the formulation of spinodal topologies have been introduced to enhance stretching-dominated behaviour under mechanical loading, which is particularly beneficial for achieving high-stiffness structures \cite{guo2024inverse}. Beyond mechanical performance, spinodal topologies are also gaining traction in multi-physics applications, such as tunable piezoelectric and pyroelectric properties \cite{shi20243d}, sound absorption \cite{wojciechowski2023additively}, and the modelling of trabecular bone for biomimetic designs \cite{vafaeefar2023morphological}, as well as in mass transport systems \cite{roding2022inverse}.

Generating spinodal topologies involves using a phase-field model to simulate the phase separation process, typically modelled using Cahn-Hilliard equations \cite{seol2003computer}. An alternative time-saving strategy involves utilising Gaussian Random Fields (GRFs), allowing for the efficient generation of spinodal-like topologies, referred to as `spinodoids'. This method efficiently creates a vast design space of anisotropic topologies. 
        
While the ability to efficiently generate the design of various anisotropies is a significant advancement, the challenge lies in identifying the optimal parameters within the expansive space, especially in the context of EA. Conventional design approaches often rely on intuition and trial-and-error methods, which can be time-consuming and computationally intensive \cite{yu2018mechanical}. 

To overcome these limitations, researchers have turned to optimisation techniques, such as topology optimisation, to search for the optimal distribution of material within a predefined domain. This approach, often coupled with computational homogenisation techniques, aims to improve the macroscopic elastic properties of the structure \cite{bendsoe2003topology, sigmund2013topology, coelho2008hierarchical}. These methods rely on an iterative approach, whereby gradients are evaluated for each design within the optimisation cycle, which can be arduous \cite{dong2019149}. Furthermore, such techniques may produce geometries with sharp angles and irregular shapes, posing challenges during fabrication, leading to imperfections \cite{sigmund1995tailoring}. To expedite the optimisation process, data-driven-based topology optimisation approaches utilising surrogate models have been increasingly utilised, enabling a faster and more efficient search for optimal properties \cite{zheng2021data, white2019multiscale}. However, such approaches often focus on maximising stiffness with linear elastic material properties, which may not fully capture real-world behaviour. Efforts to address these limitations include the incorporation of plasticity, and fracture mechanics, though this may further increase computational costs \cite{tauzowski2021topology, kang2017topology}. 

% Inverse design
Building on the concept of data-driven optimisation, structures can be inverse-designed by starting with desired performance characteristics and working backwards to generate optimal designs. This approach leverages machine learning techniques such as neural networks to capture complex, non-linear relationships \cite{bastek2022inverting}, convolutional neural networks (CNNs) for handling spatially varying designs, and generative adversarial networks (GANs) for exploring complex design spaces \cite{challapalli2021inverse, zheng2023deep}. However, many of these methods are highly data-dependent, requiring large, high-quality datasets that can be costly to generate, making it difficult to acquire the necessary complex non-linear data due to computational demands. To mitigate this challenge, physics-informed machine learning has been implemented, where experimental data is incorporated to tailor complex structures for large-deformation scenarios \cite{thakolkaran2024experimentinformedfinitestraininversedesign}.

% Multi-objective
This challenge becomes even more pronounced in real-world applications, such as automotive crashworthiness or aerospace structures, where focusing solely on single-objective optimisation—such as independently maximising energy absorption (EA) or minimising peak force (PF)—is often inadequate. A balance between multiple conflicting objectives must be achieved simultaneously. In response to these multifaceted optimisation problems, researchers have turned to multi-objective optimisation (MOO) techniques, allowing the identification of optimal solutions that meet diverse requirements. Past implementations to solve such a problem included the use of evolutionary algorithms \cite{deb2011multi} such as the Non-dominated Sorting Genetic Algorithm (NSGA-II), coupled with an external surrogate model, such as the  Radial Basis Function \cite{lanzi2004multi}, polynomial response surface method \cite{xu2016crash}, and the use of Gaussian Processes (or kriging) \cite{bessa2018design,deng2019multi}. Other implementations include the use of particle swarm-based optimisation \cite{yildiz2012multi}, and hybrid approaches using the VIKOR method \cite{zhang2019hybrid, peng2023performance}. Recently, neural networks have also been utilised as surrogates for MOO problems \cite{baykasouglu2020multi, kappe2024multi}. 

% Again you talked about the challenges of the demand for large dataset, this is somehow a repetition of what your mentioned before. 
However, these methods necessitate the implementation of regression models as surrogates to map the input-output relationship, specifically the structure-property relationship, and predict the objectives accurately. Yet, achieving precise predictions relies on utilising a sufficiently large dataset, ideally sourced from cheap-to-evaluate functions. Nonetheless, capturing the realistic behaviour of architected structures under loads requires implementing complex material models (plasticity, fracture, etc.), consequently increasing the cost of evaluation. 

% Gaussian process
To address such challenges, the increased computational cost of data generation can be alleviated by utilising Bayesian optimisation (BO) in the design process. BO minimises the number of points needed for sampling, which is particularly beneficial when dealing with costly-to-evaluate functions \cite{shahriari2015taking, snoek2012practical, brochu2010tutorial, bessa2019bayesian, raßloff2024inversedesignspinodoidstructures}. An additional benefit lies in its ability to construct a computationally efficient surrogate model by combining the probabilistic nature of BO with Gaussian Processes, as well as providing an efficient means to balance exploration of unknown regions within design space and exploitation of current information. This integration allows designers to evaluate optimal designs economically even with additional constraints \cite{rasmussen2006gaussian}. The first open-source data-driven framework \cite{bessa2017framework,van2024f3dasm} introduced a modular approach where (1) design of experiments, (2) data generation, (3) machine learning and (4) optimisation modules are synchronised to discover new materials and structures. If the data generation step is time consuming, the design problem can be reduced to a parametric shape optimisation problem, and data scarce machine learning models such as Gaussian processes can be used as surrogate models that are then robustly optimised by gradient-based or gradient-free optimisers \cite{bessa2018design,bessa2019bayesian}. Furthermore, the machine learning and optimisation modules can be combined by using Bayesian optimisation \cite{shin2022spiderweb}, including multiple objectives \cite{kuszczak2023bayesian}, and even in the presence of multi-fidelity data \cite{cupertino2024centimeter}. 

Building on these advantages, multi-objective Bayesian Optimisation (MOBO) extends the capabilities of BO to handle multiple conflicting objectives. The current advancements in MOBO focus on improving scalability and efficiency in high-dimensional spaces by enhancing sampling strategies \cite{daulton2022multi}. These include leveraging deep Gaussian Processes (DGPs), which stack multiple Gaussian Processes to better capture complex dependencies in the data \cite{hebbal2023deep}, and combining Bayesian statistics with neural networks to form Bayesian Neural Networks (BNNs), which provide scalable and flexible surrogate models \cite{springenberg2016bayesian}. Furthermore, enhanced acquisition functions, such as Expected Hypervolume Improvement (EHI) and Predictive Entropy Search for Multi-Objective Optimisation (PESMO), along with their advanced extensions \cite{daulton2020differentiable, suzuki2020multi, hernandez2016predictive}, have significantly improved the ability to balance exploration of the search space and exploitation of known promising regions, thereby accelerating convergence to the Pareto front and enabling efficient design space exploration.

However, many real-world problems involve non-linear, non-convex constraints, including manufacturing limitations or safety criteria. Existing MOBO frameworks often struggle to effectively incorporate or optimise under these constraints. To address this, our research introduces a gradient-based filtering approach within the MOBO process to identify and remove structures prone to densification. By dynamically filtering out infeasible solutions early, the framework ensures convergence to robust, manufacturable designs while reducing computational cost. Building on this, the proposed framework leverages MOBO to design spinodoid topologies that incorporate elastic-plastic material behaviours. Data is generated using finite element method (FEM) simulations, which predict post-yield performance metrics such as PF, and EA. This enables a comprehensive evaluation of the structural response and facilitates the optimisation of design parameters to enhance EA while reducing PF.

The novelty of this study lies in three key advancements, offering a comprehensive solution for optimising energy-absorbing structures: (1) The proposed optimisation framework significantly enhances MOBO by efficiently handling multiple conflicting objectives, enabling comprehensive exploration of the design space and balancing trade-offs that traditional methods often struggle to address. Leveraging pre-existing data reduces computational costs, requiring fewer simulations and achieving faster convergence. Additionally, the framework incorporates real-world constraints, such as manufacturing tolerances and fabrication defects by integrating complex material constitutive behaviours into FEM simulations, leading to robust and reliable designs. The introduction of a gradient-based filtering mechanism to identify and exclude parameter sets that result in structures that undergo densification which may degrade the performance of designs. (2) This study applies MOBO to optimise structures exhibiting highly non-linear crush behaviours, a challenging domain that has not been extensively explored. By using FEM simulations to evaluate elastic-plastic material responses, the framework optimises critical performance metrics, such as PF and EA. (3) The study pioneers the application of spinodoid topologies for crush energy absorption, demonstrating their tunability of anisotropy and energy absorption. This novel use of spinodoids leverages their unique structural characteristics to enhance energy-absorbing capabilities, making them particularly suitable for crashworthiness applications. The strategy opens up new possibilities for utilising spinodoid designs in diverse engineering disciplines.

The outline of this paper is as follows: (i) Design parameters used to generate the spinodoid cellular topologies have been introduced. (ii) ABAQUS is utilised to construct FEM model for assessing the mechanical performance of spinodoid cellular topologies. In addition, the influence of design parameters on the objectives has been demonstrated. (iii) Lastly, the MOBO framework with various methods and analysis of the Pareto fronts obtained through different approaches, namely, scalarisation and hypervolume-based MOBO methods as well as a traditional MOO method, namely, NSGA-II has been introduced.

\section{Methodology}

\subsection{Design parameters of Spinodoid cellular structures}
\label{section: Design parameters of Spinodoid cellular structures}
The design of spinodal topologies is inspired by the process of phase separation in polymer blends, known as spinodal decomposition \cite{zheng2021data, cahn1965phase} which can be modelled by the Cahn-Hill equation, as follows

\begin{equation}
    \frac{\partial c}{\partial x} = D \nabla^2(c^3-c-\omega\nabla^2c)
    \label{equ: cahn-hill equation}
\end{equation}
where, $c$ is the concentration of the two phases indicated by $c = \pm 1$ along spatial position $x$, $D$ is a diffusion coefficient, and $\omega$ is related to the transition region between phases \cite{cahn1958free}. However, simulating the phase separation process using Eq.~(\ref{equ: cahn-hill equation}) can be expensive, thus a means to generate smooth spinodal-like topologies is utilised that modified Cahn's GRF-based approach. A Gaussian random field of the form

\begin{equation}
    \varphi(\boldsymbol{x}) = \sqrt{\frac{2}{N}}\sum_{i = 1}^{N >> 1}\cos(\lambda\textbf{n}_i\cdot \boldsymbol{x} + \gamma_i),
    \label{equ: gaussian random field}
\end{equation}

can be used to approximate spinodal phase decomposition. Where $\varphi(\boldsymbol{x})$ represents the changes in concentration of one of the phases at a given multidimensional position $\boldsymbol{x}$, within a domain $\Omega$. In addition, $\textbf{n}_i \sim \mathcal{U}(S^2)$, and $\gamma_i  \sim \mathcal{U}([0, 2\pi))$ are the directions and the angles of the $i{\text{th}}$ wave vector, respectively, sampled from uniform probability distributions with zero mean. Furthermore, $S^2 = \{ \textbf{k} \in \mathbb{R}^3 : ||\textbf{k}|| = 1 \}$, represents the unit sphere in three dimensions, and $N$ is the number of waves, with $\lambda > 0$ being the wave number. It is important to emphasise that the variable $\lambda$ directly influences the microstructure, hence serving as an additional control parameter used to generate the topology.

Assuming the principal directions of mobility are aligned with the Cartesian basis \{$\hat{\mathbf{e}}_1$, $\hat{\mathbf{e}}_2$, $ \hat{\mathbf{e}}_3$\}, the resulting anisotropic topologies of spinodal structures, $\textbf{n}_i$ in Eq.~(\ref{equ: gaussian random field}) can be approximated by using a non-uniform orientation distribution function, parameterised by the following equation, as first introduced in \cite{kumar2020inverse}. This formulation provides an anisotropic extension to Cahn's GRF solution.

\begin{equation}
    \textbf{n}_i \sim u(\{ \textbf{k} \in S^2:(|\textbf{k} \cdot \hat{\mathbf{e}}_1 | > \cos\theta_1)\oplus (|\textbf{k} \cdot \hat{\mathbf{e}}_2 |> \cos\theta_2)\oplus(|\textbf{k}\cdot\hat{\mathbf{e}}_3|>\cos\theta_3)\})
\end{equation}
Therefore the wave vectors are constrained to conical angles $\{\theta_1, \theta_2, \theta_3 \} $. Furthermore, to distinguish the two phases (solid and void), a level set $\varphi_0$ can be applied to the phase field \cite{soyarslan20183d},

\begin{equation}
  \psi (\textbf{x}) =
    \begin{cases}
      1 & \text{if} \ \varphi(\textbf{x}) \leq \varphi_0 \\
      0 & \text{if} \ \varphi(\textbf{x}) > \varphi_0\\
    \end{cases}       
\end{equation}

whereby a value of 0 or 1 represents the presence of void or solid, respectively. The level set is defined as $\varphi_0 = \sqrt{2}\text{erf}^{-1}(2\rho - 1)$. The variable $\rho$ represents the relative density of the solid phase. Combining all the parameters, the design space can be characterised $\Theta = \{\rho, \lambda, \theta_1, \theta_2, \theta_3 \}$, such that $\rho \in [0.3, 0.6]$, here the range of relative density has been constricted to avoid generating structures that are discontinuous and close to a solid cube, the cone angles $\theta_{i}^{i = 1,2,3} \in [0, \pi/2]$ and $\neg(\theta_1 = 0^\circ \wedge \theta_2 = 0^\circ \wedge \theta_3 = 0^\circ)$, and $\lambda$ being a non-zero parameter. 

\begin{figure}[!h]
    \centering
    \includegraphics[width=1\textwidth]{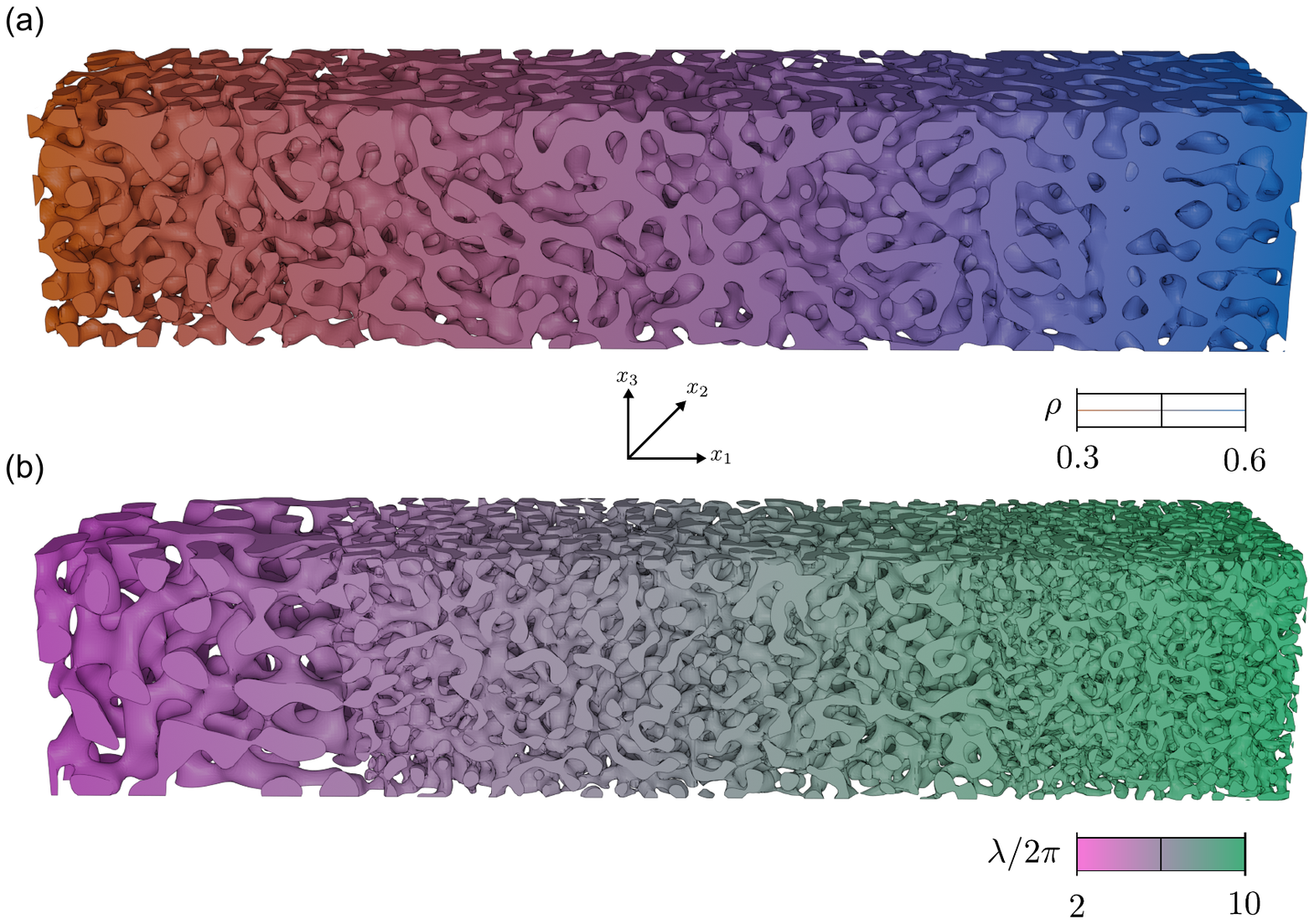}
    \caption{Visualisation of changes in two design parameters using graded spinodoids. Both structures were produced with $\theta_1 = 90^\circ, \theta_2 = 0^\circ, \theta_3 = 0^\circ$. (a) A linearly graded generated with an increasing relative density from left to right. The left side has the lowest relative density, starting at 0.3, the middle portion has a relative density of 0.45, and the right side reaches a relative density of 0.6. (b) A structure illustrating an increase in the wave number from left to right. The left side has the lowest wave number of $4\pi$, the right has the highest at $20\pi$, and the middle portion has an intermediate value of $12\pi$. It should be noted that these structures serve as a visualisation tool and do not represent the structures being optimised.}
    \label{varying_relative_density_wave number}
\end{figure}

To demonstrate the effects of varying the design parameters, several topologies have been generated. Fig.~\ref{varying_relative_density_wave number} illustrates the impact of adjusting the relative density and the wave number, showcased through two linearly graded spinodoid structures. Progression from left to right, these structures indicate an increase in the respective parametric values. However, it is worth noting that these graded structures are not indicative of the structures generated during the optimisation, rather, they serve as visualisation aids. Nonetheless, Fig.~\ref{varying_relative_density_wave number}(a), shows the significance of relative density, which quantifies the level of porosity present within the structure normalised against an equivalent-sized solid structure (with $\rho = 1$), consequently, an increase in the relative density results in topologies resembling a solid cube. Furthermore, Fig.~\ref{varying_relative_density_wave number}(b) serves as a tool to ensure the separation of scales from microscale to macroscale, facilitating the homogenisation of structural properties \cite{deng2024ai}. 

\begin{figure}[!ht]
    \centering
    \includegraphics[width=1\textwidth]{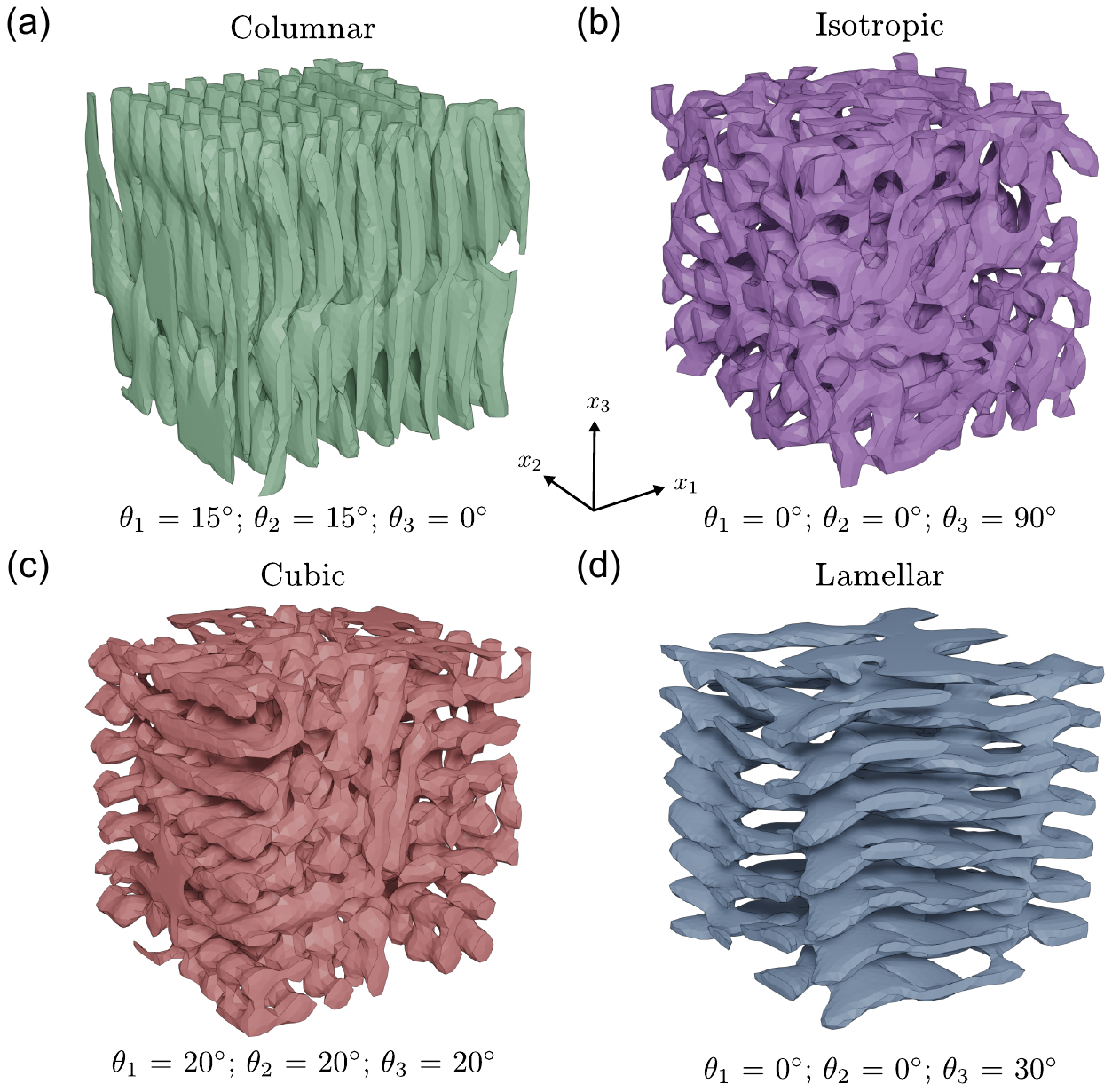}
    \caption{Four unique topologies generated through a unique choice of $\theta$, serving as benchmarks for FEM validation and showcasing the robustness of the design space. The topologies were generated assuming $\rho = 0.3$, and $\lambda = 15\pi$. (a) The anisotropic `Columnar' topology features uniformly distributed, interconnected column-like structures oriented along the direction of loading. (b) The `Isotropic' topology is produced with relatively high conical angles. (c) The `Cubic' structure arises from prescribing relatively small conical angles of equal magnitude. (d) The `Lamellar' structure exhibits anisotropy in the direction perpendicular to the loading direction.}
    \label{fig:4 spinodoid types}
\end{figure}

To illustrate the influence of the three cone angles $\theta_{i}^{i = 1,2,3}$, four distinct structures were created, each showcasing a unique combination of these angles. This approach demonstrates the extensive parametric design space offered by spinodoid topologies, with these structures being depicted in Fig.~\ref{fig:4 spinodoid types}. The design space is capable of producing anisotropic structures, as exemplified by  Fig.~\ref{fig:4 spinodoid types}(a) and (d). These structures are generated as a result of imposing small cone angle values, with at least one angle being set to zero. These configurations exhibit stiffness favouring at least one of the principal directions. For instance, the `Columnar' structure in  Fig.~\ref{fig:4 spinodoid types}(a) arises from a choice of two small, non-zero angle values, which results in a network of interconnected pillar-like structures. Another form of anisotropic structure, termed `Lamellar', is demonstrated by  Fig.~\ref{fig:4 spinodoid types}(d), where one non-zero cone angle value produces a structure featuring several interconnected layers, with the layer count being proportional to the wave number. \textcolor{black}{Demonstration of controlling the anisotropy of structures through a choice of cone angles can be found in \ref{appendix:Controlling the anisotropy of topologies}, which also highlights the symmetry in the design space, where the selected values of $\theta$ merely alter the orientation of the topology.} Conversely, Fig.~\ref{fig:4 spinodoid types}(c) showcases the formation of a `Cubic' topology through the use of three non-zero cone angles of equivalent magnitude, which produces a structure with identical stiffnesses in all principal directions. Meanwhile, `Isotropic' structures, as shown in  Fig.~\ref{fig:4 spinodoid types}(b), emerge from any combination of angles featuring sufficiently large values of cone angles.

\subsection{FEM setup and material characterisation}

The spinodoid topologies with dimensions $L \times L \times L$ (where $L= 40$ mm) were generated using MATLAB, and FEM simulations were performed using the ABAQUS Explicit solver. \textcolor{black}{Tetrahedral elements were used for meshing, and the structures were compressed along the $x_3$ direction. Due to the symmetry of design parameters, $\theta$, compression in a single direction was sufficient to represent the behaviour across the entire design space, as illustrated in \ref{appendix:Controlling the anisotropy of topologies}.} The simulation setup included two rigid plates: one acting as the loader, which moved until a maximum strain of 50\% was achieved, and the other serving as a stationary anvil. In addition, general contact was defined to simulate interactions between the platens and the structure as well as self-contact, enabling densification and preventing self-interaction during deformation.

Since the spinodoid topologies were fabricated using additive manufacturing, directional dependencies were inherently introduced. These dependencies arose due to the layer-by-layer printing process, which affects the mechanical response depending on the orientation of the layers relative to the loading direction. Capturing this anisotropy in the simulation required an accurate representation of the material properties. This effect was taken into account by utilising engineering constants to describe the linear elasticity in orthotropic materials. This involved calculating three moduli in each principal direction $E_1, E_2, E_3$; Poisson's ratio $\nu_{12}, \nu_{13}, \nu_{23}$; and the shear moduli $G_{12}, G_{13}, G_{23}$. These constants are then used to construct the elastic compliance, which is as follows

\[
\begin{Bmatrix}
\varepsilon_1 \\
\varepsilon_2 \\
\varepsilon_3 \\
\gamma_{12} \\
\gamma_{13} \\
\gamma_{23}
\end{Bmatrix}
=
\begin{bmatrix}
\frac{1}{E_1} & -\frac{\nu_{12}}{E_2} & -\frac{\nu_{13}}{E_3} & 0 & 0 & 0 \\
-\frac{\nu_{12}}{E_1} & \frac{1}{E_2} & -\frac{\nu_{23}}{E_3} & 0 & 0 & 0 \\
-\frac{\nu_{13}}{E_1} & -\frac{\nu_{23}}{E_2} & \frac{1}{E_3} & 0 & 0 & 0 \\
0 & 0 & 0 & \frac{1}{G_{12}} & 0 & 0 \\
0 & 0 & 0 & 0 & \frac{1}{G_{23}} & 0 \\
0 & 0 & 0 & 0 & 0 & \frac{1}{G_{13}} \\
\end{bmatrix}
\begin{Bmatrix}
\sigma_1 \\
\sigma_2 \\
\sigma_3 \\
\tau_{12} \\
\tau_{13} \\
\tau_{23}
\end{Bmatrix}
\]

 \textcolor{black}{In addition, Hill anisotropic plastic model was used to describe the anisotropic plastic deformations, exhibited as a result of printing direction during manufacturing. The quadratic Hill yield model \cite{hill1948theory} is as follows:}

\begin{equation}
    \textcolor{black}{\sqrt{F(\sigma_{22}-\sigma_{33})^2+G(\sigma_{33}-\sigma_{11})^2+H(\sigma_{11}-\sigma_{22})^2+2L\sigma^2_{23}+2M\sigma^2_{31}+2N\sigma^2_{12}} = f(\mathbf{\sigma})}
\end{equation}

\begin{equation}
    \textcolor{black}{F = \frac{1}{2}\left[ \frac{1}{R_{22}^2} + \frac{1}{R_{33}^2} - \frac{1}{R_{11}^2}\right] \qquad L = \frac{3}{2R^2_{23}}}
\end{equation}

\begin{equation*}
    \textcolor{black}{G = \frac{1}{2}\left[ \frac{1}{R_{33}^2} + \frac{1}{R_{11}^2} - \frac{1}{R_{22}^2}\right] \qquad M = \frac{3}{2R^2_{13}}}
\end{equation*}

\begin{equation*}
    \textcolor{black}{H = \frac{1}{2}\left[ \frac{1}{R_{11}^2} + \frac{1}{R_{22}^2} - \frac{1}{R_{33}^2}\right] \qquad N = \frac{3}{2R^2_{12}}}
\end{equation*}

\textcolor{black}{where $R_{11} = \frac{\sigma_{11}^y}{\sigma^0}, R_{22} = \frac{\sigma_{22}^y}{\sigma^0}, R_{33} = \frac{\sigma_{33}^y}{\sigma^0}, R_{12} = \frac{\sigma_{12}^y}{\tau^0}, R_{13} = \frac{\sigma_{13}^y}{\tau^0}, R_{23} = \frac{\sigma_{23}^y}{\tau^0}$, with $\tau_0 = \frac{\sigma_0}{\sqrt{3}}$. $R_{ij}$ represents the anisotropic yield stress ratios, with $\sigma^y_{ij}$ being the yield stress values when only $\sigma_{ij}$ has been applied. Here, $\sigma_0$ is a user-defined yield stress reference.}

\begin{figure}[!h]
  \centering
   \includegraphics[width=1\textwidth]{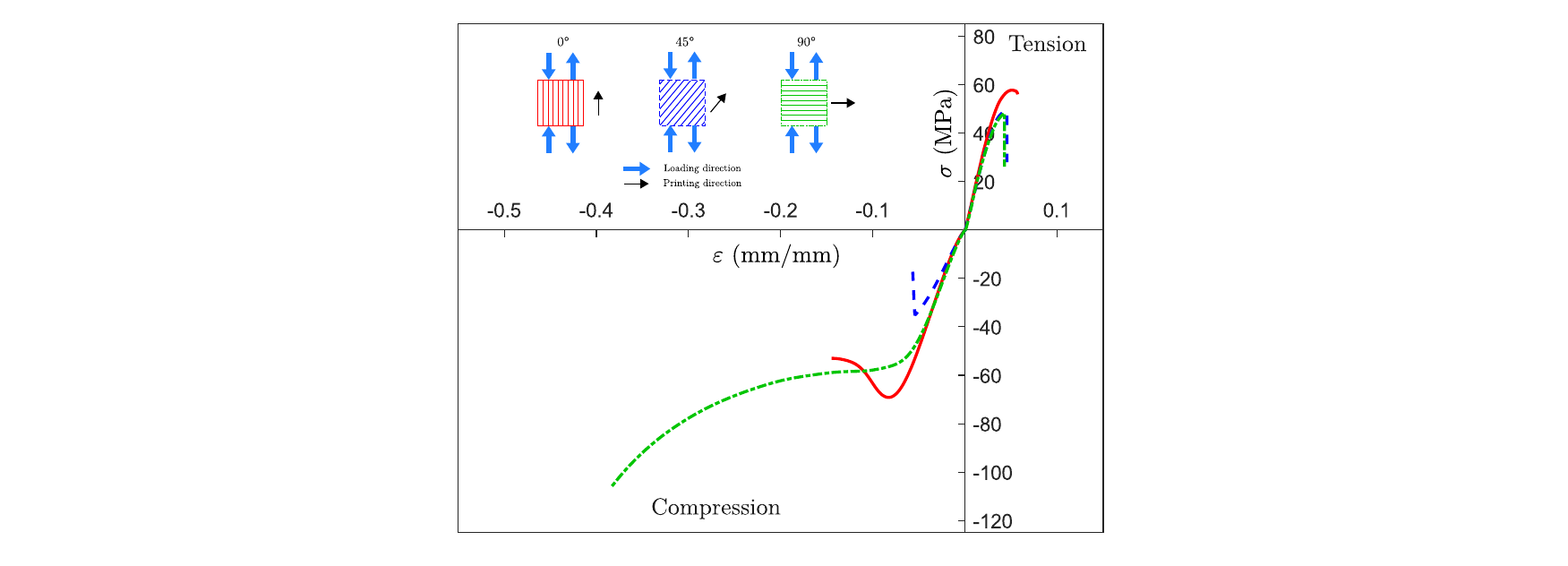}
   \caption{\label{fig:PET-G experiments tension and compression}Results of material characterisation tests for Polyethylene terephthalate glycol (PET-G) under compression and tension are presented. Negative values indicate compression, while positive values indicate tension. The solid red line, dashed blue line, and dash-dotted green line represent specimens printed at $0^\circ$, $45^\circ$, and $90^\circ$, respectively. The subfigures in the top left illustrate the general printing directions used for manufacturing the test samples, with black arrows indicating the printing angle from the horizontal and blue arrows showing the loading directions.}
\end{figure}

To account for these effects, material properties were obtained through material characterisation tests conducted in three directions: along the printed layers, orthogonal to it, and at a $45^\circ$ angle in the same plane to obtain the shear behaviour. Tensile and compressive tests were performed on \textcolor{black}{sample type 5A of the ISO-527 standards} and $20 \times 20 \times 20$ $\text{mm}^3$ cubes, respectively, with the smaller cube size selected to avoid requiring a high-capacity load cell for compression tests. Both quasi-static compression and tensile tests were conducted at a strain rate of $0.1~s^{-1}$ using an INSTRON-5967 machine. Test specimens were fabricated via fused deposition modelling on an UltiMaker S5, printed at three orientations: $0^\circ$, $45^\circ$, and $90^\circ$, representing the directions along the loading axis to obtain material properties. The results of these tests have been graphed in Fig.~\ref{fig:PET-G experiments tension and compression}. 

\textcolor{black}{Material properties were extracted from the experimental data to define the constitutive model for the simulations. The parameters for the Hill anisotropic plasticity model were determined based on stress-strain data obtained from the material characterisation tests. The Young's modulus values were assigned as $E_1 = E_{0^\circ}$, reflecting the behaviour in the plane parallel to the build plate, but perpendicular to the loading direction, while $E_2 = E_3 = E_{90^\circ}$ corresponds to the plane orthogonal to the printing direction, but parallel to the loading direction. Shear moduli were approximated using the relation $G_{12} = G_{13} = G_{23} = E_{45^\circ}/2(1+\nu)$, where $E_{45^\circ}$ is the modulus obtained from the $45^\circ$ sample, with the assumption of uniform Poisson's ratio across orientations at a value of 0.3. Yield strengths were extracted using the 0.2\% offset method: $\sigma_{11}^y = \sigma_{0^\circ}^y$ were obtained for in-plane direction, $\sigma_{22}^y = \sigma_{33}^y = \sigma_{90^\circ}^y$ for out of plane direction, and $\sigma_{12}^y = \sigma_{45^\circ}^y$ was inferred from the $45^\circ$ orientation for the shear strength to represent shear strength was based on its alignment with the principal material directions in the Hill model. In addition, the reference yield stress to obtain the yield stress ratio was set to be $\sigma_0 = \sigma_{11}^y$. The selection of directions for applying these material properties was based on the orientation of the spinodoid geometry slices used for 3D printing the structure, in this case, the geometries were sliced along the $x_1 - x_2$ plane i.e., printed perpendicular to the direction of loading.}

 \textcolor{black}{After determining the material parameters required to construct the Hill anisotropic model using the characterisation tests, the accuracy of the fitted parameters was evaluated through an initial validation test. Specifically, FEM simulation was compared with experimental results from one of the material characterisation tests to ensure the validity and robustness of the model. A $90^\circ$ cube under compression was selected as the benchmark due to its printing orientation, which closely aligned with the fabricated spinodoid topologies, ensuring a representative comparison. Furthermore, the cube geometry was particularly suitable for this validation because the primary performance assessment of the spinodoid topologies in this study is under compressive loading conditions. The use of this setup allowed a direct and meaningful comparison between simulation and experiment, providing confidence in the accuracy of the material model. While the $90^\circ$ cube is not isotropic, its geometry and orientation were specifically chosen to reflect the intended application of the material. This ensures the validation is representative of real-world conditions, particularly the compressive behaviour of spinodoid topologies. The detailed results of this validation test are presented in \ref{appendix:Comparison of material model with material characterisation test}.}

\subsection{Experimental validation}

To validate the FEM model, the four spinodoid structures described in Fig.~\ref{fig:4 spinodoid types} were employed. Similar to the testing samples, the spinodoid structures of dimensions $20 \times 20 \times 20$ mm$^3$ were sliced using UltiMaker Cura, and ultimately additively manufactured using UltiMaker S5. These prototypes were printed using PET-G, with PVA as the support material, which was later removed by immersing the printed samples in water for an extended duration.

Furthermore, the tests were conducted on samples with a relative density of 0.3 and a wave number of $24\pi$. A smaller dimension of spinodoid structures was chosen, along with low relative density to offset the densification, and by extension, not require a higher capacity load cell. Compressive testing was conducted using a quasi-static strain rate of 0.1 $s^{-1}$. This strain rate was carefully chosen to closely match the conditions of the material characterisation test, minimising the influence of strain-rate variations on the results. To ensure the separation of scales, a larger wave number was utilised as a result of using a smaller-sized structure, wave numbers larger than $24\pi$ would result in geometries with features that cannot be resolved by the 3D printer, even with layer height set to a minimal. 

\begin{figure}[!h]
  \centering
   \includegraphics[width=1\textwidth]{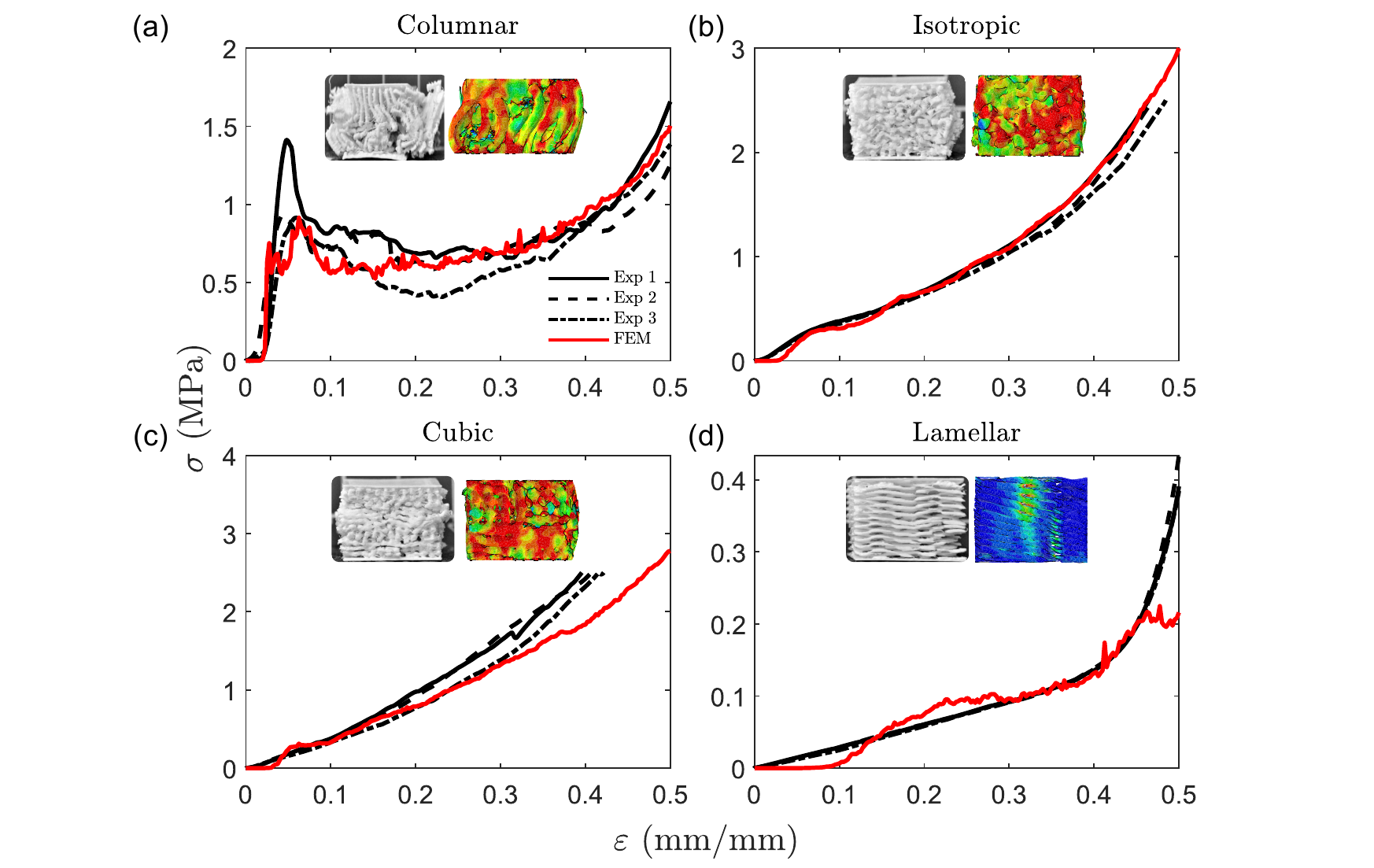}
   \caption{\label{fig:PET-G experimental validation}Comparison of stress-strain behaviour of structures generated using the conical angles described in Fig.~\ref{fig:4 spinodoid types}, undergoing crushing with a relative density of $\rho = 0.3$ and a higher wave number of $\lambda = 24\pi$. The results were obtained from both experiments and FEM analysis. The three different black lines represent results from each repeated experiment, while the solid red line represents data from FEM analysis. The subfigures above each plot show a visual comparison between the 3D-printed structures (left) and the contour plots from FEM analysis (right).}
\end{figure}

Fig.~\ref{fig:PET-G experimental validation} shows the results of the compressive testing of the spinodoid structures. The compressive tests were conducted on three samples for each type of topology to ensure repeatability as well as to consider the variation in crushing behaviour resulting from manufacturing defects. In the figure, the black lines show the experimental results, while the red lines show the results extracted from the FEM simulations. \textcolor{black}{The material constants used to build the constitutive material model were derived from compressive testing data. This choice was made because the spinodoid prototypes were primarily subjected to compressive loading during their intended applications, ensuring that the mechanical behaviour of PET-G aligns more closely with the working conditions.} 

\textcolor{black}{However, directly using the material constants obtained from the material characterisation tests caused an overprescription of stiffness within the structures, resulting in a mismatch between experimental and simulation results. This was remedied by reducing the material constants to values that were 20\% of the original (i.e. $0.2\cdot E$, where $E$ is the value of the originally calculated material modulus, with a similar penalty being applied to the material's yield strength). Introducing this artificial reduction in the stiffness constants accounts for any structural defects and artefacts created during the manufacturing of the prototypes, which may weaken the structural integrity of the structures as a result of improper bonding between adjacent layers.} 

\textcolor{black}{For `Lamellar'-type structures, a unique stiffness penalty was applied to account for their distinctive geometry. These structures, defined by layers oriented perpendicular to the loading direction, required a reduction to 60\% of the initially calculated material constants (i.e., 0.6$\cdot$ E). This adjustment specifically targeted topologies where $0 \leq \theta_{1,2} \leq 5$ and $0 \leq \theta_3 \leq 45$, as they exhibited similar layer-like characteristics. Again, an equivalent penalty was applied to the yield strength for that particular orientation.}

\textcolor{black}{The variation in stiffness correction factors across different topologies highlights differences in defect levels caused by the printing orientation. Geometries with their primary features aligned parallel to the print bed generally display fewer defects, as the larger surface area allows subsequent layers to form stronger bonds. In contrast, structures with smaller features are more prone to defects. Additionally, the printing orientation significantly influences the mechanical behaviour of the structures highlighted by the need to use anisotropic material model in FEM simulations. For instance, `Columnar'-type structures with columns aligned parallel to the print bed require less reduction in material parameters compared to those with columns oriented perpendicular to the bed. This disparity stems from the printer's inability to resolve small features, which significantly affects the structural integrity of the spinodoid prototypes unless a nozzle with a very small diameter is utilised. As a result, applying varying stiffness correction factors based on geometry enables the consideration of real-life printing defects.}

\textcolor{black}{Nonetheless, the difference between the experiment and the FEM simulations can be a result of a number factors. Firstly, this could be a result of fracture being a pervasive mode of failure, which is not considered in the current material model, where only elasticity and anisotropic plasticity are considered. This is especially apparent in the case of `Columnar'-type structure. Additionally, the difference can be caused by the inherent assumptions made when calculating the material parameters for the anisotropic Hill model. This difference can be clearly seen in \ref{appendix:Comparison of material model with material characterisation test}, whereby the initial elastic strain with elastic constants being directly calculated from the particular stress-strain response can lead to a lower stiffness than initially prescribed, which could be a result of the initial FEM configuration. This could be one of the reasons why there is a small difference in the initial elastic response between the experiment and FEM simulations for the spinodoid structures. Nonetheless, a mixture of mass scaling and time scaling were also utilised to reduce computation time, which can also contribute to the initial non-linear response in the stress-strain response obtained from FEM simulation at small strains, which has been demonstrated in \ref{appendix:Effect of time-scaling and mass-scaling on stress-strain behaviour}. On the other hand, the disparity in large strain scale (i.e. $\varepsilon > 0.4$) for some structures could be a result of reduced porosity in the 3D-printed spinodoid prototypes as a result of manufacturing defects such as stringing, prohibiting water from permeating the structure, allowing the PVA support material to dissolve, resulting in faster densification to occur at high strain values especially in the case of `Cubic' structure due to its unique geometry. Furthermore, the discrepancy between experimental results and FEM simulations can also be attributed to the fact that material parameters were derived solely from compression tests. Given that PET-G exhibits significantly different behaviour under tension and compression, addressing this disparity would require the implementation of advanced material models, such as those defined through User-Defined Material Models (UMAT/VUMAT).} 

\subsection{Effect of design parameters on performance indicators}
\label{subsection:Effect of design parameters on performance indicators}

EA and PF are two common performance indicators employed to assess the crashworthiness of structures. Following Fig.~\ref{fig:energy_abs_peak_force_diagram}, the EA is calculated by integrating the F-D curve. The PF, denoted by $P_{max}$, is identified at the end of the linear elastic region, but at the beginning of the progressive crushing zone. Mathematically, the following equations can be used to extract the values of EA and PF,

\begin{equation}
    \text{EA} = \int_0^{\delta_{max}}Pd\delta \qquad \text{PF} = \max_{0 \leq \delta \leq 0.2}( P(\delta)) 
\end{equation}

where the EA is obtained by integrating the Force, $P$, and, Displacement, $\delta$, curve, between the interval $[0, \delta_{max}]$, where $\delta_{max}$ is the maximum displacement the structure has been crushed. Additionally, the PF is extracted by finding the maximum value of $P$, between the displacement intervals $0 \leq \delta \leq 0.2L$. This range has been chosen after studying the most common area at which the highest value of force is found for several structures. For better comparability, the performance indices have been normalised using the EA and PF values obtained from simulating the behaviour of a solid cube block of the same dimensions and material under compressive load. Mathematically, this is expressed as $\overline{\text{EA}} = \text{EA}/\text{EA}_s$ and $\overline{\text{PF}} = \text{PF}/\text{PF}_s$, where $\text{EA}_s$ and $\text{PF}_s$ are the energy absorption and peak force values, respectively, of the solid cube. 

\begin{figure}[!h]
  \centering
   \includegraphics[width=1\textwidth]{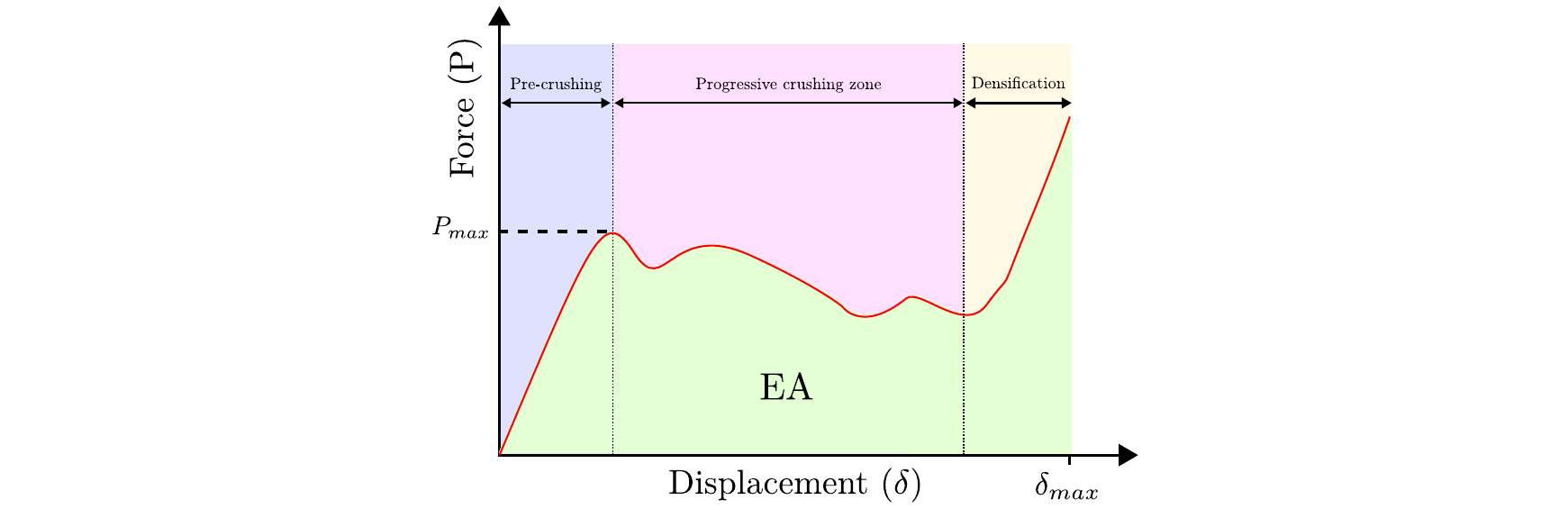}
   \caption{\label{fig:energy_abs_peak_force_diagram}The Force-Displacement (F-D) diagram illustrates the typical behaviour of cellular structures under compressive loading, which can be divided into three stages: i) The pre-crushing stage involves linear elastic deformation occurring until peak force has been attained, followed by ii) the force plateauing as a result of a combination of plastic deformation, and fracture, this is called the progressive crushing zone, iii) the final phase involves the structure undergoing densification signified by an exponential increase in the crushing load required for further deformation. }
\end{figure}

\begin{figure}[!h] % !htbp
  \centering
   \includegraphics[width=1\textwidth]{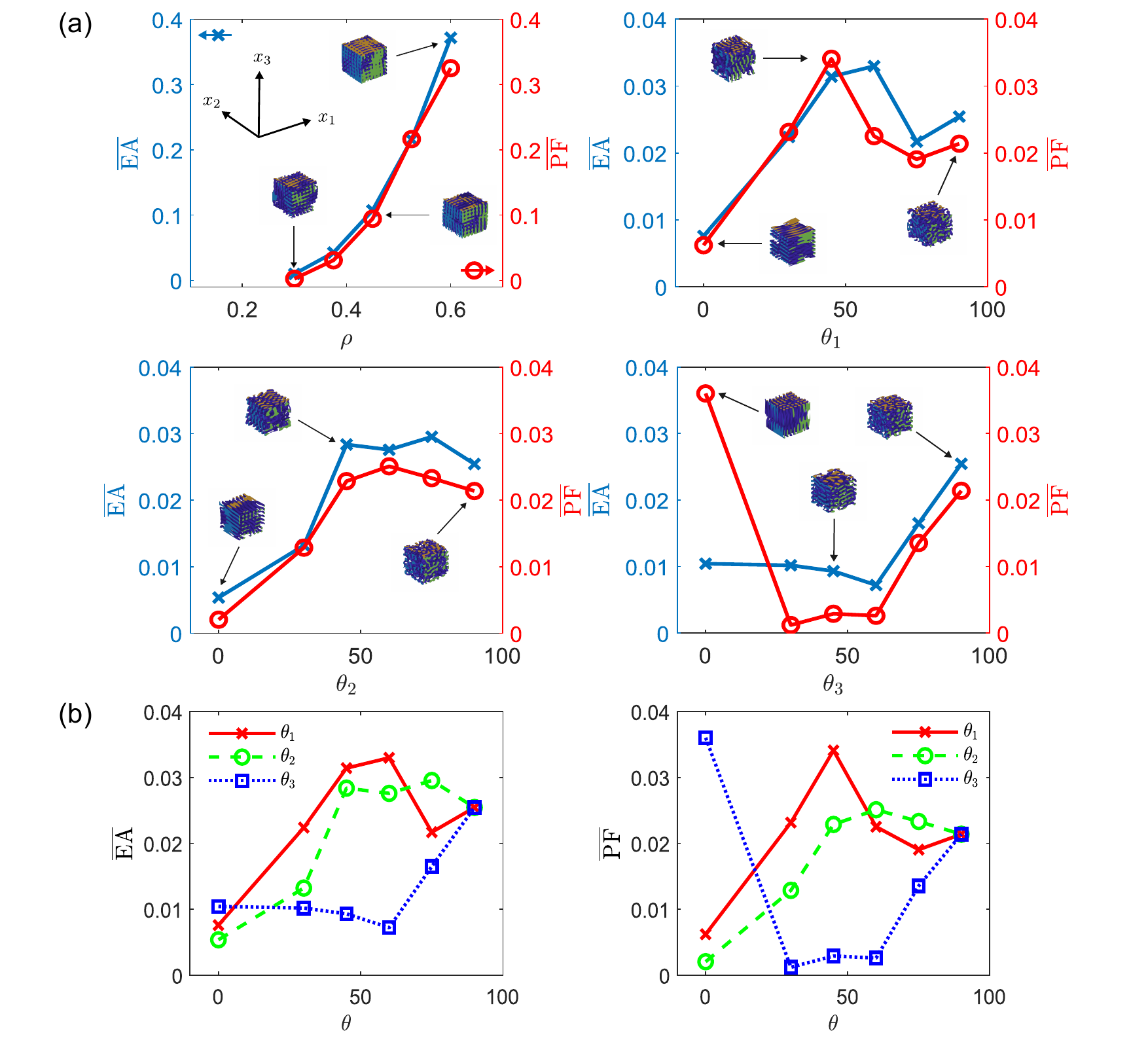}
   \caption{\label{fig:effect_of_parameters_03_15_15_15}(a) Visualisation of the one-at-a-time analysis, demonstrating how varying each individual design parameter affects the performance indicators. The initial design parameters were set at $\rho = 0.3$ and $\theta_1 = \theta_2 = \theta_3 = 15^\circ$. Starting from the top-left and moving clockwise, the results show the effect of varying only the relative density, whilst keeping $\theta_1 = \theta_2 = \theta_3 = 15^\circ$, with $\rho$ values ranging from 0.3 to 0.6. For $\theta_1$, the relative density was held constant at $\rho = 0.3$ and $\theta_2 = \theta_3 = 15^\circ$. The same approach was applied for $\theta_2$ and $\theta_3$. Each subplot includes the topologies generated for specific design parameters. (b) The subplots display the impact on the performance indicators - EA on the left and PF on the right - when varying the different conical angles individually. \textcolor{black}{It should be noted that these structures were also crushed along the $x_3$ direction.}}
\end{figure}

Using the FEM simulations, performance indicators were extracted for a series of Design of Experiments (DoEs), wherein input design parameters were generated by quasi-random sampling of the Sobol' Sequence \cite{bessa2017framework}. This methodology ensures that any set of DoEs effectively cover the span of the available design space. Here, a mesh resolution of 30 and a wave number of $\lambda = 15\pi$ have been chosen (See \ref{appendix:Effect of wave number on performance indicators}, and \ref{appendix:Effect of mesh resolution on stress-strain behaviour of spinodoid structures} for the choice of those specific values). The results of the one-at-a-time analysis have been presented in Fig.~\ref{fig:effect_of_parameters_03_15_15_15}(a), which shows how individual parameters affect the performance indicators, assuming only one parameter is being varied at a time, with initial design being $\rho = 0.3$ and $\theta_1 = \theta_2 = \theta_3 = 15^\circ$. An increase in relative density leads to a rise in both EA and PF, as expected due to the greater volume of solid material. However, the relationship between $\theta$ and the performance indices is complex. For instance, an increase in the performance indicators is observed for $\theta_1$ and $\theta_2$ until a value of 45$^\circ$ has been reached, with a drop in performance indicators for an increase in cone angles beyond 45$^\circ$. In contrast, an increase in performance indices is observed for $\theta_3$ values exceeding 60$^\circ$, whereas similar values of EA and PF are obtained for values below 60$^\circ$. When plotting all cone angles together, as shown in Fig.~\ref{fig:effect_of_parameters_03_15_15_15}(b), $\theta_1$ and $\theta_2$ exhibit similar behaviour, whereas $\theta_3$ shows distinct variations. At a cone angle of 90, EA and PF converge to similar values, which can be attributed to isotropic structures generated at high cone angles. 

The plots in Fig.~\ref{fig:effect_of_parameters_03_15_15_15}(a) were created by varying one design parameter while keeping the other three fixed at arbitrary values. This method provides insight into the behaviour of structures for those particular parameter settings but does not capture the full range of possibilities across the entire design space, where different values could be assigned to the fixed parameters. For a comprehensive understanding of the structure-property relationship, it is vital to consider not just the variation of individual parameters but also their interactions, This is where global sensitivity measures can be utilised. The key purpose of employing global sensitivity analysis is to identify influential model parameters that drive the model output by quantifying their influence. This is used primarily to simplify complex models by way of reducing the dimensions, and by extension, an efficient allocation of resources \cite{saltelli2008global}. This study employs the variance-based sensitivity index also known as Sobol' sensitivity analysis \cite{saltelli2010variance, sobol2001global}, using the open-source Python libraries SALib \cite{herman2017salib} integrated in f3dasm \cite{bessa2017framework,van2024f3dasm}. Sobol' sensitivity analysis quantifies the influence of each parameter and their interactions on the output by decomposing its variance and measuring each parameter's contribution to the total variance. The sensitivity indices used to define these contributions numerically include the first-order index (S1), which is a measure of the contribution of a single standalone parameter input to the output variance; the second-order index (S2) measures the contribution of the interaction between two parameter inputs on the output variance; and the total-order index (ST) measures the output variance caused by a parameter input, which includes the first-order effects and higher-order interactions \cite{saltelli2010variance}. Computing the sensitivity indices involves calculating the variances using analytical integrals, but it may not be possible to solve these integrals. Alternative methods involve using Monte Carlo (MC) approximations of the integrals. These approximations are produced by sampling a set of input parameters with $n$ number of samples, with approximation accuracy improving with higher values of $n$. Two matrices ($\mathbf{A}$ and $\mathbf{B}$) of dimensions $n \times p$-matrices are created by sampling the Sobol' sequence with $p$ input dimensions. Cross-sampling values from the two matrices result in requiring a total of $n\cdot(p+2)$ evaluations of the objective function \cite{saltelli2002making}. As an example, consider two points sampled using the Sobol' sequence within the spinodoid design space with four input parameters $(\rho, \theta_1, \theta_2, \theta_3)$, the resulting combinations are shown in Table \ref{tab:cross-sampling_sobol_sequence_table}. Subsequently, these points are utilised by MC estimators to approximate the variance.
\begin{table}[ht]
\centering
\begin{tabular}{|c|c|c|c|}
    \hline
    $\rho$ & $\theta_1$ & $\theta_2$ & $\theta_3$ \\
    \hline
    \cellcolor{red!15} 0.327 & \cellcolor{red!15} 52.3 & \cellcolor{red!15} 36.697 & \cellcolor{red!15} 89.159 \\
    \hline
    \cellcolor{blue!15} 0.587 &  \cellcolor{red!15} 52.3 & \cellcolor{red!15} 36.697 & \cellcolor{red!15} 89.159 \\
    \hline
    \cellcolor{red!15} 0.327 & \cellcolor{blue!15} 57.096 & \cellcolor{red!15} 36.697 & \cellcolor{red!15} 89.159 \\
    \hline
    \cellcolor{red!15} 0.327 & \cellcolor{red!15} 52.3 & \cellcolor{blue!15} 48.099 & \cellcolor{red!15} 89.159 \\
    \hline
    \cellcolor{red!15} 0.327 & \cellcolor{red!15} 52.3 & \cellcolor{red!15} 36.697 & \cellcolor{blue!15} 16.343 \\
    \hline
    \cellcolor{blue!15} 0.587 & \cellcolor{blue!15} 57.096 & \cellcolor{blue!15} 48.099 & \cellcolor{blue!15} 16.343 \\
    \hline
\end{tabular}
\caption{An example of cross-sampling involving taking two primary points sampled from the Sobol' sequence in four dimensions ($\rho, \theta_1, \theta_2, \theta_3$) and creating various combinations by selecting individual elements from each dimension of the sampled points. This results in a set of mixed points. In the table, each row represents a unique point to be evaluated. The first (red) and last (blue) rows correspond to the two original points from the Sobol' sequence. The intermediate rows, containing a mix of red and blue elements, represent the combinations of elements taken from the two primary points. The set of points are then evaluated using the model to form the sensitivity indices.}
\label{tab:cross-sampling_sobol_sequence_table}
\end{table}

\begin{figure}[!h]
  \centering
   \includegraphics[width=1\textwidth]{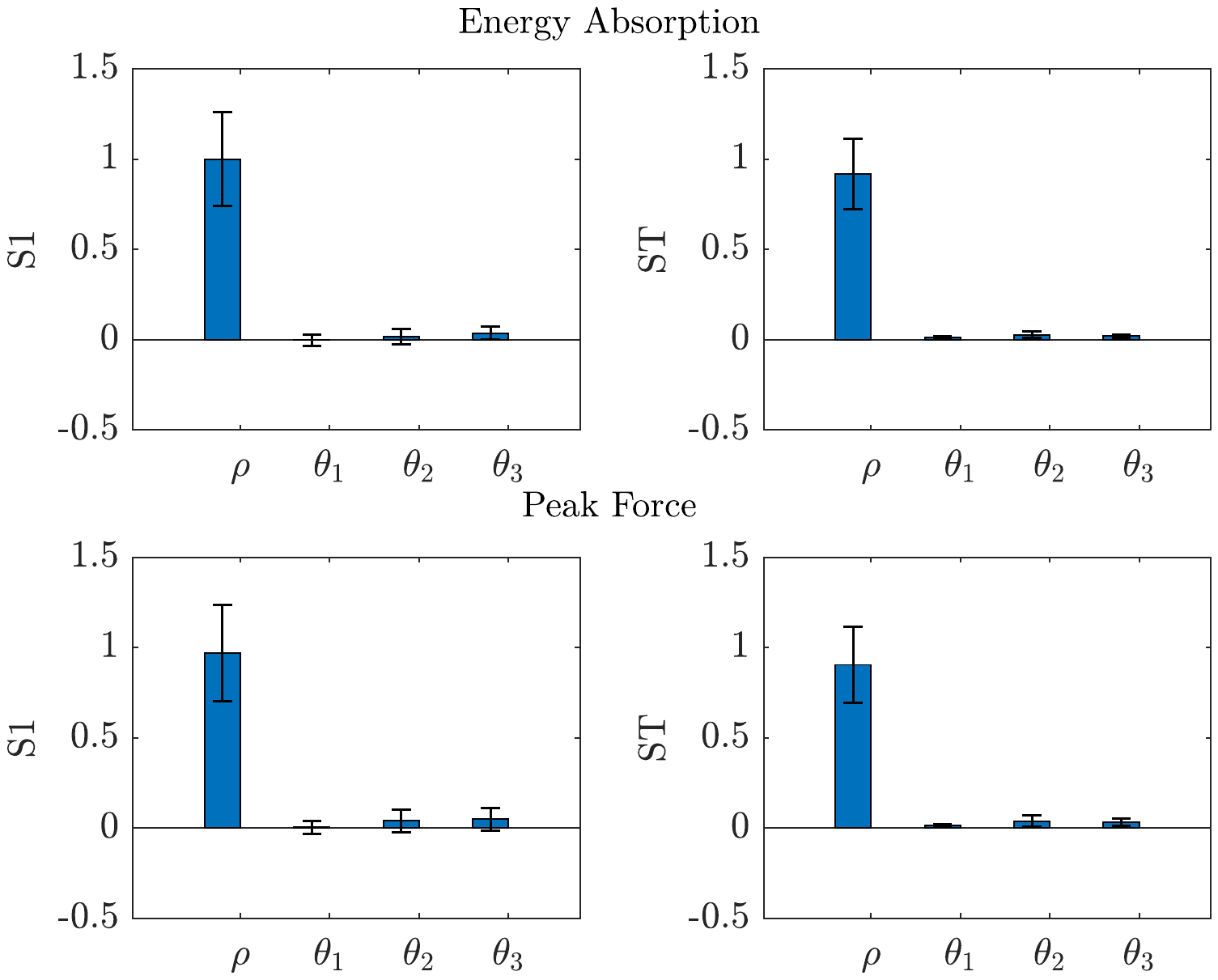}
   \caption{\label{fig:sensitivity_analysis}\textcolor{black}{Bar charts illustrating the results of two independent Sobol' sensitivity analyses for two performance indicators, EA and PF. These analyses were conducted using four inputs $(\rho, \theta_1, \theta_2, \theta_3)$ with 474 model evaluations and the results are represented as sensitivity indices: S1 (First-order) and ST (Total-order), along with their confidence bounds. The results reveal that $\rho$ exhibits extremely high sensitivity, while the $\theta$ inputs show nearly negligible sensitivity.}}
\end{figure}

In this study, two independent sensitivity analyses were conducted for each objective. Fig.~\ref{fig:sensitivity_analysis} shows the results of applying Sobol' sensitivity analysis to the spinodoid design space that is composed of four input parameters $(\rho, \theta_1, \theta_2, \theta_3)$. Consequently, each pair of Sobol' sequence points sampled necessitates the evaluation of six different points, as exemplified in Table \ref{tab:cross-sampling_sobol_sequence_table}. A total of 474 function evaluations were required to achieve converged sensitivity indices. The progression of these indices with the number of evaluations is depicted in \ref{fig:sensitivity_analysis_4var_convergence} with four design parameters, although further increasing the number of evaluations introduces numerical instabilities. The bar charts in Fig.~\ref{fig:sensitivity_analysis} indicate a significant relationship between relative density and the performance indicators, with both EA and PF increasing as relative density increases. This is evidenced by the sensitivity indices S1, and ST, which are approximately one for relative density, while the indices for the remaining design parameters are close to zero. These results indicate that relative density is the most influential parameter on the performance indicators, while the other input parameters contribute very little. However, this does not imply that the different input parameters should be disregarded, as they play a critical role in influencing the anisotropy of structure, which is a key determinant of the performance of topologies under compression. 

To gain a deeper understanding of the behaviour of conical angles and their influence on performance indicators, an additional Sobol' sensitivity analysis was conducted with three input parameters $(\theta_1, \theta_2, \theta_3)$, while keeping the relative density constant at $\rho = 0.3$. Reducing the number of input variables decreases the number of evaluations required for each pair of sampled Sobol' sequence points. Instead of the six total cross-sampled points shown in Table \ref{tab:cross-sampling_sobol_sequence_table}, using three input parameters results in a combination of five points.

\begin{figure}[!h]
  \centering
   \includegraphics[width=1\textwidth]{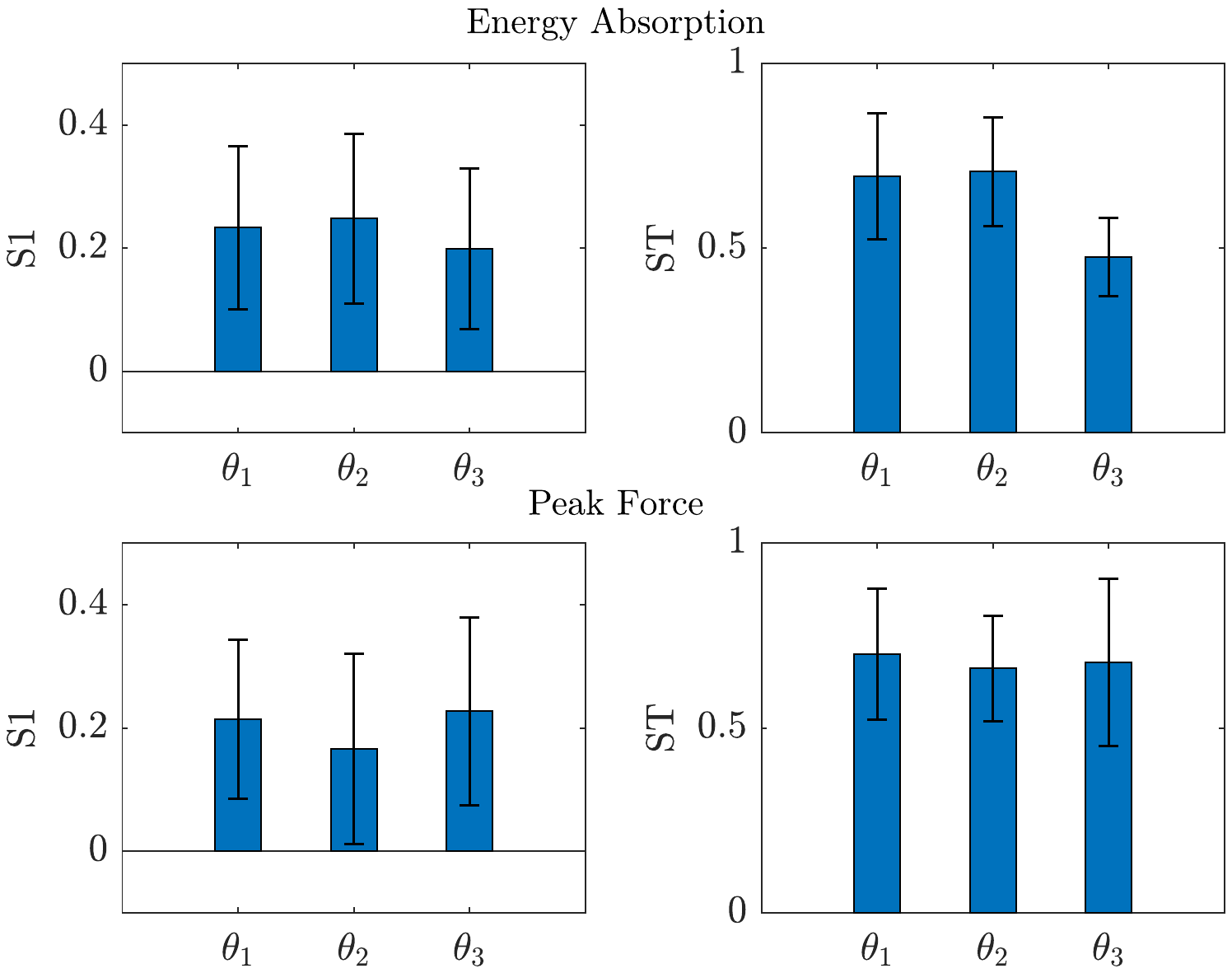}
   \caption{\label{fig:sensitivity_analysis_3var}\textcolor{black}{Bar charts depicting the outcomes of two separate Sobol' sensitivity analyses for the performance indicators EA and PF. These analyses were performed using three inputs $(\theta_1, \theta_2, \theta_3)$ across 1,500 model evaluations while keeping the $\rho$ constant at $\rho = 0.3$. The results, presented as sensitivity indices—S1 and ST—include corresponding confidence bounds, showing an even distribution of sensitivities among the inputs $\theta$.}}
\end{figure}

A total of 1,500 points were used to evaluate the Sobol' sensitivity indices, increasing in increments of 5 points. The evolution of the sensitivity indices as the number of Sobol' sequence samples increases is illustrated in \ref{fig:sensitivity_analysis_3var_convergence}. Converged values of S1 and ST were obtained at approximately 1,200 evaluations, beyond which additional data had minimal effect on the indices. However, the sensitivity indices, along with their confidence bounds, are shown in Fig.~\ref{fig:sensitivity_analysis_3var} with a total of 1,500 evaluations. Unlike the previous sensitivity analysis, markedly different behaviour is observed. Focusing on the first-order sensitivity index, it is evident that all three inputs directly influence the performance indicators, with approximately 20\% of the output variance being explained by the variance in a single input alone.

On the other hand, ST indices indicate an output variance of 70\% of the output variance is attributed to most of the input parameters for both objectives. It should be noted that the total-order index is a metric that combines both the direct effects and the interaction effects among inputs. Consequently, the difference between ST and S1 represents the output variance explained through input interactions. This means that approximately 50\% of the output variance is contributed by interactions for most inputs. Intuitively, this suggests that a maximal change in the objectives would be observed if a combination of $\theta$ were varied. Therefore, it is essential to utilise all conical angles, as this would allow for a broader design space to be explored with greater variability in performance indicators.

\section{Multi-Objective Bayesian Optimisation (MOBO)}

Bayesian optimisation (BO) is a sequential global optimisation strategy used to tackle expensive-to-evaluate black-box functions. This optimisation strategy is based on Bayes' theorem, which provides a way to update probabilities of a hypothesis (or model), H, given evidence (data, or observations), E, is proportional to the likelihood of E given H, multiplied by the prior probability of H. Mathematically,

\begin{equation}
    \text{P}(\text{H}|\text{E}) = \frac{\text{P}(\text{E}|\text{H})\text{P}(\text{H})}{\text{P}(\text{E})}
\end{equation}

In the context of BO, the prior represents the initial assumptions (such as smoothness, noise in observations, etc) about the objective function, $f$. The likelihood here quantifies how likely the observed data, O, is, given the prior, whereby the observed values are sampled data points of the true objective function. The combination of the prior with the evidence allows us to obtain the posterior distribution. 

In MOBO, the task involves concurrently optimising multiple objective functions. Mathematically, this can be expressed as 
\begin{equation}
\text{max}(f_1(\hat{\textbf{x}}), f_2(\hat{\textbf{x}}), ..., f_p(\hat{\textbf{x}})) \; \;\;\; \hat{\textbf{x}} \in \mathbb{R}^q
\end{equation}
where $p$ is the number of objective functions, $f_i(i=1,...,p)$ are the objective functions, and $\hat{\textbf{x}}$ are the independent design parameters in $q$ dimensions. This assumes that the problem requires the maximisation of the objectives. Conversely, the problem can be transformed into a minimisation problem by negating the objective functions. In reality, these objectives may be conflicting, leading to a situation where it is improbable to maximise all objectives simultaneously, giving rise to the concept of Pareto optimality. A solution is considered Pareto optimal if no other solution would make any objective function better without compromising at least one other objective. By considering Pareto optimality, the trade-offs between conflicting objectives can be navigated, ultimately identifying a set of Pareto-optimal solutions. In the context of MOBO, Pareto-optimal solutions can also be referred to as non-dominated solutions within the solution space. Conversely, a solution is considered dominated if there exists another solution that surpasses it in at least one objective without worsening any other objective. The approaches to solve MOBO problems can be classified depending on whether weights are assigned to the objectives beforehand (known as scalarisation methods) or not (known as hypervolume (HV) methods). Regardless of the methods used, in MOBO, each objective function is approximated by an independent GP, which is used to represent the posterior distribution.  

\subsection{Gaussian Processes}
A GP defines a probability distribution over functions, such that for any set of inputs in a continuous domain, the corresponding output values can be modelled as a joint Gaussian distribution \cite{rasmussen2003gaussian}. They are characterised by a mean function $m(\textbf{x})$, and a covariance function $k(\textbf{x}, \bar{\textbf{x}})$, defined in terms of input vectors \textbf{x}, and $\bar{\textbf{x}}$. For simplicity, the mean function can identically be zero, without losing generality \cite{books/lib/RasmussenW06}. 

\begin{equation}
    f(\textbf{x}) \sim \mathcal{GP}(m(\textbf{x}), k(\textbf{x},\textbf{x}*))
\end{equation}

GPs can be trained using a given set of observations, $O = \left \{ \textbf{x}_i, \text{y}_i \right \}^N_{i = 1}$, where $\textbf{x}_i \in \mathbb{R}^q$, given $N$ number of observations, and in a $q$ dimensional space. \textcolor{black}{ In addition, assuming noisy observations, $\text{y}_i = f(\textbf{x}_i) + \epsilon_i \in \mathbb{R}$, where the noise, $\epsilon \sim \mathcal{N}(0, \bar{\sigma^2_n})$, is modelled as mean-zero Gaussian noise with fixed variance $\bar{\sigma^2_n}$. This approach simplifies the noise modelling by treating the results from the FEM simulations as the `ground truth'. However, in practice, the noise term may not have a mean of zero ($\mu \neq 0$), as it encompasses two types of uncertainty: aleatoric (data) uncertainty and epistemic (model) uncertainty. Aleatoric uncertainty arises from random variations in observations, such as sensor noise or experimental imperfections, and is inherently irreducible. Epistemic uncertainty, on the other hand, stems from limitations or approximations in the computational models or objective function calculations, which may not fully capture real-world conditions due to idealised assumptions or incomplete knowledge of the system. As a result, this work focuses on aleatoric uncertainty and models the noise as mean-zero Gaussian noise ($\epsilon \sim \mathcal{N}(0, \bar{\sigma^2_n})$) for simplicity. To better address epistemic uncertainty, the noise model could be refined using prior knowledge, such as FEM validation results (e.g., Fig.~\ref{fig:PET-G experimental validation}), to determine the noise parameters (mean and variance) more accurately.}  

Consequently, the Gaussian distribution can then be obtained using 
\begin{equation}
    \textbf{y} \sim \mathcal{N}(\mu, \mathbf{K}_{ij})
\end{equation}

where $\mu_i = m(\textbf{x}_i), i = 1, ..., n$, and $\mathbf{K}_{ij} = k(\textbf{x}_i, \textbf{x}_j), i,j = 1, ..., n$, forming the prior distribution. This distribution is independent of the training data but assumes a smooth objective function. Next, updating the prior requires the joint Gaussian distribution to be formed, assuming $\textbf{y} = \left \{\text{y}_i \right \}^N_{i = 1}$ is the observed output of the training dataset, let $\textbf{y}_* = \left \{\text{y}_i^* \right \}^N_{i = 1}$ be the outputs from the test set, with $\textbf{X}$ being the input for the training set, and $X^*$ being the input for the test set.

\begin{equation}
    \begin{bmatrix}
        \mathbf{y}\\
        \mathbf{y}_*
    \end{bmatrix}
    \sim
    \mathcal{N}\left(
    \mathbf{0},
    \begin{bmatrix}
        K_X = k(\textbf{X}, \textbf{X}) + \bar{\sigma^2_n} \textbf{I} & k(\textbf{X}, X^*)\\
        k(\textbf{X}, X^*)^T & k(X^*, X^*)
    \end{bmatrix}
    \right)
\end{equation}

The predictive distribution over the test points is then given by

\begin{equation}
    \textbf{y}_*|\textbf{y} \sim \mathcal{N}(k(\textbf{X}, X^*)^TK_X\textbf{y}, k(X^*, X^*) - k(\textbf{X}, X^*)^TK_X^{-1}k(\textbf{X}, X^*))
\end{equation}

The mean and variance at a given input location $\textbf{x}^*$, with a training set $X$ can then be obtained using

\begin{align}
    \mu(\mathbf{x}^*) &= k(X, \mathbf{x}^*)^T (K_X^{-1} \mathbf{y}) \\
    \bar{\sigma^2}(\mathbf{x}^*) &= k(\mathbf{x}^*, \mathbf{x}^*) - k(X, \mathbf{x}^*)^T (K_X^{-1} k(X, \mathbf{x}^*))
\end{align}

It is essential to choose a covariance function that is representative of the objective function, as it directly influences the output. One such function is the radial basis function (RBF), also known as the squared exponential
\begin{equation}
    k(\textbf{x}', \textbf{x}'') = \theta_0^2 \cdot \exp\left(\ - \frac{\| \textbf{x}' - \textbf{x}''\|^2}{2\ell^2}\right)
\end{equation}

$\ell$ is the length scale, which controls the smoothness of the kernel function. The hyperparameter $\theta_0$ controls the amplitude of the kernel function, whereby it determines the amount of variation introduced by the function being modelled. These are kernel parameters that are optimised by the GP model by maximising the likelihood over the observed dataset, which requires the use of gradient-based optimisers such as L-BFGS-B or Adam. The numerator, $\| \textbf{x}' - \textbf{x}''\|^2$, represents the Euclidean distance between the input points $\textbf{x}'$ and $\textbf{x}''$.  Other kernel functions include the Exponential, Matern kernel functions, etc.

To be able to guide the sequential optimisation, acquisition functions play a pivotal role in effectively navigating the design space and finding the global optimum of the objective function. One example of such a function is the probability of improvement (PI), whereby a point is selected based on the highest probability of improvement over the current maximum \cite{wilson2018maximizing}. Assuming that the function to be maximised is $f(\textbf{x})$ and the current maximum is $f(\textbf{x}^+)$, thus the `improvement' can be defined as

\begin{equation}
    \text{PI}(\textbf{x}) = \text{P}(f(\textbf{x}) > f(\textbf{x}^+)) = \Phi\left ( \frac{\mu(\textbf{x}) - f(\textbf{x}^+) }{\bar{\sigma}(\textbf{x})}\right )
\end{equation}

where $\Phi$ is the cumulative distribution function of the standard normal distribution, and $\bar{\sigma}$ is the standard deviation. By maximising the value of the acquisition function, the best samples can be found that optimise the objective values and reduce the uncertainty. This is an iterative process, allowing the objective space to be efficiently explored by considering the accumulated data.  

\subsection{Scalarisation methods}

With the increased computational effort as a result of an increasing number of objectives, scalarisation methods provide an effective alternative. These methods attempt to convert multiple objectives into a single objective function effectively. Some of the scalarisation techniques include using the weighted sum method \cite{marler2010weighted}, the weighted norm method, used in multi-objective Evolutionary Algorithm based on Decomposition (MOEA/D) \cite{zhang2007moea}, using the Chebyshev function and its extension the Augmented Chebyshev function that is normally used in efficient global optimisation algorithms (ParEGO) \cite{knowles2006parego}. This paper focuses on utilising the ParEGO method to obtain a set of Pareto optimal solutions.

ParEGO assigns weights randomly with a uniform distribution, in our case, sampling a uniform simplex. The ParEGO method works by applying the augmented Chebyshev scalarisation, which is given by:

\begin{equation}
    g(\textbf{y})=\max_i[w_i|f_i-z_i^*|]+\alpha\sum_{i=1}^k|f_i-z_i^*|
\end{equation}
where $w_i$ are weights that are randomly assigned to the $i$th objective, $z_i^*$ is an ideal/utopian objective vector that is unattainable. As the objectives are normalised to values between [0, 1], $z_i^*$ becomes a vector of zeros, and $\alpha = 0.05$. In a bi-objective optimisation problem, $k = 2$. The scaled function $g(\textbf{y})$ is then minimised, with GPs being used to approximate the scaled function. 

An alternative method is the weighted sum, whereby random weights are assigned to the objectives and combined to form a singular objective function. Mathematically, this can expressed as

\begin{equation}
    g(\mathbf{y}) = \sum_{i=1}^{k}w_if_i
\end{equation}

which is similar to the formulation of the Chebyshev scalarisation, but this method also considers the magnitude of all objectives.

Subsequently, both scalarisation techniques are used with a singular GP as a surrogate and an acquisition function such as the Expected Improvement (EI). \textcolor{black}{EI is a widely used acquisition function and one of the earliest methods developed for efficient global optimisation tasks \cite{jones1998efficient, mockus1974bayesian}. However, in this work, an extended version of the EI function is employed to address the issue of vanishing gradients that can arise when optimising improvement-based acquisition functions \cite{ament2023unexpected}. It is important to note that the choice of acquisition function is largely heuristic, serving as a strategy to navigate the design space and map it onto a belief space. This choice offers a practical and efficient approach to navigating the design space, particularly in cases like this, where the influence of design variables on the objective functions (energy absorption and peak force) varies significantly.} In this study, the log-EI function has been utilised, which is an extension of the EI function, expressed as 

\begin{equation}
    \text{logEI}(\textbf{x}) = \log(\mathbb{E}(\max(f(\textbf{x}) - f(\textbf{x}^+),0)))
    \label{equ:log ei}
\end{equation}

This is derived from the probability of improvement acquisition function. Finding the expectation of the results (i.e.~multiplying by the probability of the outcomes) in the formulation of the expected improvement acquisition function. The logarithm function ensures numerical stability, which is especially important in automatic differentiation.

\subsection{Hypervolume methods}
\label{subsection: hypervolume methods}
\begin{figure}[!ht]
  \centering
   \includegraphics[width=1\textwidth]{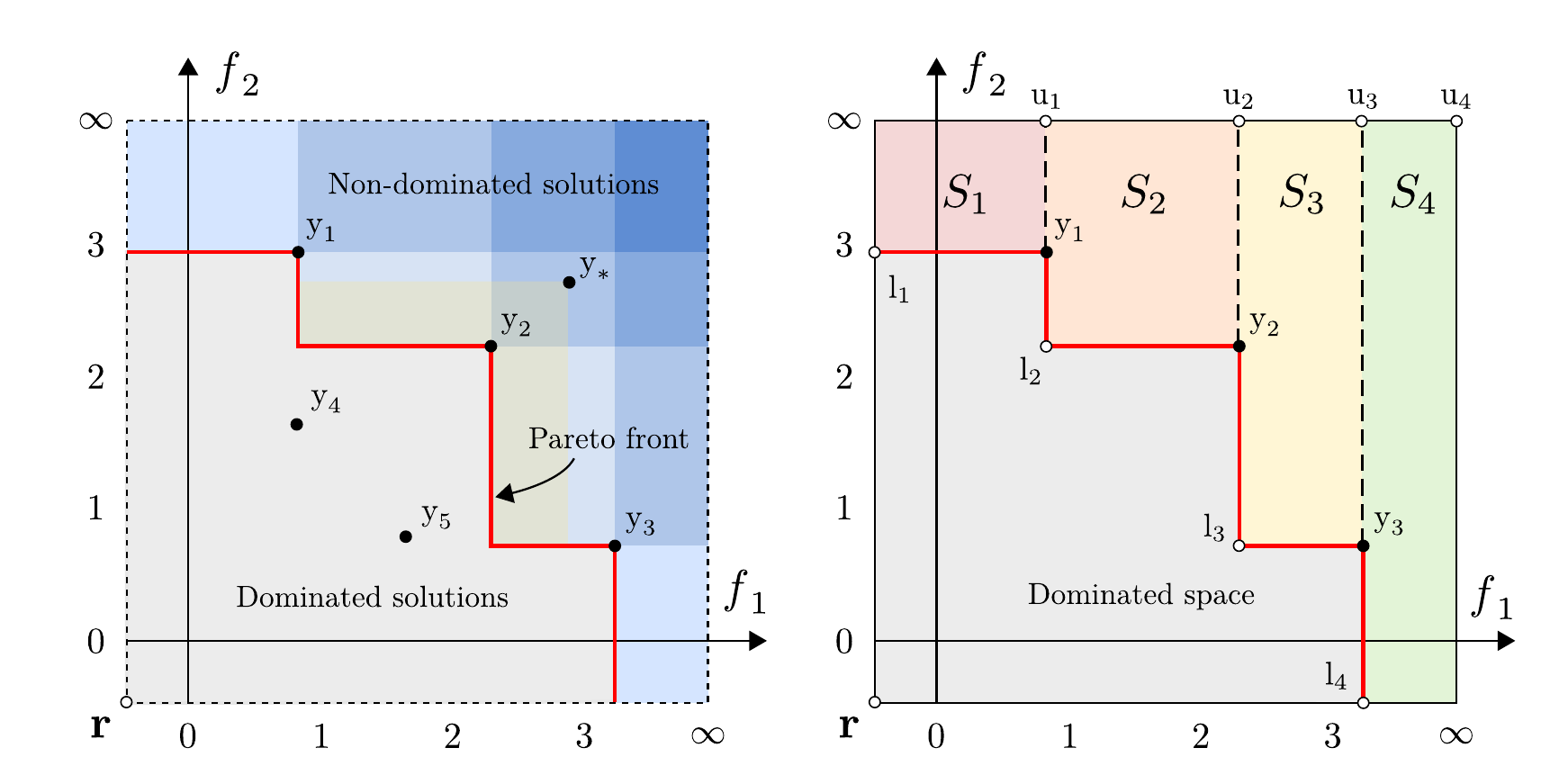}
   \caption{\label{fig:pareto_front_and_hv_explanation} (a) Illustration of Hypervolume Improvement (HVI) in a two-dimensional objective space with objectives $f_1$, and $f_2$. The Pareto front (Red line) $\mathcal{P}$ consists of a set of evaluated points $\{ \text{y}_1,  \text{y}_2, \text{y}_3\}$, which are the non-dominated solutions. The dominated solutions, represented by $\{ \text{y}_4,  \text{y}_5\}$, lie within the dominated solutions region. The yellow region indicates the HVI resulting from evaluating a new point $\text{y}_*$. (b) Visualisation of fast partitioning of the non-dominated space for efficient Expected Hypervolume Improvement (EHVI) computation in 2-D using box decomposition. Each disjointed rectangle is labelled $S_i$, and is defined by its upper and lower vertices $u_i$ and $l_i$. Finally, $\mathbf{r}$ refers to the reference point from which the hypervolume indicator is calculated.}
\end{figure}

However, for MOBO, more complex acquisition functions are required to traverse the design space. While the computation of these MOBO acquisition functions is expensive, it allows effective exploration across the Pareto front. In the hypervolume approaches, the Hypervolume Indicator (HV) is utilised, which is used to evaluate the quality of a Pareto front approximation set, $\mathcal{P}$. This approach measures the size of the dominated space bounded from below by a reference point \textbf{r} \cite{zitzler1999multiobjective}. This reference point is generally chosen by the user, in a manner that satisfies the condition that it is dominated by all objectives, i.e.~a nadir point. Mathematically, the HV of a set $\mathcal{P}$, where $\mathcal{P} = \{ \textbf{y}_{(1)},...,\textbf{y}_{(i)}\} \subset \mathbb{R}^d$, is defined as the $d$-dimensional Lebesgue measure, $\lambda_d$, where $d$ corresponds to the number of objectives, and $i$ corresponds to the number of Pareto optimal points, here a bi-objective problem is being solved, hence $d = 2$, consequently, $\textbf{y} = \{f_1, f_2\}$ with $f_1$ and $f_2$ being two objectives in question:
\begin{equation}
    \text{HV}(\mathcal{P}) = \lambda_2(\cup_{\textbf{y} \in \mathcal{P}}[\textbf{r}, \textbf{y}_i])
\end{equation}
Where r and $\textbf{y}_i$ represent the vertices of a rectangle. By extension, the Hypervolume Improvement (HVI) can be calculated, which considers the improvement in the HV for a new observation, $\textbf{y}_*$:
\begin{equation}
    \text{HVI}(\mathcal{P}, \textbf{y}, \textbf{r}) = \text{HV}(\mathcal{P} \cup  \textbf{y}_*, \textbf{r}) - \text{HV}(\mathcal{P}, \textbf{r})
\end{equation}
Fig.~\ref{fig:pareto_front_and_hv_explanation}(a) shows the improvement in the hypervolume as a result of observing a new value in the objective space. However, to consider the magnitude of the improvement, the expectation of HVI must be calculated, allowing the Expected Hypervolume Improvement (EHVI) to be calculated. This is defined as:

\begin{equation}
    \text{EHVI}(\mathbf{\mu}, \mathbf{\sigma}, \mathcal{P}, \textbf{r}) = \int_{\mathbb{R}^2}\text{HVI}(\mathcal{P}, \textbf{y}, \textbf{r}) \cdot \phi_{\mu,\bar{\sigma}}(\textbf{y}) d\textbf{y}
    \label{equ:EHVI}
\end{equation}
Where $\phi_{\mu,\bar{\sigma}}$ represents the probability density function for a multivariate normal distribution with mean values of $\mu \in \mathbb{R}^d$ and standard deviation of $\bar{\sigma} \in \mathbb{R}^d_+$, as shown by Yang et al.~\cite{yang2019multi}. However, the analytical calculation for the HVI can be computationally expensive. To remedy this, a fast partitioning method can be used, which partitions the non-dominated space into $n + 1$ disjointed rectangles $S_k$ (in 2D), in tandem with piece-wise integration \cite{emmerich2016multicriteria}. Each disjoint rectangle is defined by a pair of upper and lower vertices, $\text{u}_i$, and $\text{l}_i$, respectively, as shown in Fig.~\ref{fig:pareto_front_and_hv_explanation}(b). By adding two additional points: $\textbf{y}_0 = (\text{r}_1, \infty)^T$, and $\textbf{y}_{n+1} = (\infty, \text{r}_2)^T$, where $\textbf{r} = (r_1, r_2)$, then the disjoint rectangles can be defined as 
\begin{equation}
    S_i = \left ( \left ( f_2^{(i)}, -\infty)^T, ( f_1^{(i-1)}, -f_2^{(i)})^T \right ) \right ) \; \; i = 1,...,n+1
\end{equation}

Defining a function $\Delta$, which denotes the subset of objective vectors $\textbf{y}  \in \mathbb{R}^2$, $\Delta(\textbf{y}, \mathcal{P}, \textbf{r})$ are dominated by a vector $\textbf{y}$ but are not dominated by $\mathcal{P}$, but dominate the reference point $r$, as shown by the grey region in Fig.~\ref{fig:pareto_front_and_hv_explanation}, the HVI in 2-D is then given by:

\begin{equation}
    \text{HVI}(\mathcal{P}, \textbf{y}, \textbf{r}) = \sum_{i=1}^{n+1}\lambda_2[S_i \cap \Delta(\textbf{y}, \mathcal{P}, \textbf{r})]
    \label{equ:disjoint rectangles}
\end{equation}
Eqs.~(\ref{equ:EHVI}) and (\ref{equ:disjoint rectangles}) can then be combined for the efficient computation of EHVI acquisition function:

\begin{equation}
    \text{EHVI}(\mathbf{\mu}, \mathbf{\sigma}, \mathcal{P}, \textbf{r}) = \int_{f_1=-\infty}^{\infty}\int_{f_2=-\infty}^{\infty}\sum_{i=1}^{n+1}\lambda_2[S_i \cap \Delta(\textbf{y}, \mathcal{P}, \textbf{r})] \cdot \phi_{\mu,\bar{\sigma}}(\textbf{y}) d\textbf{y}
\end{equation}
Further simplifications can be found in \cite{yang2019multi}.

\section{MOBO Framework}
\label{sec:mobo framework}
The MOBO framework comprises two key components: a Python wrapper that leverages the BoTorch package to perform BO with pre-built functions \cite{balandat2020botorch}, and a MATLAB-ABAQUS interface designed for generating and meshing spinodoid topologies, followed by FEM analysis. Within this framework, multiple MOBO-based methods have been implemented, each addressing the objectives in distinct ways: the hypervolume-based EHVI method, two scalarisation-based approaches - ParEGO and the weighted-sum method - and a comparison with random sampling using the Sobol' sequence.

\subsection{BO setup}
\label{subsec:BO Framework}
\begin{figure}[!h]
  \centering
   \includegraphics[width=1\textwidth]{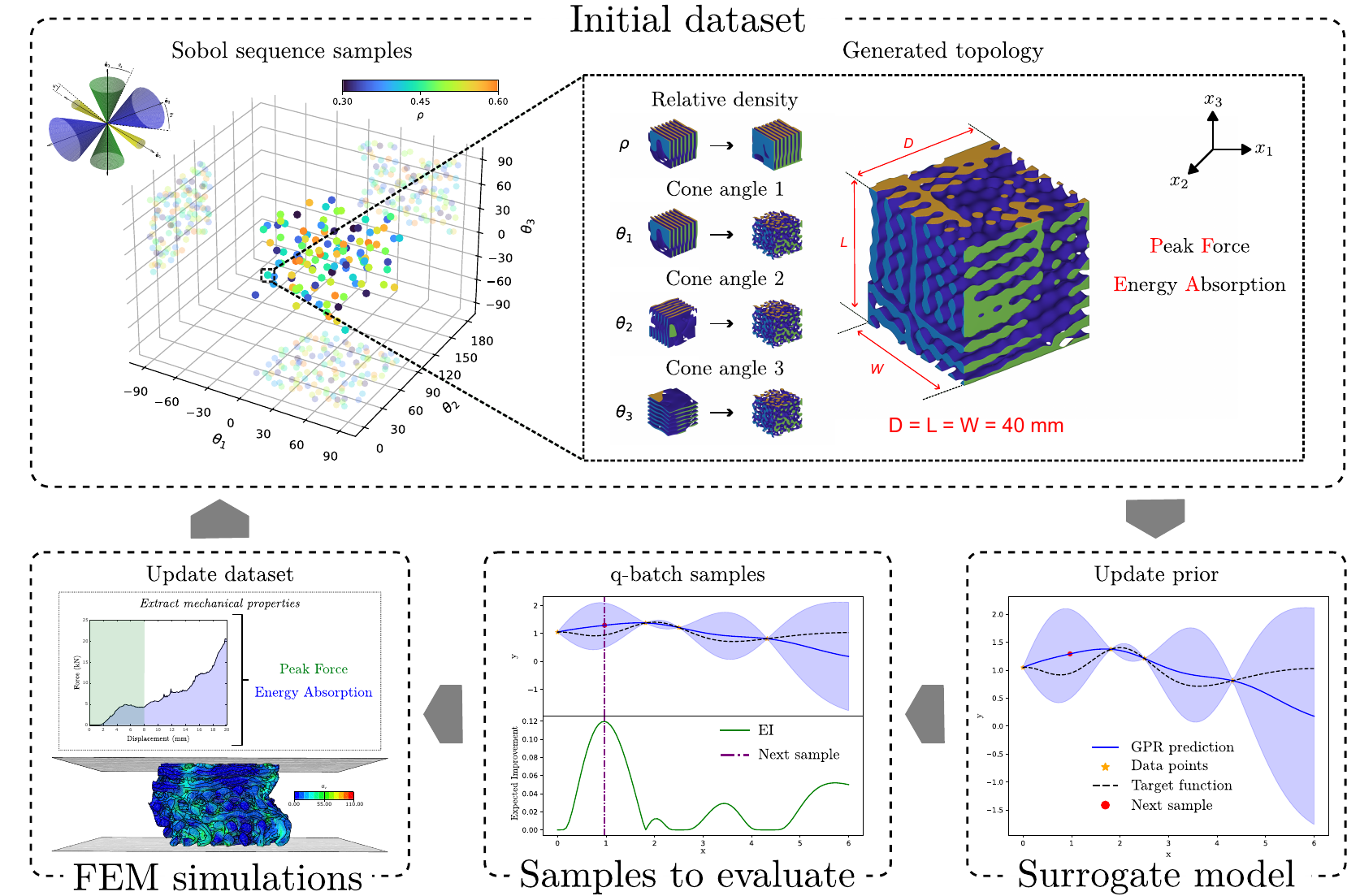}
   \caption{\label{fig:MOBO framework}A schematic of the multi-objective Bayesian optimisation (MOBO) process for optimising the spinodoid cellular structure design space involves four steps of the data-driven process \cite{bessa2017framework,shin2022spiderweb}. (1) Initial Dataset Creation: Sampling 50 points from the design space using the Sobol' sequence and evaluate them with FEM simulations to build the initial dataset. (2) Surrogate Model Update: Updating the Gaussian Process model based on the dataset to predict structural properties. (3) Identifying samples to evaluate: Using an acquisition function to identify and evaluate the most promising design points. (4) FEM Simulations: Performing FEM analysis on generated structures and extracting objectives to then update the dataset.}
\end{figure}

Fig.~\ref{fig:MOBO framework} illustrates the various steps involved in the MOBO framework to generate solutions for the given problem. Initially, independent GPs are configured for each objective using the GPyTorch package, in combination with the multi-output GP function \texttt{ModelListGP}, to establish the prior distribution. The prior is then updated using an initial dataset to form the posterior, thereby creating an initial surrogate model. However, to balance the exploration and exploitation by guiding the optimisation process, acquisition functions are used to obtain input points from the design points to evaluate next. 

BoTorch allows the use of two extensions of the EHVI and logEI (used for scalarisation methods) acquisition functions, specifically qNEHVI and qlogNEI. The qlogNEI acquisition function, which is an extension of the acquisition function as described by Eq.~(\ref{equ:log ei}), is designed to account for noise in observations. In scenarios where functions are not deterministic, a single best value cannot be determined, qlogNEI integrates over all possible values to identify the next best point for sampling \cite{ament2024unexpected}. Similarly, qNEHVI assumes that the points observed on the Pareto front represent noisy realisations of the true Pareto front. Consequently, qNEHVI and qlogNEI require integration over the uncertainty of the in-sample (observed) data introduced. In addition, the two extensions support parallel evaluation of q-batch candidate points. Each candidate is selected considering the joint improvement from multiple candidate points. Moreover, asynchronous evaluation is possible, allowing the optimisation process to proceed while awaiting the remaining results of the q-batch samples. The acquisition function is conditioned on the pending points when selecting the next point to evaluate \cite{daulton2020differentiable}. 

Additionally, these acquisition functions utilise MC methods, which improve computational efficiency. Specifically, MC samples are drawn from the posterior distribution to approximate the expectation, and subsequently, the acquisition function. It is important to note that when observations are noiseless, these acquisition functions behave similarly to conventional acquisition functions. 

In higher-dimensional spaces, acquisition functions often become extremely complex, making analytical solutions impractical. As a result, direct solutions are generally unattainable, necessitating the use of numerical methods for optimising the acquisition function. A commonly used approach is the multi-start LBFGS-B algorithm. This second-order method relies on gradient and Hessian matrix calculations to locate the maximum of the acquisition function. A simple heuristic is applied to initialise the optimisation process, selecting 10 initial restart locations from a set of 512 randomly generated samples across the search space. Here, a restart location refers to the starting point at which the optimisation algorithm begins its search, helping to avoid getting stuck at a local optimum. Furthermore, the MC-based acquisition functions were approximated using 256 MC samples, corresponding to the accuracy of the posterior approximation, whereby the higher the value is the greater the accuracy at the cost of increased computation time.

As stated in Sec.~\ref{subsection: hypervolume methods}, hypervolume-based methods such as qNEHVI require a reference point to compute the HV (shown by the point on the bottom left of Fig.~\ref{fig:pareto_front_and_hv_explanation}). A typical rule of thumb used to prescribe a value for the reference point is choosing a point that is slightly worse than the lower bound of the objectives, since the problem that is being solved assumes that objectives are required to be maximised. However, many different methods exist that can be applied to select a reference point, especially when no knowledge of the solution domain exists, especially the lower bound of the objectives. 

To demonstrate the implementation of the MOBO framework, two benchmark test problems were used: a) the Branin-Currin problem with added observation noise for each objective, and b) the ZDT1 with no additional observational noise. Both are maximisation problems. Five methods were tested, including the noiseless qEHVI function, to compare its effectiveness against its noisy counterpart qNEHVI. Additionally, a comparison was made between two different kernels - the RBF and Matern kernels - for calculating the covariance matrix. Before optimisation, six samples were generated from the design space to form the initial dataset, with the maximum number of iterations set at 200. The HV was used to assess the quality of the solutions provided by each method. The benchmark studies led to the following conclusions: 1) Regardless of the problem type, the Matern kernel outperformed the RBF kernel by capturing more intricate relationships between neighbouring points, allowing most methods to converge faster and achieve a total hypervolume closer to the maximum hypervolume. Consequently, the Matern kernel was selected for solving the spinodoid structure optimisation problem. 2) In noisy settings, the qNEHVI function outperformed qEHVI as expected. However, in the noiseless problem, both qEHVI and qNEHVI performed similarly, suggesting that qNEHVI is generally preferable. Thus, qNEHVI was one of the methods employed in the study of spinodoid cellular structures. Further details on the benchmark studies can be found in \ref{appendix:Benchmarks}.

\subsection{Spinodoid problem setup}

With the BO framework discussed and tested in Subsec.~\ref{subsec:BO Framework}, it is crucial to emphasise the second part of the MOBO process, which involves integrating the MATLAB-ABAQUS interface to complete the optimisation cycle, as visualised in Fig.~\ref{fig:MOBO framework}. The initial dataset for the BO process was created by sampling 50 data points using the Sobol' sequence, which was subsequently evaluated. The pair plot in \ref{appendix:Pair plot of initial dataset} demonstrates the relationships between each design input and objective, highlighting a positive correlation between EA and PF. 

Whilst the integration streamlines the design exploration process, achieving optimal performance also requires addressing critical structural behaviours that could compromise the functionality of the design. One such phenomenon is densification, which necessitates implementing filtering techniques to exclude undesirable designs, thereby ensuring that the optimisation process delivers practical and reliable solutions.

\subsubsection{Gradient-Based Filtering to Address Densification in Spinodoid Structures}
\label{subsubsec:Addressing Densification in Spinodoid Structures}

Densification refers to the process where a structure's density increases under compressive forces, leading to a significant rise in stiffness. This increased stiffness reduces the structure's ability to absorb energy, as it becomes incapable of deforming easily, leading to a rapid increase in the force transmitted through the structure. In crashworthiness applications, structures are designed to prolong the duration in which the energy is being absorbed. This is primarily done by dissipating energy over a longer distance. However, a densified material shortens the duration of energy absorption, resulting in rapid deceleration, which is undesired. As a consequence, it is crucial to avoid selecting designs that densify within the working conditions, in this case at strain levels below 50\%. This necessitates the implementation of a process to filter out designs that exhibit such behaviour.

\begin{figure}[!h]
  \centering
   \includegraphics[width=1\textwidth]{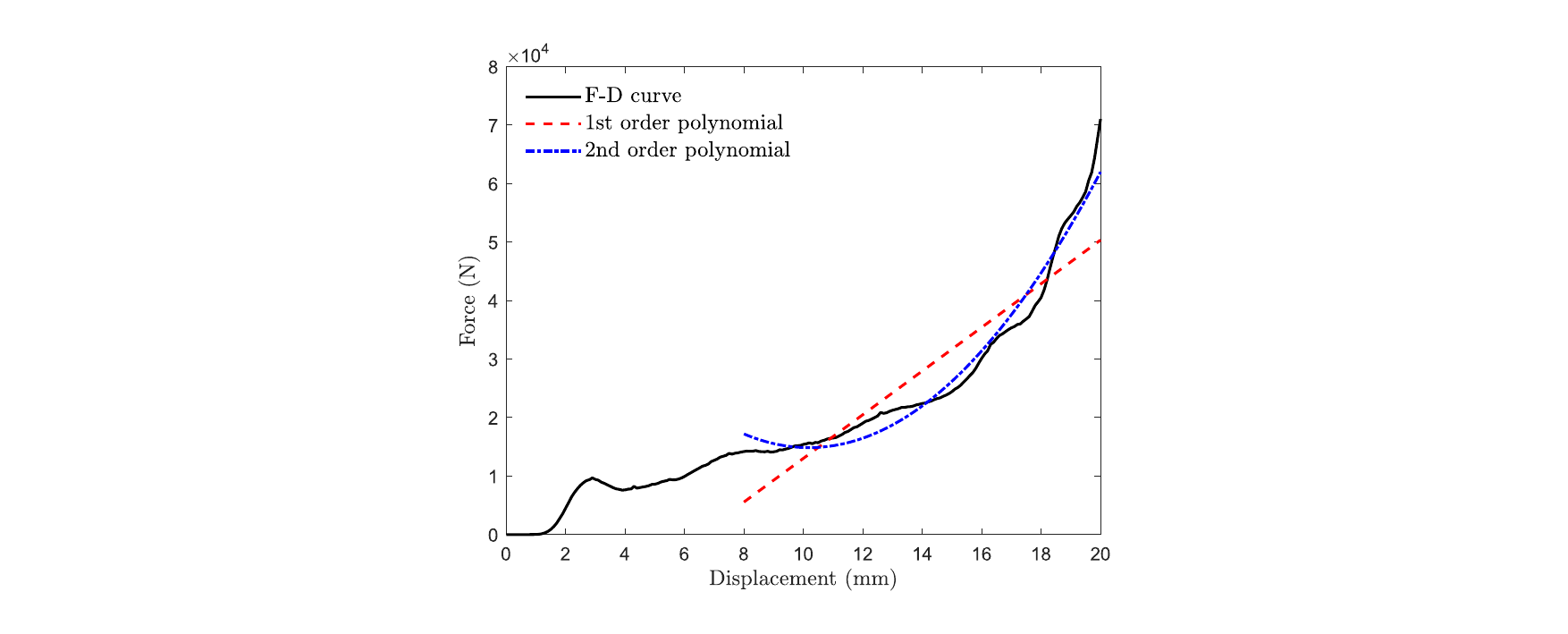}
\caption{\label{fig:comparison_1st_vs_2nd_order_polynomial}Comparison between 1st order and 2nd order polynomial lines used to fit the F-D curve.}
\end{figure}

\begin{figure}[!h]
  \centering
   \includegraphics[width=1\textwidth]{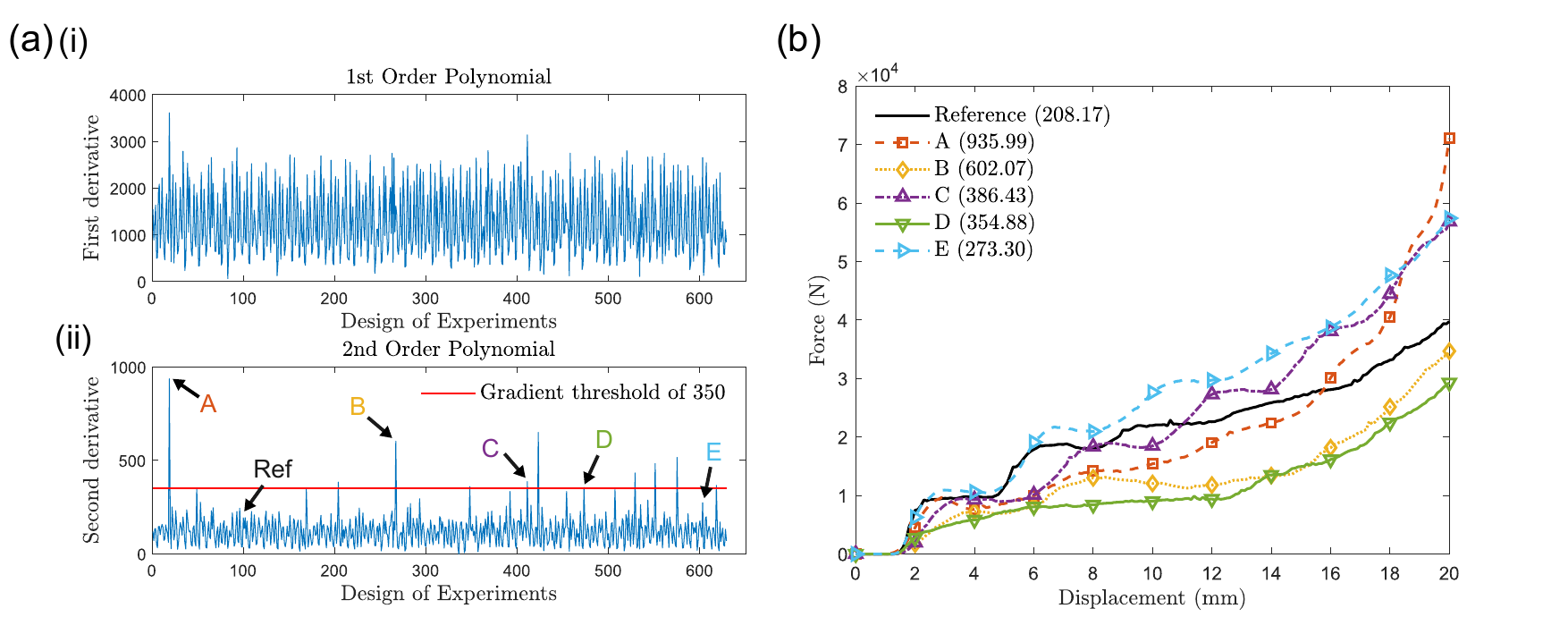}
\caption{\label{fig:doe_gradient_filtering}(a) Plots illustrating the application of a gradient-based filter to 631 Design of Experiments (DoEs), with each evaluated point represented. (i) This plot displays the first derivative applied to the F-D curve, which is fitted using a first-order polynomial. (ii) This plot presents the second derivative values obtained from fitting the F-D curve with a second-order polynomial. The red horizontal line indicates the threshold above which designs are considered to be undergoing densification. (b) This subplot showcases select F-D curves from various designs with differing second derivative values, with each curve corresponding to the annotations found in (a)(ii).}
\end{figure}

To effectively filter these designs, it is essential to first identify what constitutes densification. In this study, linear regression is used to address this issue, with polynomial curve fitting applied within the MATLAB-ABAQUS pipeline using MATLAB's built-in \texttt{polyfit} function. This function returns the coefficients for the nth-degree polynomial that best fits the y data, thus minimising the sum of squares. These coefficients are then used to recreate the fitted lines. Here, data points within the strain interval between 0.2 and 0.5 are used for curve fitting. Fig.~\ref{fig:comparison_1st_vs_2nd_order_polynomial} illustrates a comparison between first-order and second-order polynomials fitted to these data points, shown by the red and blue lines, respectively. The fit's quality is visually assessed by how closely these lines match the target behaviour, represented by the solid black line, which shows the typical behaviour of a structure undergoing densification. Densification on a Force-Displacement (F-D) curve manifests as an exponential rise in the force required to further deform the structure. The rate at which the force rises with an increasing displacement can be quantified using gradients, which has been shown in Fig.~\ref{fig:doe_gradient_filtering}.

The results of applying the first derivative to the first-order polynomial line of best fit are shown in Fig.~\ref{fig:doe_gradient_filtering}(a)(i). The first derivative here reflects the rate of change, which is the stiffness on an F-D curve. However, these values do not indicate or highlight significant peaks, which would signal areas of greatest change i.e. maximum stiffness. Additionally, the gradient values are all positive due to the elastic-plastic constitutive material model used in the FEM analysis, meaning fractures do not occur, hence stiffness cannot be reduced. As a result, the first derivative method is relatively insensitive and ineffective at detecting rapid changes in the stiffness.

To remedy this, the second derivative of the second-order polynomial was used, as it measures the rate of change in stiffness. For example, a positive second derivative would suggest that a structure is stiffening at an accelerating rate, which is indicative of densification. This has been shown in Fig.~\ref{fig:doe_gradient_filtering}(a)(ii), where certain DoEs display very high second derivative values, which were not captured by the first-order polynomial filter. To demonstrate the effectiveness of the second-order polynomial filtering, various points were selected based on their second derivative values annotated as A through E, with a reference line showing solid black representing the behaviour of a structure not undergoing densification. The corresponding F-D curves for these points are plotted in Fig.~\ref{fig:doe_gradient_filtering}(b).

To differentiate between structures undergoing densification and those that are not, a threshold value is necessary. In this case, a second derivative value of 350 was used as the cut-off, above which structures were assumed to be undergoing densification. This threshold was determined by plotting the annotated points in Fig.~\ref{fig:doe_gradient_filtering}(b) and comparing them to the reference line. Cases A to D exhibit a rapid increase in the force during the latter stages of crushing, in contrast to the reference and case E, which show a more gradual increase in force. Nonetheless, a lower threshold, closer to 274, could be chosen if stricter filtering were to be required.

It should be noted that the gradient-based filter was also applied to the initial dataset, with one of the points within the dataset deemed to have undergone densification when the filter had been applied. 

\subsubsection{Scaling Objectives}

Nevertheless, direct optimisation of EA and PF in their current form is infeasible, as the MOBO framework requires both objectives to be maximised. To address this, the following scalings were applied to the normalised objective values, resulting in a convex-shaped Pareto front

\begin{equation}
\begin{aligned}
    \overline{\text{EA}'} &= \ln(\overline{\text{EA}}) \\
    \overline{\text{PF}'} &= -\overline{\text{PF}}
    \label{equ:scaling}
\end{aligned}
\end{equation}

By applying these scalings in Eq.~(\ref{equ:scaling}), the minimisation problem is effectively converted into a maximisation problem. This approach was found to be more effective than directly maximising EA and assuming that the inverse of PF should also be maximised. The scalings enhanced the traversal across the Pareto front, optimising the effectiveness of the acquisition function.

\subsubsection{MOBO workflow}
Given that this study focuses on optimising spinodoid cellular structures for crush energy absorption applications, where the goal is to maximise energy absorption while minimising peak force, FEM analysis and problem setup are crucial components of the optimisation process. Although EA and PF are the primary objectives, additional objectives could be considered if data is available for each structure, such as material weight (by calculating volume or density) or reusability (with a more comprehensive material constitutive model that accounts for hysteresis).

The MATLAB-ABAQUS interface facilitates this process by generating a series of input files, including part geometries (topology, anvil, and loader), an assembly file detailing the positioning of each part, the material model with its specific coefficients, and a settings file defining boundary conditions and history plots. These files are created once the next evaluation point is identified. ABAQUS is then executed via the command line on MATLAB to run the FEM simulation. After the simulation is completed, MATLAB is employed for post-processing, extracting the F-D data for further analysis.

To ensure the robustness of the optimisation process, several fail-safes were implemented. If a simulation fails due to issues in generating the spinodoid structure, a random point from the Sobol' sequence is sampled as a replacement, and this is repeated until a valid output is obtained. Additionally, a constraint was introduced to prevent optimal structures from densifying within the 50\% strain interval. This was achieved by applying a gradient-based filter to the extracted F-D data, classifying designs as either undergoing 'progressive failure' or 'densification'. Structures classified as having gone densification are automatically considered dominated solutions, with output values for EA and PF set to $(-4, -0.2)$. These values have also been applied when the acquisition function recommends a design with all three conical angles being zero, as it is not possible to generate such a structure.

In this study, sequential optimisation was performed with $q=1$. The stochastic nature of spinodoid topologies, due to GRFs and random initialisations, results in varying microstructures and randomness in the measured mechanical properties, introducing noise into the observations. This makes acquisition functions that can handle noisy data particularly suitable for this problem. However, to ensure repeatability, a seed can be set for the random number generator, in this case, it was set to \texttt{rng(123)} in MATLAB, producing consistent structures and eliminating noise. Nevertheless, in this study, the reference point was determined after multiple runs with various reference points, eventually settling on $\mathbf{r} = (-10, -3)$. 

The optimisation loop was executed for a maximum of 250 iterations to ensure that a sufficiently populated Pareto front could be constructed, which also served as the stopping criterion. The full optimisation history can be found in \ref{appendix:Optimisation history}. The performance of the MOBO methods was evaluated against the NSGA-II algorithm by comparing the final hypervolume encompassed by the non-dominated solutions. The NSGA-II algorithm was implemented using the Python-based \texttt{pymoo} package, with the initial population derived from the initial dataset. The algorithm was run for 5 generations, resulting in a total of 250 evaluations ($50 \times 5 = 250$) \cite{blank2020pymoo}. Simulated Binary Crossover (SBX) was used to generate offspring, while Polynomial Mutation was applied to introduce small variations in gene values, thus promoting diversity in subsequent generations. It should be noted that NSGA-II was used as a black box; further tuning of the crossover and mutation rates may improve the results. Specific hardware information, which was used to run the optimisation tasks on the cluster can be found in \ref{appendix:HPC hardware specifications}.

\section{Results}

\begin{figure}[!ht]
  \centering
   \includegraphics[width=1\textwidth]{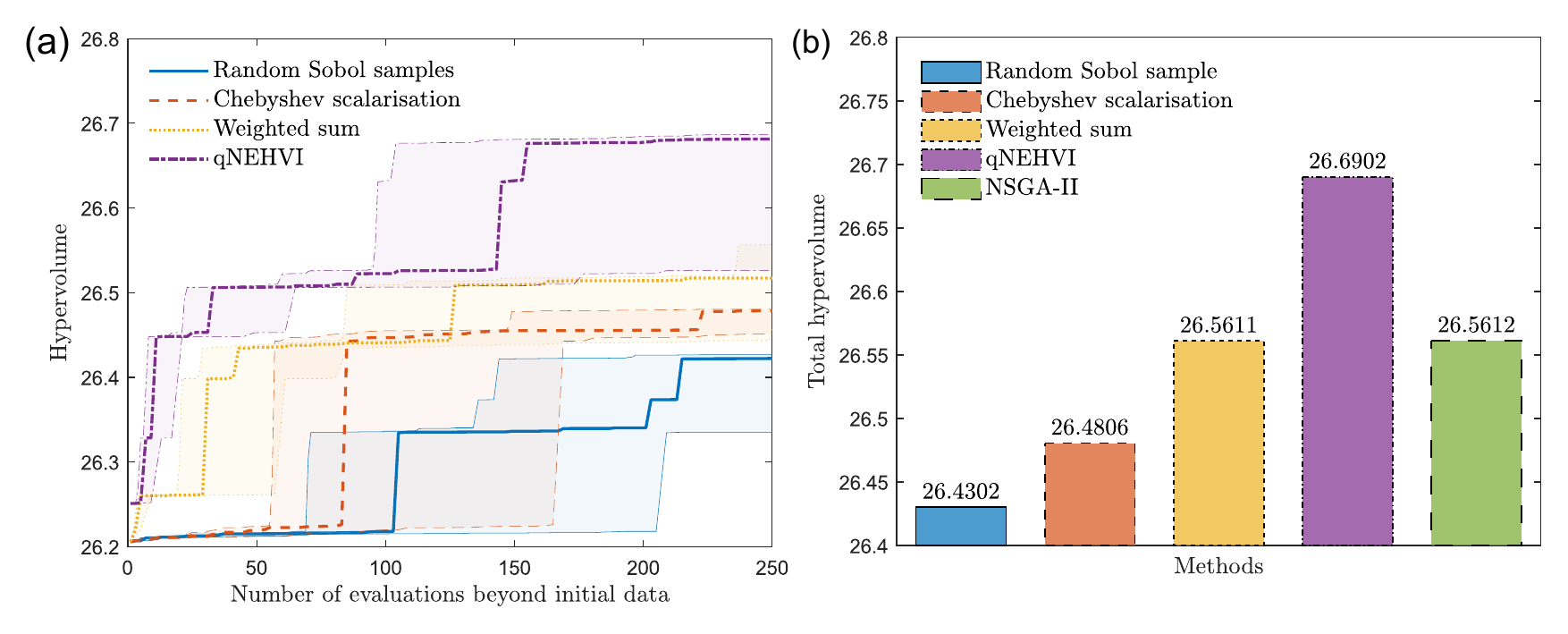}
   \caption{\label{fig:optimisation_results_hypervolume}(a) Comparison of the hypervolume achieved by each MOBO method at each iteration of the optimisation process, starting from the dataset used to initialise the Gaussian Process surrogate model. The comparison continues until reaching the 250-iteration threshold. The hypervolume values for each method include the upper bound, median, and lower bound. The thicker lines depict the median values, while the thinner lines above and below represent the upper and lower bounds, respectively. (b) A bar chart comparing the final hypervolume obtained by various MOBO methods and the traditional NSGA-II method after 250 iterations.}
\end{figure}

Following the MOBO framework setup described in Sec.~\ref{sec:mobo framework}, four MOBO methods were employed to obtain a set of Pareto-optimal designs, with NSGA-II - a traditional method - serving as a benchmark. Fig.~\ref{fig:optimisation_results_hypervolume}(a) presents the hypervolume of the non-dominated solutions, calculated at each iteration beyond the initial dataset, for all implemented MOBO methods. The qNEHVI method outperforms other MOBO methods, as evidenced by a rapid rise in hypervolume after a few iterations, with a further significant increase at around 150 iterations. This suggests qNEHVI discovered optimal designs not found by other methods, resulting in a superior Pareto front. The weighted sum method also showed a strong early performance, with a steep increase in hypervolume up to 50 iterations, followed by gradual improvements until around 125 iterations, and a slight increase thereafter. Chebyshev scalarisation performed similarly to the weighted sum method but lacked the initial rapid improvement, showing a gradual increase in hypervolume starting around 80 iterations. As expected, all three methods outperformed the random Sobol' sequence sampling.

Fig.~\ref{fig:optimisation_results_hypervolume}(b) displays the final hypervolume calculated after 250 iterations, including results from NSGA-II. While NSGA-II achieved a higher hypervolume than random Sobol' sequence sampling, the weighted sum, and Chebyshev scalarisation methods, it was outperformed by the qNEHVI method. The poorer performance of Chebyshev scalarisation can be attributed to its suitability for non-convex Pareto fronts, unlike the convex Pareto front observed in this study. The weighted sum method, being straightforward, is better suited to handling convex Pareto fronts but lacks the targeted hypervolume improvement and exploration-exploitation balance of qNEHVI. A random combination of weights in the weighted sum and Chebyshev scalarisation methods may lead to overexploitation of certain regions, affecting the quality of the surrogate model. In the case of Augmented Chebyshev, the method might overly focus on regions near the ideal point, resulting in insufficient exploration. NSGA-II, while competitive, may not explore the objective space efficiently if mutation and crossover rates and other hyperparameters are not fine-tuned, and it often requires many generations to converge to the true Pareto front, which can be computationally expensive. In contrast, qNEHVI is more effective overall, requiring minimal hyperparameter tuning and specifically targeting hypervolume improvement.

\begin{figure}[!ht]
  \centering
   \includegraphics[width=1\textwidth]{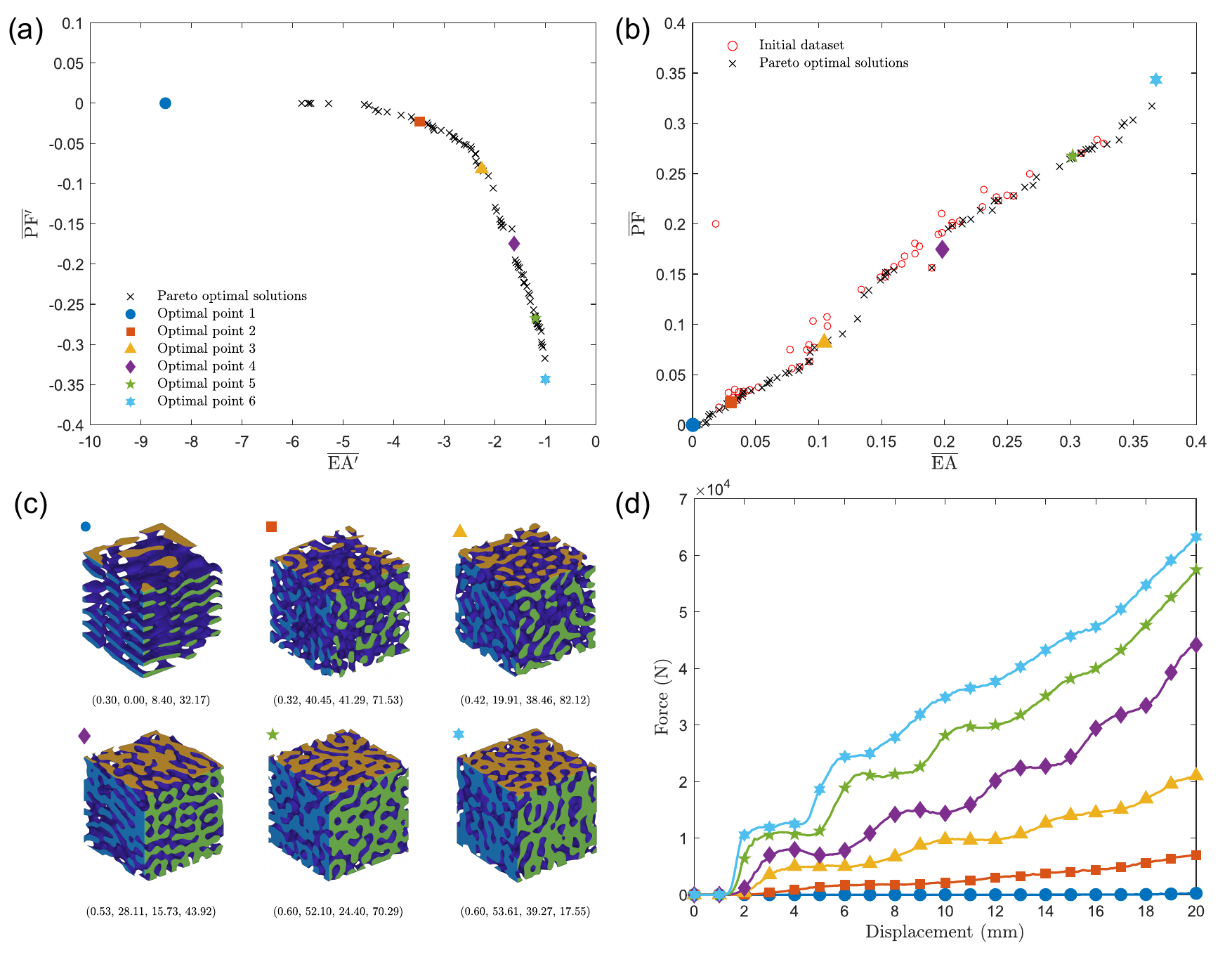}
   \caption{\label{fig:pareto_optimal_solutions}The pareto-optimal solutions obtained using the hypervolume-based qNEHVI method are visualised alongside selected example points, indicated by the coloured markers. Subplot (a) illustrates the scaled objectives, adjusted according to Eq.~(\ref{equ:scaling}), while subplot (b) presents the unscaled results in the normalised objective plane with points in red showing the initial dataset. Subplot (c) shows the topologies corresponding to the respective objective values, represented by the differently coloured markers in (a) and (b). Finally, subplot (d) depicts the F-D curves of the chosen structures undergoing compression.}
\end{figure}

Fig.~\ref{fig:pareto_optimal_solutions}(a) illustrates the set of Pareto-optimal solutions derived from the qNEHVI method. The Pareto front includes 78 designs, each contributing uniquely to the overall hypervolume, Fig.~\ref{fig:pareto_optimal_solutions}(b) maps the rescaled Pareto-optimal points onto the original normalised objective plane, also displaying the initial dataset for comparison. Compared to the initial dataset, the MOBO method has identified several optimal designs that exhibit reduced PF while preserving the EA capability. Moreover, qNEHVI has uncovered previously undiscovered optimal points at the higher end of the EA range, particularly in the upper-right region of the normalised PF-EA graph. Six optimal points were selected to demonstrate the robustness of these designs, with Fig.~\ref{fig:pareto_optimal_solutions}(c) showcasing the generated topologies, featuring various microstructures with differing relative densities. The F-D behaviour of each selected design under compression is plotted in Fig.~\ref{fig:pareto_optimal_solutions}(d), revealing distinct performance characteristics. 

\textcolor{black}{For instance, the reduction in normalised PF for the optimal points, such as the purple diamond (optimal point 4), corresponds to a decrease from 0.19 to 0.17, which translates to an approximate reduction in peak force of 1,610 N when normalised by the peak force of 80,476 N from compressing a cube of size $40 \times 40 \times 40\ \text{mm}^3$. Even small reductions in normalised PF can have meaningful implications, particularly in applications where performance or safety is sensitive to force thresholds. For example, in structural or collision scenarios, lower PF values can reduce material stress, enhance energy dissipation, and improve durability or safety. While the magnitude of the reduction may seem modest, such improvements often contribute to achieving more optimal designs or meeting stringent performance criteria. This highlights the value of even incremental advancements achieved through optimisation.}

The absence of high PF values in the optimal solutions highlights the effectiveness of the optimisation process in balancing competing objectives. The optimal designs predominantly feature geometries with mixed values of $\theta$, which lack distinct characteristics or a clear correlation with conical angles. This outcome reflects the inherent flexibility of the MOBO approach, which identifies a diverse set of Pareto-optimal solutions by varying design variables to achieve desired trade-offs between EA and PF. As a result, the optimisation process systematically navigates the Pareto frontier, favouring designs that maximise energy absorption while keeping peak forces within acceptable limits.

These are examples of bending-dominated structures, where loads are primarily resisted through bending and flexural mechanisms. While these structures tend to have lower stiffness and strength due to localised deformations that can lead to early failure through buckling or yielding, the bending mechanisms enable stress distribution across the structure. This allows the load to be spread over a larger area as the structure deforms more easily, effectively delaying catastrophic failure, which results in such structures having higher EA. It should also be noted that these solutions were deemed to not have undergone densification, which is also evident in the F-D curves of the selected cases.

\begin{figure}[!ht]
  \centering
   \includegraphics[width=1\textwidth]{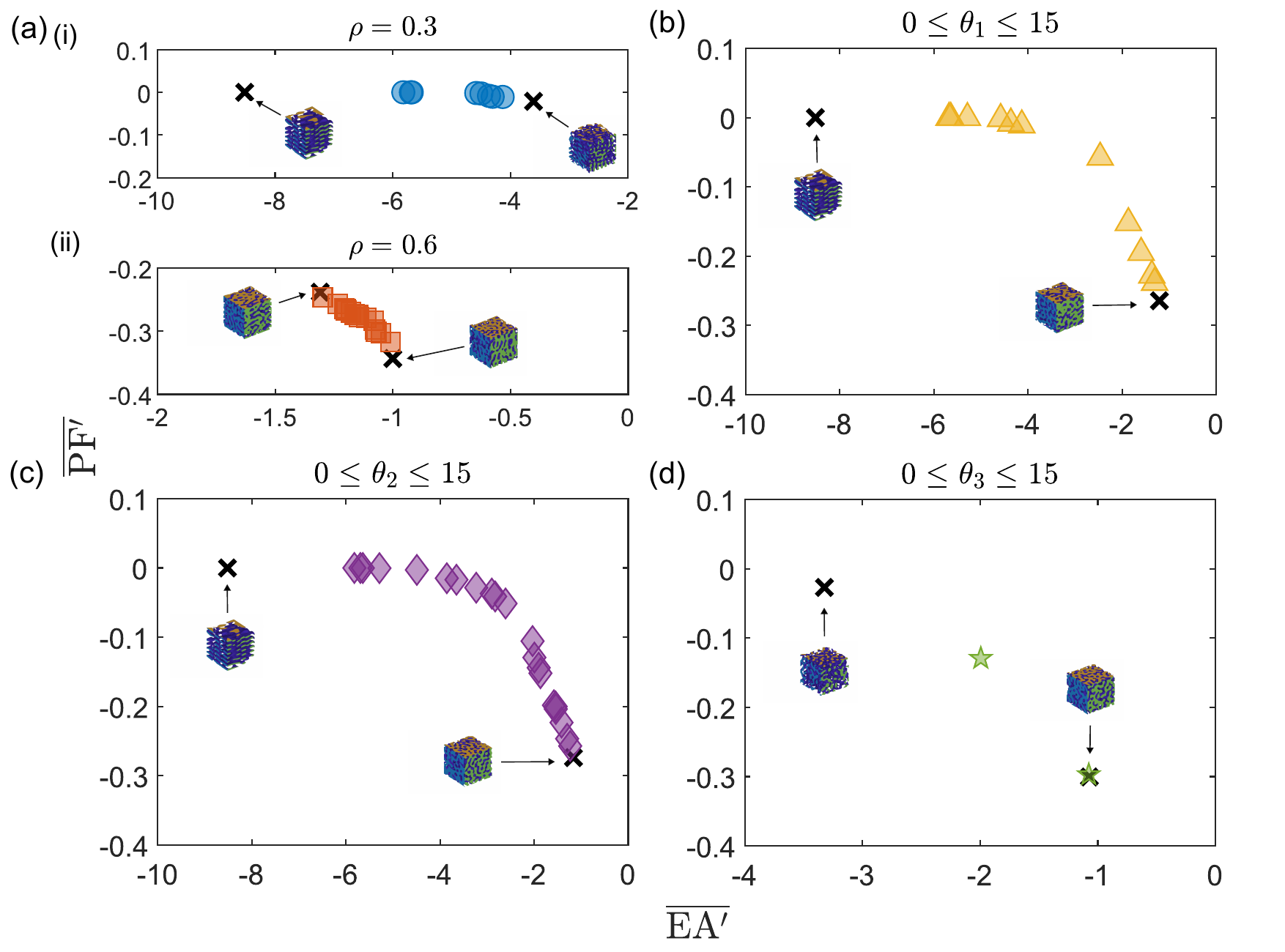}
   \caption{\label{fig:conditional_optimal_solutions_figure}Illustration of subplots showcasing a set of solutions generated by applying various design constraints to the Pareto-optimal solutions obtained using the qNEHVI method. (a) Displays solutions with (i) low relative density, $\rho = 0.3$, (ii) high relative density, $\rho = 0.6$. Additional constraints are also applied to achieve different anisotropies by limiting one of the cone angles, $0 \leq \theta \leq 15$: (b) Conical angle $\theta_1$, (c) Conical angle $\theta_2$, and (d) Conical angle $\theta_3$.}
\end{figure}

Further design constraints can be imposed on the Pareto-optimal solutions to provide greater flexibility depending on the application. The impact of applying such constraints is shown in Fig.~\ref{fig:conditional_optimal_solutions_figure}. For instance, if the application requires the structure to be as lightweight as possible, designs with low relative density may be preferred. Conversely, if maximising EA is the priority, previous insights into the structure-property relationship indicate that EA scales with relative density, with the highest EA achievable at the upper bound of relative density, specifically at $\rho = 0.6$. Additionally, suppose varying degrees of anisotropy are required. In that case, the conical angles can be constrained to smaller values, such as $0 \leq \theta \leq 15$, which influences anisotropy as described in  \ref{appendix:Controlling the anisotropy of topologies}. Imposing these constraints produces a subset of optimal designs that exhibit different relationships between the objectives EA and PF. However, a notable scarcity of solutions is observed for $\theta_3$ compared to the other cases. This is because structures with anisotropy aligned parallel to the loading direction tend to be stretching-dominated, leading to high stiffness and consequently high PF. These characteristics make such structures less ideal, leading to their exclusion from the Pareto-optimal solutions, i.e. these are part of the dominated solutions that do not contribute to the total hypervolume. The few solutions that do emerge are associated with higher values for conical angles $\theta_1$ and $\theta_2$, which enhance interconnectivity between individual structural elements, transforming the design into bending-dominated structures that are more suitable for EA applications. Subsequently, the choice would be left to the user based on their requirements, depending on the application.

\section{\textcolor{black}{Optimisation with three variables}}

\textcolor{black}{The primary approach to obtaining Pareto-optimal solutions in the spinodoid design space involves optimising all four design variables simultaneously. This ensures globally optimal results without introducing any prior assumptions, allowing for a comprehensive exploration of the design space. Such an approach also captures complex interactions between variables, leading to a more diverse and well-populated Pareto front.} 

\textcolor{black}{However, scenarios exist where efficiency gains can be achieved by leveraging insights from the sensitivity analysis highlighted in Subsec.~\ref{subsection:Effect of design parameters on performance indicators}. Specifically, the highly sensitive parameter, $\rho$, can be fixed prior to the optimisation based on performance criteria. For example, to obtain a set of Pareto-optimal solutions targetting the lowest weight, $\rho = 0.3$ could be set before generating the initial dataset. This strategy reduces computational cost in two ways. First, this strategy reduces the dimensionality of the problem, allowing the optimisation process to focus on the remaining, less sensitive variables, $\theta$'s allowing convergence to occur faster. Second, the main contributor to the increased optimisation time is incurred when evaluating the next point recommended by the acquisition function. As the computation time for simulations scales with the relative density, a low value for $\rho$ would speed up the computation time to complete a single iteration i.e. time taken to conduct FE analysis on a single point, thus significantly reducing the time to complete the optimisation task. However, the opposite is also true, if a large value of $\rho$ is chosen, the computation time will significantly increase due to a greater number of elements required to mesh the geometry as well as the difference in mechanical behaviour.} 

\begin{figure}[!ht]
  \centering
   \includegraphics[width=1\textwidth]{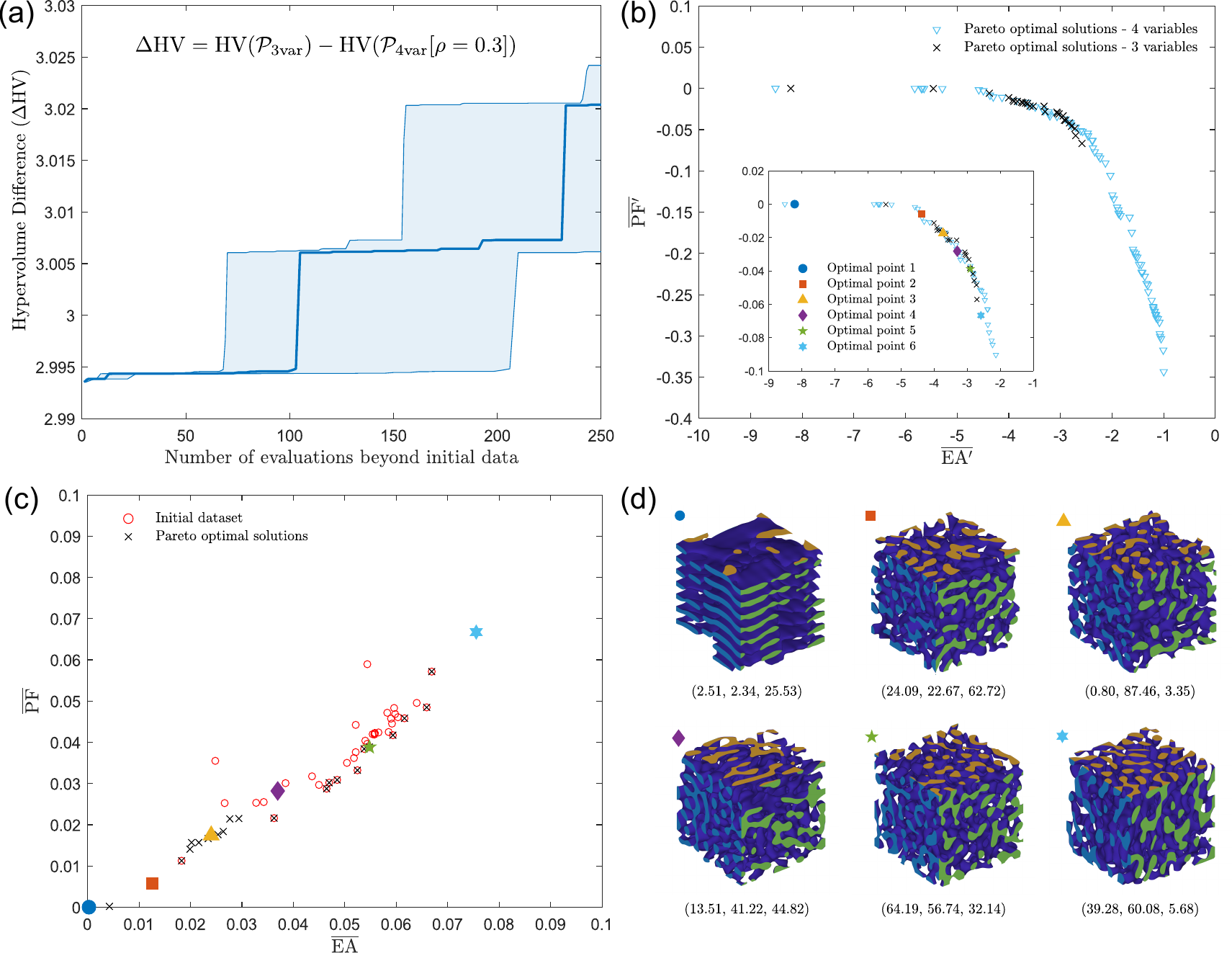}
\caption{\label{fig:pareto_optimal_solutions_3var_v2}\textcolor{black}{Pareto-optimal solutions obtained from optimisation with three variables, $\theta$'s, conducted with a fixed relative density of $\rho = 0.3$ using the qNEHVI method. Subplot (a) represents the difference in hypervolume between the Pareto-optimal solutions filtered post-optimisation to include only those with $\rho = 0.3$, derived from a four-variable optimisation (as shown in Fig.~\ref{fig:conditional_optimal_solutions_figure}(a)(i)), and the Pareto-optimal solutions obtained from a three-variable optimisation. The hypervolume difference is quantified as $\text{HV}(\mathcal{P}_{\text{3var}}) -\text{HV}(\mathcal{P}_{\text{4var}}[\rho = 0.3])$. This comparison is evaluated over 250 iterations, with the solid central line representing the median and the thinner lines at the top and bottom indicating the 25th and 75th percentiles, respectively. Subplot (b) presents Pareto-optimal points scaled using Eq.~(\ref{equ:scaling}) obtained from three-variable and four-variable optimisation, with example geometries highlighted by markers of different colours shown in the insert. Subplot (c) displays the unscaled, normalised objectives, where black cross markers denote Pareto-optimal points, and red circular markers highlight the initial dataset. Subplot (d) visualises the topologies corresponding to specific objective values, as indicated by their respective markers. The values shown beneath each image represent the $\theta$'s required to generate the structures at $\rho = 0.3$.}}
\end{figure}

\textcolor{black}{Following the successful implementation of the qNEHVI method with four variables, a three-variable optimisation was conducted with the relative density fixed at $\rho = 0.3$ prior to the optimisation. The Sobol' sequence was again used to generate an initial dataset of 50 points, and the optimisation process was run for 250 iterations. Detailed optimisation history is provided in \ref{appendix:Optimisation history with three variables}. The evolution of the hypervolume difference across the 250 iterations is shown in Fig.~\ref{fig:pareto_optimal_solutions_3var_v2}(a). This figure compares the hypervolume of the three-variable optimisation ($\rho = 0.3$ fixed) with that of the four-variable optimisation, isolating all Pareto-optimal solutions with $\rho = 0.3$ i.e. $\text{HV}(\mathcal{P}_{\text{3var}}) -\text{HV}(\mathcal{P}_{\text{4var}}[\rho = 0.3])$, where $\text{HV}(\mathcal{P}_{\text{3var}})$ is the hypervolume of Pareto-optimal solutions obtained from three-variable optimisation, with $\rho = 0.3$ fixed prior to the optimisation and $\text{HV}(\mathcal{P}_{\text{4var}}[\rho = 0.3])$ is the hypervolume of Pareto-optimal solutions obtained from the four-variable optimisation, but filtered \emph{a posteriori} to include only those solutions satisfying $\rho = 0.3$ (Illustrated in Fig.~\ref{fig:conditional_optimal_solutions_figure}(a)(i)). The positive difference observed in the initial iterations indicates that the three-variable optimisation outperforms the conditional solutions from the four-variable optimisation from the very first iteration. By focusing on the less sensitive parameters, the three-variable approach exploits correlations between the $\theta$'s and the objectives more effectively, leading to better hypervolume performance from early on. Furthermore, the three-variable method continues to demonstrate improvements as the optimisation progresses, highlighting its ability to efficiently refine the design space.} 

\textcolor{black}{The scaled Pareto-optimal solutions obtained from the three-variable optimisation are shown in Fig.~\ref{fig:pareto_optimal_solutions_3var_v2}(b), with 26 points forming the Pareto-front. Similar to the four-variable optimisation, six optimal points on the Pareto front are marked with different coloured symbols. Unlike the Pareto front in Fig.~\ref{fig:pareto_optimal_solutions}(a), the three-variable solutions occupy a smaller portion of the objective space due to the lower relative density, with $\overline{\mathrm{PF'}}$ ranging from -0.08 to approximately 0, and $\overline{\mathrm{EA'}}$ ranging from -9 to -2. Despite the reduced objective space, the three-variable optimisation discovers new solutions by varying the $\theta$'s, achieving improved EA values while maintaining similar PF levels, which four-variable optimisation found by varying the $\rho$ that improves the EA. In some cases, however, the four-variable optimisation yields solutions that outperform those from the three-variable optimisation.}

\textcolor{black}{Fig.~\ref{fig:pareto_optimal_solutions_3var_v2}(c) presents the unscaled Pareto-optimal points in the normalised $\overline{\mathrm{PF}}$-$\overline{\mathrm{EA}}$ plane alongside the initial dataset. The initial Sobol' sampling caused the evaluated objectives to cluster in the top-right region of the objective space, making it densely populated and reducing uncertainty in the Gaussian process regression (GPR) model in that region. In contrast, the sparsely explored bottom-left region exhibited higher uncertainty, which the acquisition function identified as an opportunity to improve the hypervolume. This led to a focused exploration in the bottom-left region of the $\overline{\mathrm{PF}}$-$\overline{\mathrm{EA}}$ plane, while the optimisation process effectively leveraged the initially sampled points in the top-right, incorporating them into the Pareto-optimal solutions, as indicated by the overlapping initial data and Pareto optimal solution markers. Illustration of topologies that result in objective values shown by the coloured markers have been shown in Fig.~\ref{fig:pareto_optimal_solutions_3var_v2}(d). Similar to the optimal design parameters obtained from four-variable optimisation, the structures generated here are also bending-dominated. These optimal solutions feature geometries with mixed values of $\theta$, leading to configurations that inherently facilitate rotational movement or the generation of bending moments under axial loading. This behaviour contrasts sharply with stretching-dominated structures, such as the `Columnar' topology, which rely on direct axial force transfer. The bending-dominated designs resemble isotropic, bone-like structures, characterised by interconnected, irregular patterns that distribute loads over multiple pathways, enabling higher flexibility and energy absorption. Such geometries reflect nature-inspired design principles, where complex structural arrangements optimise mechanical performance by balancing strength, stiffness, and ductility. In conclusion, by predefining the relative density, the optimisation process more effectively maps the structure-property relationship between the cone angles and the objectives. This contrasts with the four-variable optimisation, where improvements in the Pareto front were primarily driven by initially adjusting the density to increase or decrease the objectives. The predefined density approach enables a more detailed and precise mapping between $\theta$'s and the objectives, as well as reducing the overall computation time from an average of 10-14 days as reported for the four-variable case down to approximately two days to complete optimisation up to 250 iterations, primarily driven by the lower cost of evaluating each recommended sample point.}

\textcolor{black}{It should also be noted that due to the heavy influence of $\rho$ in four-variable optimisation, the acquisition function tends to prioritise $\rho$ and recommend values at the extremes of its design space, while the $\theta$'s are sampled in a more stochastic manner without specifically targetting the extreme ends of their ranges. This behaviour reflects the sensitivity disparity between $\rho$ and $\theta$'s, where $\rho$ drives most of the changes in the objectives, leading to less focused exploration of the $\theta$'s. To address this imbalance, standardisation was applied to approximately normalise the variables based on their order of magnitude, ensuring that they are of comparable scale. While the exact bounds of the variables are not fully known in the context of optimisation, this scaling helps mitigate the disparity in sensitivity and encourages more balanced exploration across all variables. The evolution of the recommended sample points over the course of the three-variable and four-variable optimisation can be found in \ref{appendix:Evolution of recommended design parameters}. Consequently, the four-variable optimisation process may require additional iterations to thoroughly explore and optimise the less sensitive variables, further contributing to the increased computational cost.}

\section{Discussion}
This work successfully demonstrates the application of the MOBO framework for crush energy absorption, highlighting its ability to balance multiple objectives and generate Pareto-optimal solutions, ultimately accelerating the selection of optimal structures. This versatile framework has the potential to be adapted for various other fields. For example, in aerospace engineering, where weight reduction, stiffness, and energy absorption are critical, it could be used to design lightweight structures tailored to specific mission profiles. Similarly, in biomedical engineering, the framework could optimise implants to maximise mechanical comparability with bone tissue while minimising material usage. In architectural design, it could assist in balancing structural integrity material efficiency, and aesthetics based on user preferences.

Despite its successes, there are several opportunities for improving the MOBO framework. Enhancements could be made to both the FEM simulation and the optimisation process. For instance, the FEM analysis could be refined by incorporating damage initiation criteria and damage evolution to better capture fracture and progressive failure, Additionally, implementing the Hill yield model could allow for the use of material constitutive models based on short-fibre composite filaments, which are often employed in applications requiring high stiffness-to-weight ratios. Since car crashes are dynamic events and involve varying speeds, strain rate dependency could be modelled using approaches such as the Johnson-cook plasticity model, thereby increasing the model's complexity and improving the representation of topologies under load. Furthermore, the framework could be expanded to include additional objectives, such as manufacturing constraints or multi-material considerations, enabling more nuanced and application-specific designs.

\textcolor{black}{On the optimisation side, the MOBO framework can be significantly enhanced by incorporating a more efficient sampling strategy that explicitly accounts for uncertainties observed in the experimental and FEM simulations. By acknowledging the inherent uncertainties in regions with large errors or inconsistent results (such as Columnar or Lamellar structures), these areas can be targetted with denser sampling. This approach would allow for more accurate predictions of key metrics—such as energy absorption and peak force—where the uncertainty is higher, while reducing the computational cost in regions where the model already performs well (e.g., Isotropic structures). This strategy aligns with the broader concept of uncertainty-aware optimisation, where sampling density is adaptively adjusted based on the observed uncertainty in the predictions. By focusing computational resources on regions with higher uncertainty, the optimisation process becomes more robust and efficient, potentially leading to more precise solutions in fewer iterations.} 

Furthermore, the optimisation process can be further accelerated by implementing parallel evaluation of multiple q-batch recommendations using array jobs on an HPC cluster. This would leverage the parallelism of acquisition functions, which can manage several evaluations simultaneously. By running multiple simulations or experiments in parallel, different regions of the design space can be explored at the same time, drastically reducing the time required to converge to optimal solutions. This parallelisation would not only accelerate the overall optimisation but also allows for more comprehensive exploration of the design space, especially when combined with the targeted sampling of uncertain regions. The combination of efficient sampling based on uncertainty and parallel evaluation of batch recommendations would lead to a much faster and more accurate optimisation process. Additionally, integrating these improvements within a multi-fidelity optimisation framework—which uses both experimental data and FEM simulations of varying accuracy levels—could further increase the robustness and efficiency of the process. Lower-fidelity simulations could be used in combination with high-fidelity experiments to guide the search for optimal solutions, balancing the trade-off between computational cost and model accuracy.

Future research could explore applying MOBO-based optimisation to macro-scale structures, such as large-scale models generated using spinodoid microstructures, where the reduced number of required evaluations may make this approach feasible. Additionally, leveraging BO to develop surrogate models for microscale structures could facilitate their integration into a concurrent modelling framework, where acquisition functions are tuned to explore regions of high uncertainty, rather than focusing solely on exploiting areas for more accurate surrogates.

\section{Conclusion}

This study developed a multi-objective Bayesian optimisation framework to optimise the spinodoid design space for crush energy absorption applications. This framework enables the identification of designs that achieve a balance between competing parameters, specifically focusing on energy absorption and peak force. The process involved several key steps: First, spinodoid topologies were generated using MATLAB, with ABAQUS employed for FEM analysis. Material characterisation tests were conducted on 3D-printed PET-G samples to determine the coefficients for the anisotropic Hill material constitutive model. Various spinodoid structures were fabricated and tested to validate the FEM model.

To evaluate the impact of design parameters on the objectives, two studies were conducted a one-at-a-time analysis, which measured the effect of each parameter on energy absorption and peak force, and a global Sobol' sensitivity analysis with input parameters $(\rho, \theta_1, \theta_2, \theta_3)$. The results showed that all parameters significantly influenced the objectives, with a trivial relationship between relative density, $\rho$, and the objectives observed. The conical angles had less impact individually, but were more influential when varied in combination.

Additionally, various MOBO-based methods were applied to solve the multi-objective problem, and the results were compared with those from the traditional NSGA-II method. After 250 iterations, the hypervolume-based qNEHVI method was found to outperform the others. The Pareto-optimal designs generally exhibited a lower peak force while maintaining energy absorption capacity, while avoiding densification through a gradient-based filter. The study demonstrated that combining FEM analysis with Bayesian optimisation is a powerful approach for solving multi-objective problems for crush energy absorption while minimising the need for costly function evaluations. 

\section{Acknowledgement}
W. Tan acknowledges the financial support from the EPSRC New Investigator Award (grant No. EP/V049259/1). We also thank Dr. Siddhant Kumar for his valuable discussions and assistance throughout this project.

\section{Data availability}
The data and codes needed to reproduce and evaluate the work of this paper are available in the GitHub repository, once this manuscript is published: \\
(https://github.com/MCM-QMUL/MOBO).

\bibliographystyle{elsarticle-num}
%\bibliography{library}
\bibliography{Ref}
%% Text of bib

\newpage
\appendix
\setcounter{figure}{0}    
\section{Appendices}

\captionsetup[figure]{labelformat=simple, labelsep=colon}
\renewcommand{\thefigure}{\thesection.\arabic{figure}}
\counterwithin{figure}{section}
\renewcommand{\figurename}{}
\renewcommand{\thefigure}{\thesubsection.\arabic{figure}}
\counterwithin{figure}{subsection}

% % Redefine the table name to be empty
% \renewcommand{\tablename}{}

% % Customize the table caption format and separator
% \captionsetup[table]{labelformat=simple, labelsep=colon}

% Include section and subsection numbers in table numbering
\renewcommand{\thetable}{\thesubsection.\arabic{table}}
\counterwithin{table}{subsection}

\subsection{Controlling the anisotropy of topologies}
\label{appendix:Controlling the anisotropy of topologies}

\begin{figure}[ht]
    \centering
    \includegraphics[width=1\textwidth]{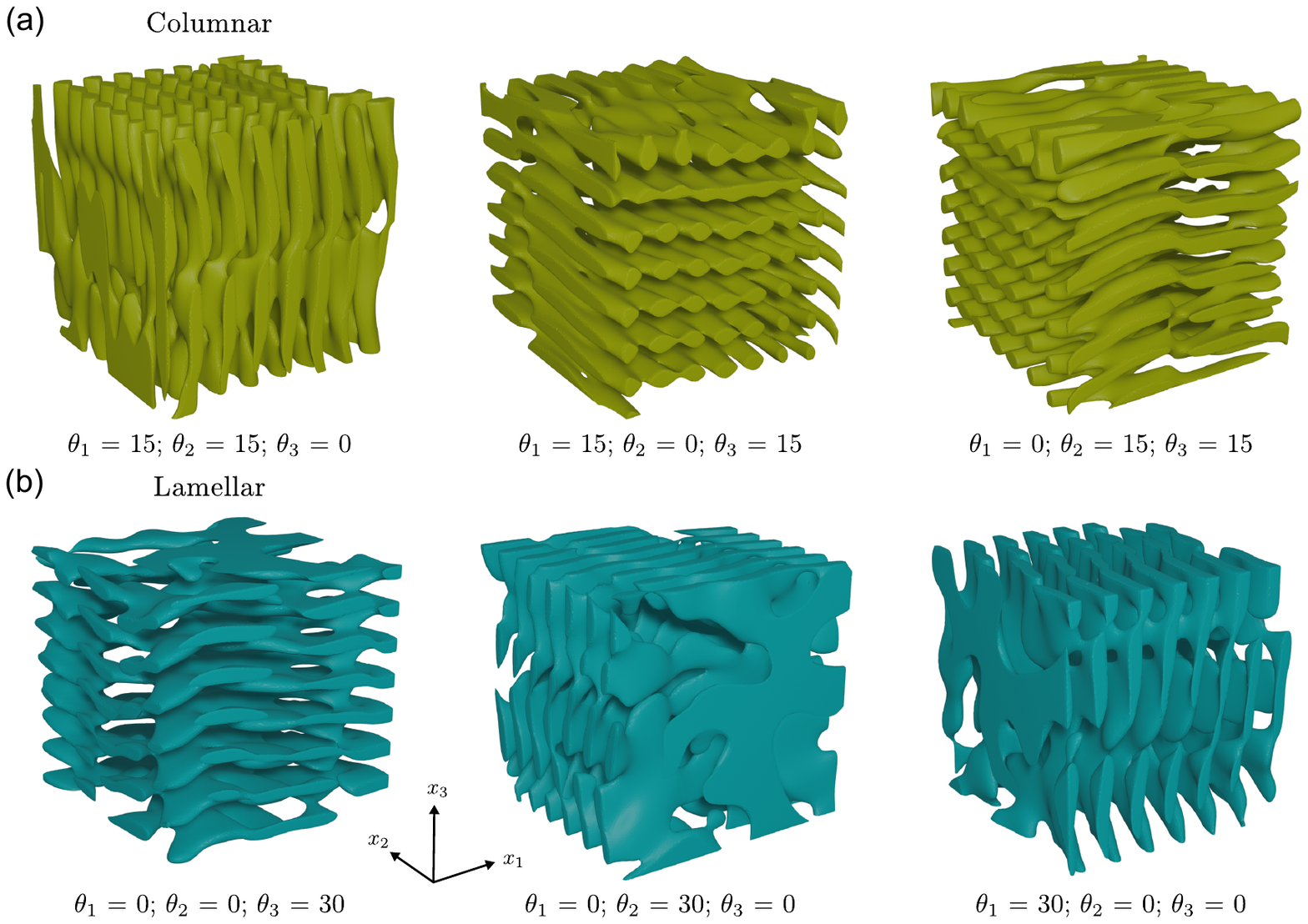}
    \caption{Demonstration of how the anisotropy of spinodoid structures can be altered by adjusting the cone angles for (a) `Columnar', and (b) `Lamellar' topologies.}
    \label{fig:effect_of_cone_angles}
\end{figure}

The anisotropy of these structures can be adjusted, as demonstrated in Fig.~\ref{fig:effect_of_cone_angles}, whereby altering the non-zero angles results in the varying direction of anisotropy. This is clearly illustrated in the examples \ref{fig:effect_of_cone_angles}. For instance, the anisotropy of `Columnar' structure can be changed by choosing the cone angles that do not lie along the direction of interest to be non-zero. In contrast, the `Lamellar' structure follows a different principle, where the direction of anisotropy is determined by aligning the desired anisotropic direction with the single non-zero cone angle.

\newpage

\subsection{\textcolor{black}{Comparison of material model with material characterisation test}}
\label{appendix:Comparison of material model with material characterisation test}

\textcolor{black}{To compare the FEM simulations with the material characterisation tests, three material models were utilised. The simplest was the isotropic elasticity model, which considered only linear elastic behaviour. Simulations were also conducted using an isotropic elastic-plastic model, incorporating plastic strain to simulate the post-yield behaviour of the cube under compression. Finally, the Hill anisotropic model was tested against these conventional models. This model accounts for anisotropy arising from the fabrication process while also capturing the post-yield behaviour. The same FE analysis setup was used for both the initial validation test and the one used to model the behaviour of spinodoids under compressive loading.}

\textcolor{black}{The comparison between the results from the FEM analyses with various material models and the experimental data revealed slight differences in the initial elastic region of the stress-strain curve. However, in the plastic deformation region, the Hill anisotropic model closely followed the experimental data, outperforming the isotropic elastic-plastic model. As the majority of energy absorption occurs within this plastic region, it is crucial to construct a FEM model that accurately replicates this behaviour.}

\textcolor{black}{The observed discrepancies between the Hill model and the experimental results can be attributed to multiple factors. First, the simplicity in calculating the coefficients for Hill's anisotropic yield criterion might have contributed to the mismatch. These coefficients were derived from limited experimental data, which may not fully capture the complex anisotropic properties of the material. Additionally, the assumptions made during the formulation of the yield stress ratios-such as uniform anisotropy or idealised deformation modes-might not align perfectly with the actual material behaviour. Variability introduced during the fabrication process, such as microstructural heterogeneities, residual stresses, or variations in layer bonding during additive manufacturing, can also lead to deviations.}

\textcolor{black}{To reduce these discrepancies, more sophisticated calibration techniques, such as inverse modelling or optimisation-based fitting, could be employed to refine the Hill model parameters. Incorporating additional experimental tests, such as biaxial or shear tests, would also provide a more complete representation of the material's anisotropic behaviour. Finally, accounting for fabrication-induced effects in the FEM model, such as layer-specific anisotropy or residual stress distributions, could improve agreement between the simulation and experiment.}

\begin{figure}[ht]
    \centering
    \includegraphics[width=1\textwidth]{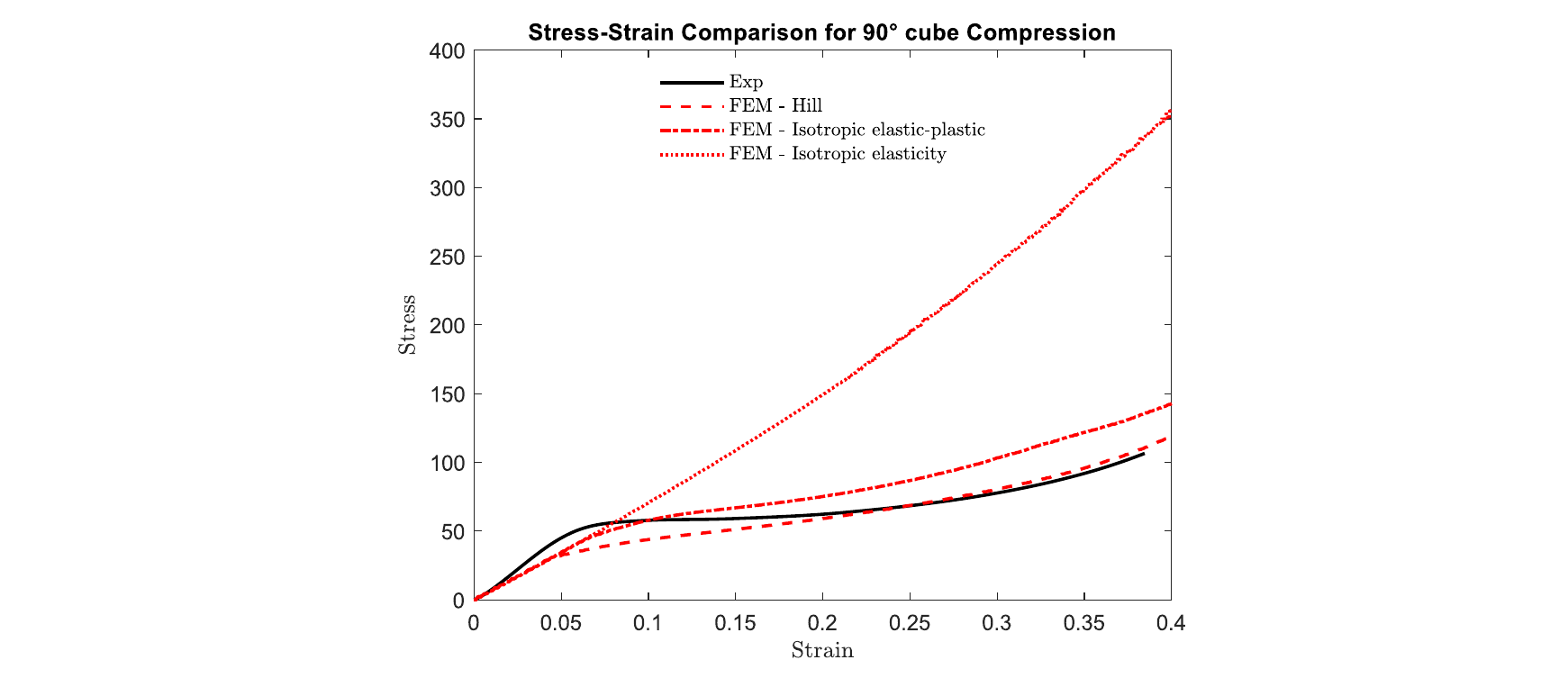}
    \caption{\textcolor{black}{Comparison of the mechanical response between the compressive testing conducted on a 3D printed cube of dimensions $20 \times 20 \times 20$ mm$^3$, which was fabricated at $90^\circ$ orientation to the loading direction (shown by solid black line), and an equivalent FE model, which was setup with various constitutive material models (shown by various red coloured lines).}}
    \label{fig:cube_90degree_compression_fem_vs_exp_v2}
\end{figure}

\subsection{Effect of wave number on performance indicators}
\label{appendix:Effect of wave number on performance indicators}

\subsubsection{Energy Absorption}

Convergence is observed in the value of EA with an increase in the wave number due to homogenisation of the structure. The majority of the structures show a plateau beyond a wave number of, $\lambda = 12 \pi$ apart from the case of `Lamellar' structure. However, the magnitude of EA is minuscule in comparison to the rest of the test cases due to the nature of the structure, whereby a plateau in the EA may not be observed until extreme values of wave number or a higher relative density are utilised. As a result, a value of $\lambda = 15 \pi$ was chosen given the dimensions of the structures being optimised, whilst taking into account the equivalence in length scales in experiments and simulations. Nonetheless, in a scenario where the sole objective is to maximise EA, regardless of artefacts and discontinuities being generated in the structures, a low wave number should be chosen. Overall, the wave number is a homogenisation parameter, hence it should be considered as a constant within the context of the optimisation.

\begin{figure}[!h]
  \centering
   \includegraphics[width=1\textwidth]{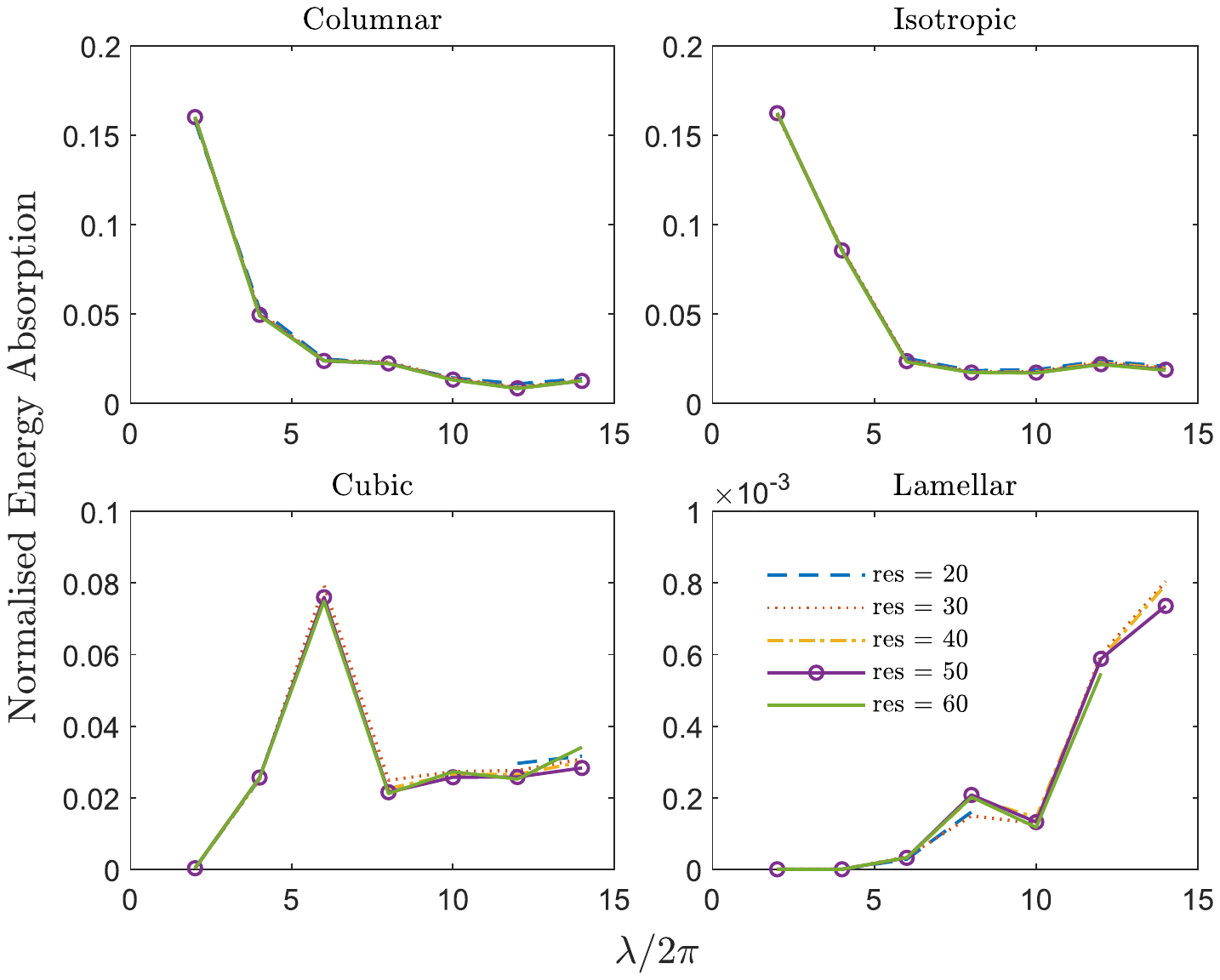}
   \caption{\label{fig:energy_absorption_vs_wavenumber}
   A demonstration of the effect of varying the wave number on the EA using four test structures generated with the cone angles is shown in Fig.~\ref{fig:4 spinodoid types}. The structures have dimensions of \(40 \times 40 \times 40\ \text{mm}^3\) and were tested using different mesh resolutions, ranging from a coarse mesh resolution of 20 to a fine mesh resolution of 60. A combination of different line styles and markers is used to distinguish between the various mesh resolutions. }
\end{figure}

\subsubsection{Peak Force}
Very little correlation exists between the wave number and the resulting PF calculated, independent of the mesh resolution used to generate the structure. This may have occurred due to a combination of the methods used to calculate the PF, and the structures being generated, which may offset the location at which the greatest force is achieved. The wave number was chosen based solely on the values of EA, while the displacement interval in between which the PF was calculated was based on the wave number selected. This is another factor which may have resulted in high variation in the values of PF.
\begin{figure}[!h]
  \centering
   \includegraphics[width=1\textwidth]{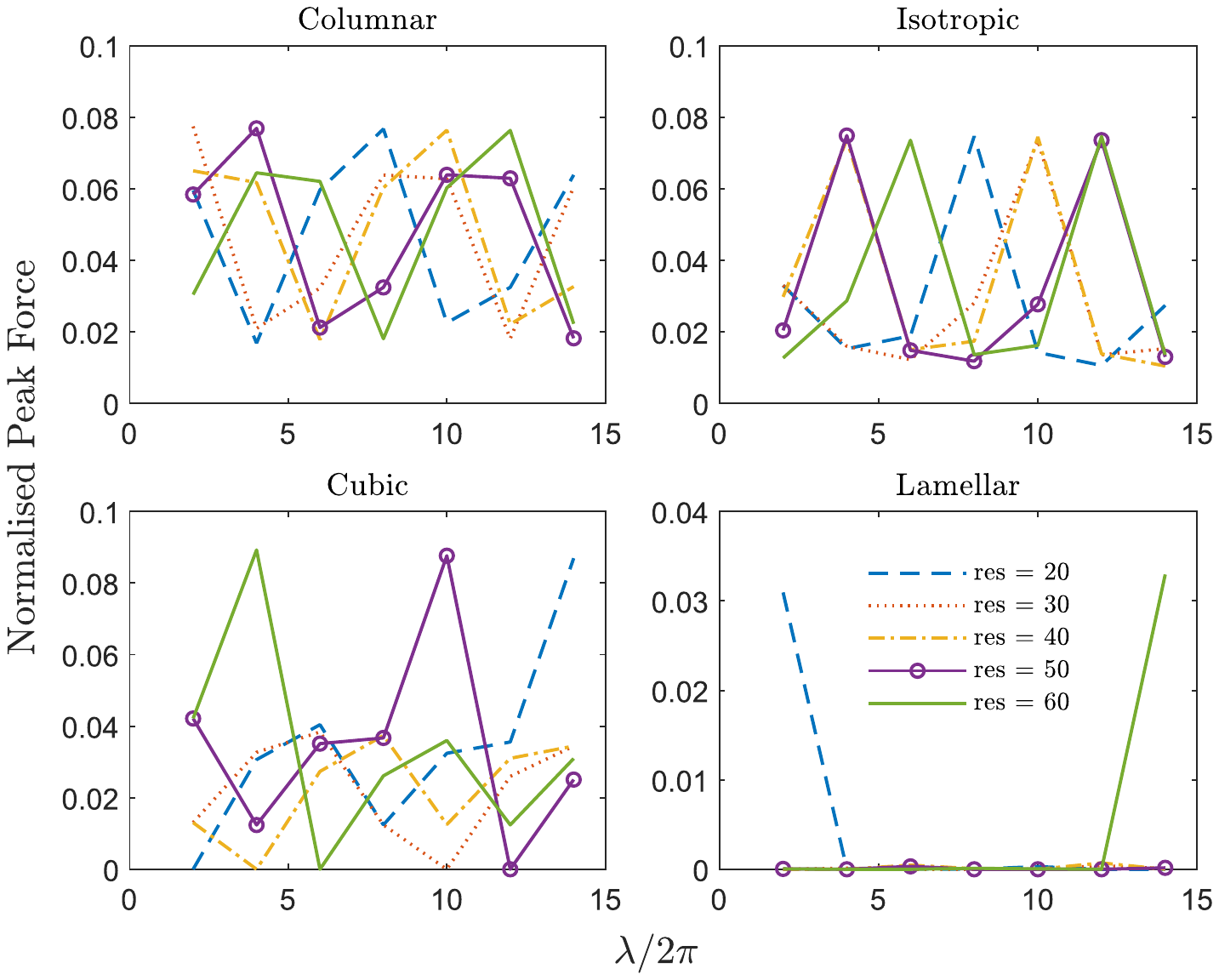}
   \caption{\label{fig:peak_force_vs_wavenumber}Similar to \ref{fig:energy_absorption_vs_wavenumber}, the figure illustrates the effect of varying the wave number on the PF using the same structures.}
\end{figure}

\newpage

\subsection{\textcolor{black}{Effect of mesh resolution on the number of elements}}
\textcolor{black}{\ref{tab:effect of resolution on mesh} illustrates the effect of mesh resolution on the number of tetrahedral elements used to discretise a structure with the parameters $\Theta = (0.3, 15\pi, 15, 15, 0)$. While the number of elements presented corresponds specifically to this structure, it is important to note that the element count will vary depending on the resolution and the geometry of the structure.}

\renewcommand{\tablename}{}

\begin{table}[ht]
\centering
\begin{tabular}{|c|c|c|}
    \hline
    Resolution & Number of elements & Average element size (mm)\\
    \hline
    20 & 18728 & 0.4398\\
    \hline
    30 &  84284 & 0.2731 \\
    \hline
    40 & 253485 & 0.2002\\
    \hline
    50 & 526626 & 0.1583\\
    \hline
    60 & 932918 & 0.1316\\
    \hline

\end{tabular}
\caption{\textcolor{black}{This example shows how the resolution changes the number of linear tetrahedral elements of type C3D4 for a structure with parameters $\Theta = (0.3, 15\pi, 15, 15, 0)$, with the average element size shown through the equivalent length found by dividing the element volume by its area.}}
\label{tab:effect of resolution on mesh}
\end{table}

\newpage

\subsection{Effect of mesh resolution on stress-strain behaviour of spinodoid structures}
\label{appendix:Effect of mesh resolution on stress-strain behaviour of spinodoid structures}

\begin{figure}[!h]
  \centering
   \includegraphics[width=1\textwidth]{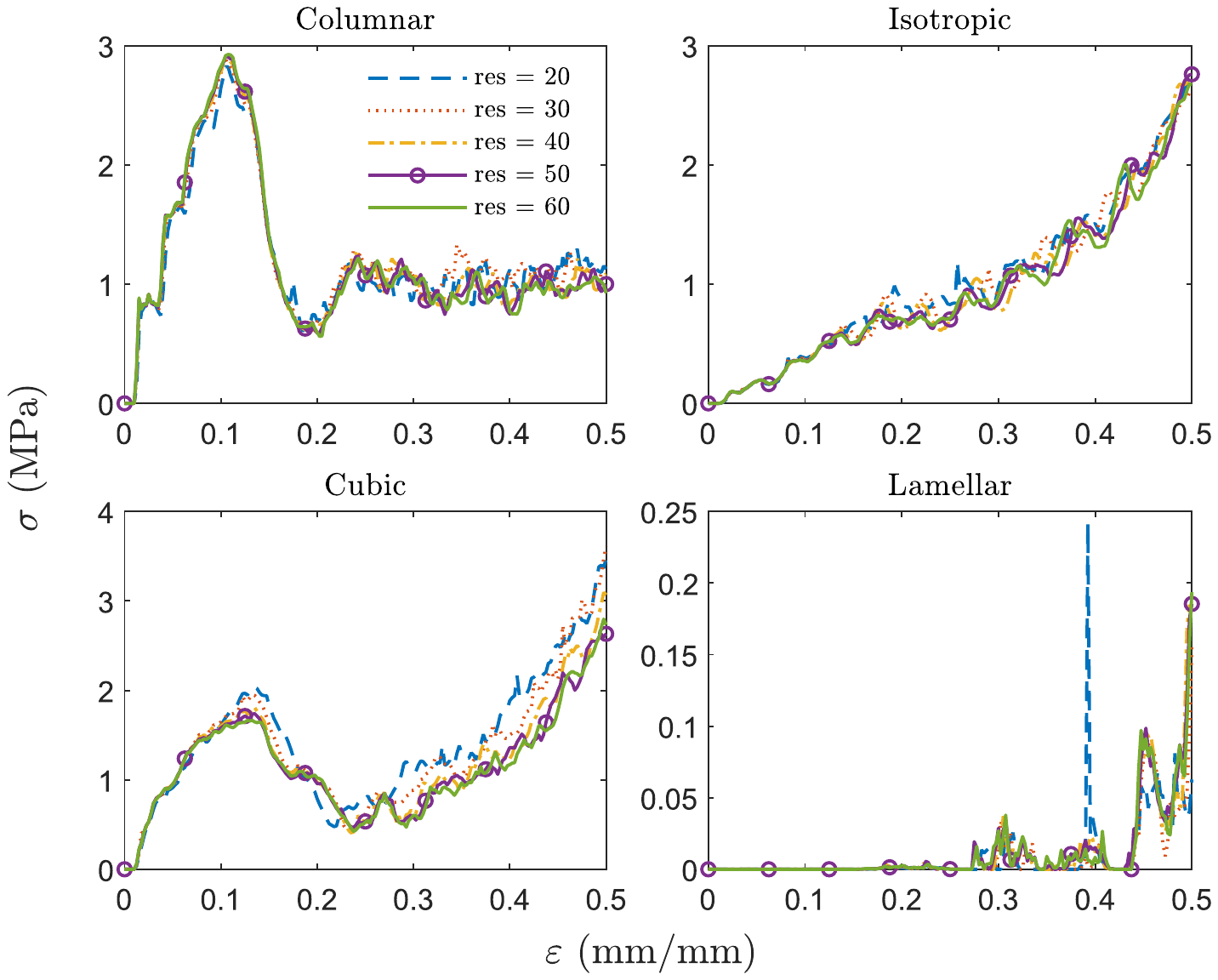}
   \caption{\label{fig:effect_of_resolution_on_stress_vs_strain_at_16pi}The subplots highlight the effect of mesh resolution on the stress-strain behaviour of four spinodoid test cases generated with angles described in Fig.~\ref{fig:4 spinodoid types}, measured at a wave number of $16\pi$. The structures were generated at various mesh resolutions, ranging from a coarse mesh resolution of 20 to a fine mesh resolution of 60. }
\end{figure}

A mesh convergence study assessed the impact of reesolution on the stress-strain behaviour of the four tested spinodoid cases. The mesh quality was controlled by varying the mesh resolution, which consequently altered the number of tetrahedral elements used to create the cellular structures. An increase in mesh resolution corresponds to a higher number of elements. \ref{fig:effect_of_resolution_on_stress_vs_strain_at_16pi} illustrates the stress-strain curves for various structures generated at different mesh resolutions, ranging from 20 to 60. Resolutions below 20 result in inadequately generated structures, while resolutions above 60 lead to excessively long simulation times. While the stress-strain behaviours of the structures are generally similar across this range, a resolution of 20 exhibited the most pronounced difference compared to higher resolutions. Beyond a resolution of 30, the variations in behaviour were negligible. Therefore, a resolution of 30 was chosen for the optimisation study as it provides a balance between computational efficiency and accuracy.

\newpage

\subsection{\textcolor{black}{Effect of time scaling and mass scaling on stress-strain behaviour}}
\label{appendix:Effect of time-scaling and mass-scaling on stress-strain behaviour}
\textcolor{black}{To reduce the solution time for ABAQUS dynamic Explicit simulations, there exists two common methods: Time scaling, and mass scaling. ABAQUS Explicit requires stable time increment $\Delta t = L^e/C_d$ to obtain a stable solution, where $L^e$ is the characteristic element length defined by the smallest element used to mesh the geometry, and $C_d = \sqrt{E/\rho}$ is the wave propagation speed, with $E$ being the Young's modulus and density of the material, respectively. For quasi-static speed at which the spinodoids were tested, the stable time dictates the number of increments required to solve the problem. Time scaling allows a reduction of the number of increments required as number of increments = $t / \Delta t$, where $t$ is the total time period. In this case, the computation time reduces as much as the time was reduced by artificially increasing the velocity. In contrast, the stable time increment can also be increased, which is called mass scaling, whereby the density is artificially increased. This decreases the computation time by the square root of the density.}

\subsubsection{\textcolor{black}{Effect of time scaling}}
\begin{figure}[!h]
  \centering
   \includegraphics[width=1\textwidth]{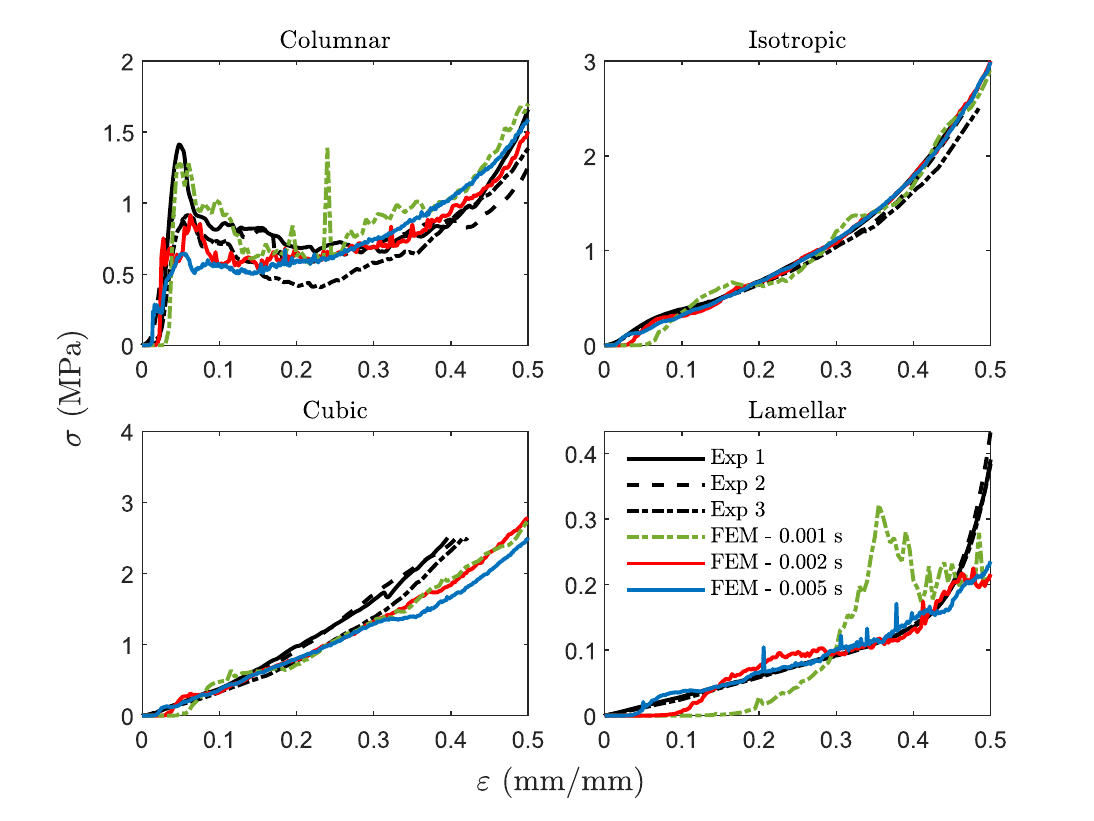}
   \caption{\label{fig:PET_G_exp_validation_time_scaling} \textcolor{black}{Plot showing the effect of time scaling for each test geometry described in Fig.~\ref{fig:4 spinodoid types} on the stress-strain response of each structure obtained from FEM simulations. The behaviour of each structure was tested under three different total time periods shown by the green, red, and blue lines indicating time periods of 0.001, 0.002, and 0.005 seconds, respectively, while the different black lines show the results from experiment.}}
\end{figure}

\textcolor{black}{The mechanical response of spinodoid structures was evaluated at three different total time periods: 0.001, 0.002, and 0.005 seconds. The analysis was conducted using the geometries described in Fig.~\ref{fig:4 spinodoid types} and compared with the corresponding experimental results, as shown in \ref{fig:PET_G_exp_validation_time_scaling}. For the `Columnar' structure, increasing the total time period led to a rapid reduction in initial stiffness, and a decrease in random fluctuations in stress values. Additionally, the time required for stress values to increase after contact between the structure and the loader measured by the stationary anvil was reduced, resulting in a better match of the small-strain response with experimental data, particularly at higher total time periods.}

\textcolor{black}{In the case of the `Isotropic' structure, the stress-strain behaviour showed minimal variation with changes in the total time period. However, the initial stiffness became more consistent with experimental results as the total time period increased. Similar trends were observed for the `Cubic' and `Lamellar' structures. For the `Cubic' structure, stress values at large strains began to deviate from experimental results at longer time periods. In contrast, for the `Lamellar' structure, the stress-strain response at high strain values aligned more closely with experimental data as the total time period increased. Based on these observations, it was essential to balance simulation accuracy and computational efficiency. At shorter total time periods, the higher loading velocity increases kinetic energy, making inertia effects more prominent. This introduces unintended strain dependency, leading to a non-linear small-strain response. However, shorter time periods also significantly reduce computation time. To achieve an optimal balance, a total time period of 0.002 seconds was selected.}

\subsubsection{\textcolor{black}{Effect of mass scaling}}

\begin{figure}[!h]
  \centering
   \includegraphics[width=1\textwidth]{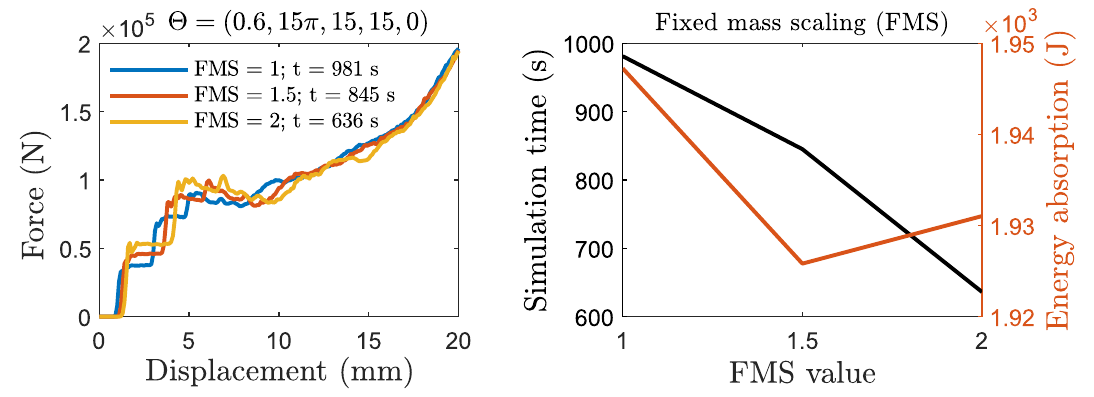}
   \caption{\label{fig:mass_scaling_v2}\textcolor{black}{Plots showing the effect of applying mass scaling to geometry with parameters $\Theta = (0.6, 15\pi, 15, 15, 0)$. The plot on the left shows the effect of three different levels of mass scaling on the mechanical response using force-displacement graph, while the plot on the right shows its effect on the simulation time and the calculated energy absorption value.}}
\end{figure}

\textcolor{black}{A structure generated using the parameters $\Theta = (0.6, 15\pi, 15, 15, 0)$ was used to test the effects of mass scaling. Three different fixed mass scaling (FMS) values were tested, showing minimal differences in the force-displacement curves, indicating no significant loss of accuracy as illustrated in \ref{fig:mass_scaling_v2}. Increasing the FMS value led to a reduction in total simulation time, with an increase of 0.5 resulting in a 14\% reduction in computation time. However, it also slightly reduced the energy absorption calculated from the force-displacement curve. Further increases in the FMS value continued to reduce simulation time, from 981 seconds at an FMS value of 1 to 636 seconds at a value of 2. Despite these performance gains, the energy absorption values showed only a partial reduction. However, artificially increasing the density (a consequence of mass scaling) reduced the wave speed, delaying the reaction force's propagation through the structure. This delay caused an initial lag in the force-displacement curve before stress values began to rise. Ultimately, an FMS value of 2 was selected, as it provided a balance between reduced simulation time and acceptable accuracy. Further increases in FMS resulted in diminished the accuracy of the mechanical response, making higher values unsuitable.}

\newpage

\subsection{Effect of number of samples on the Sobol' sensitivity indices}
\label{appendix:Effect of number of evaluations on the Sobol' sensitivity indices}

\subsubsection{Sensitivity analysis with four input variables}

The sensitivity indices were calculated at increments increasing by multiples of six, up to a maximum of 474 samples. This limit was set because adding six more samples would cause numerical instabilities, with the sensitivity indices S1 and ST approaching 1 for $\rho$, and 0 for the three conical angles. This underscores the significant impact of relative density on the performance indicators. 

\begin{figure}[!h]
  \centering
   \includegraphics[width=1\textwidth]{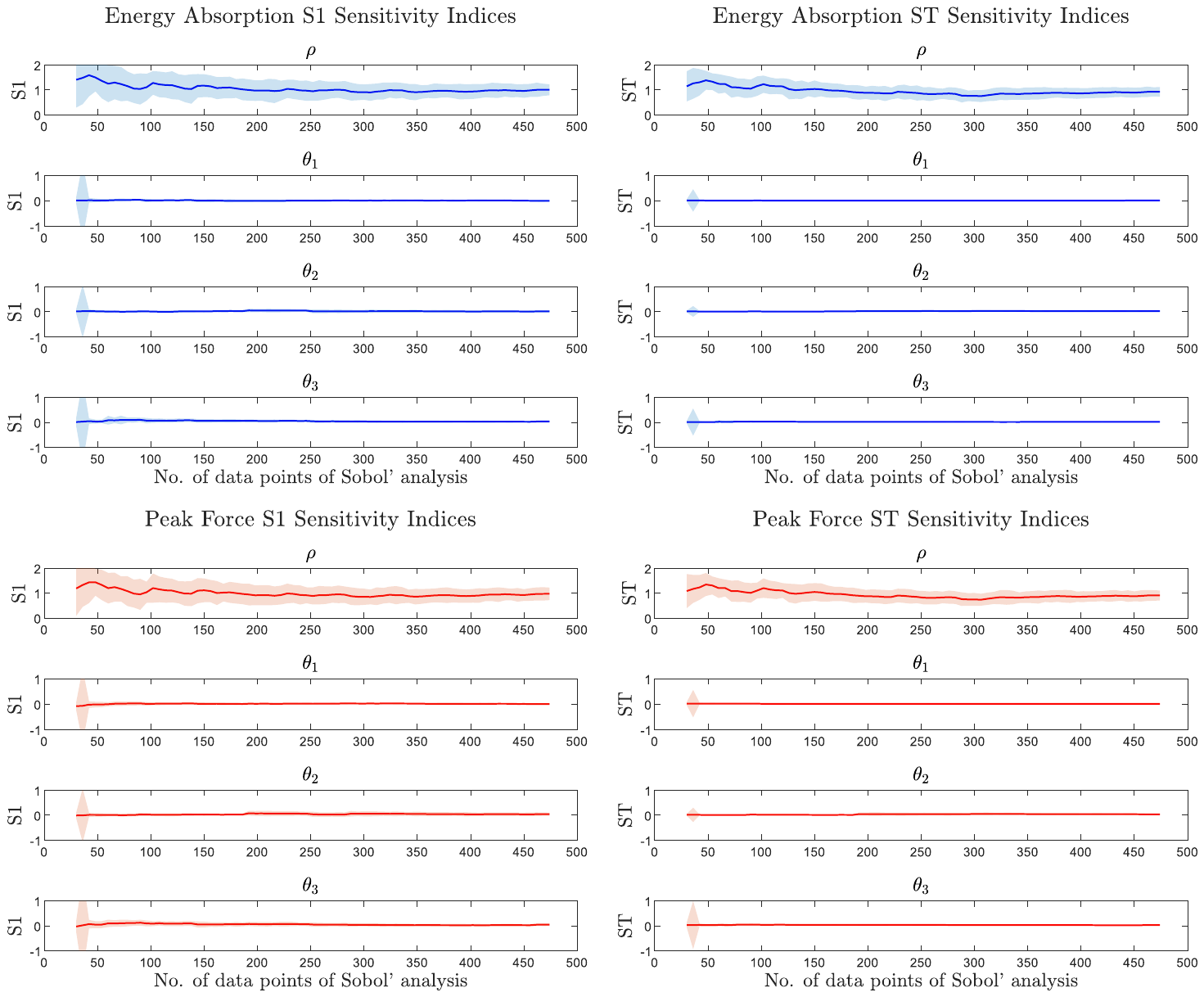}
\caption{\label{fig:sensitivity_analysis_4var_convergence}Illustration showing the convergence of sensitivity indices S1 and ST as the number of samples used to calculate the Sobol' sensitivity indices increases, with a maximum of 474 evaluations. The analysis is demonstrated for the performance indicators EA and PF. The Sobol' sensitivity analysis was conducted assuming four input parameters ($\rho, \theta_1, \theta_2, \theta_3$).}
\end{figure}

\subsubsection{Sensitivity analysis with three input variables}

The sensitivity indices were computed in increments increasing by multiples of five, up to a maximum of 1,500 samples. This analysis was based on the assumption that the relative density remained constant at $\rho = 0.3$.

\begin{figure}[!h]
  \centering
   \includegraphics[width=1\textwidth]{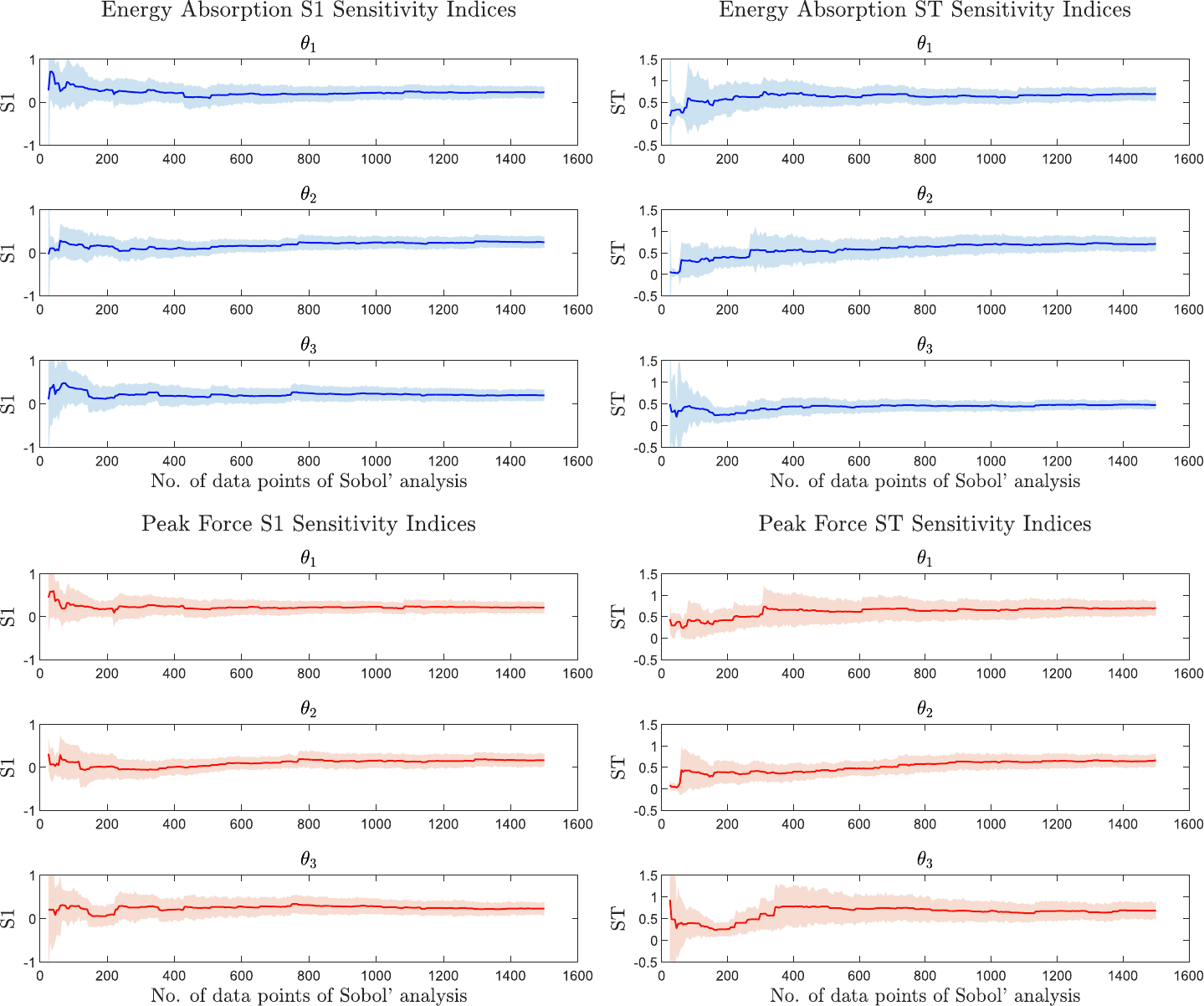}
\caption{\label{fig:sensitivity_analysis_3var_convergence}Similar to \ref{fig:sensitivity_analysis_4var_convergence}, an additional Sobol' sensitivity study was conducted with three input parameters ($\theta_1, \theta_2, \theta_3$), with a total number of samples used in calculations reaching 1,500. This study assumes a constant relative density of $\rho = 0.3$.}
\end{figure}

\newpage

\subsection{Benchmarks}
\label{appendix:Benchmarks}

\subsubsection{Benchmark optimisation history}
\begin{figure}[!h]
  \centering
   \includegraphics[width=1\textwidth]{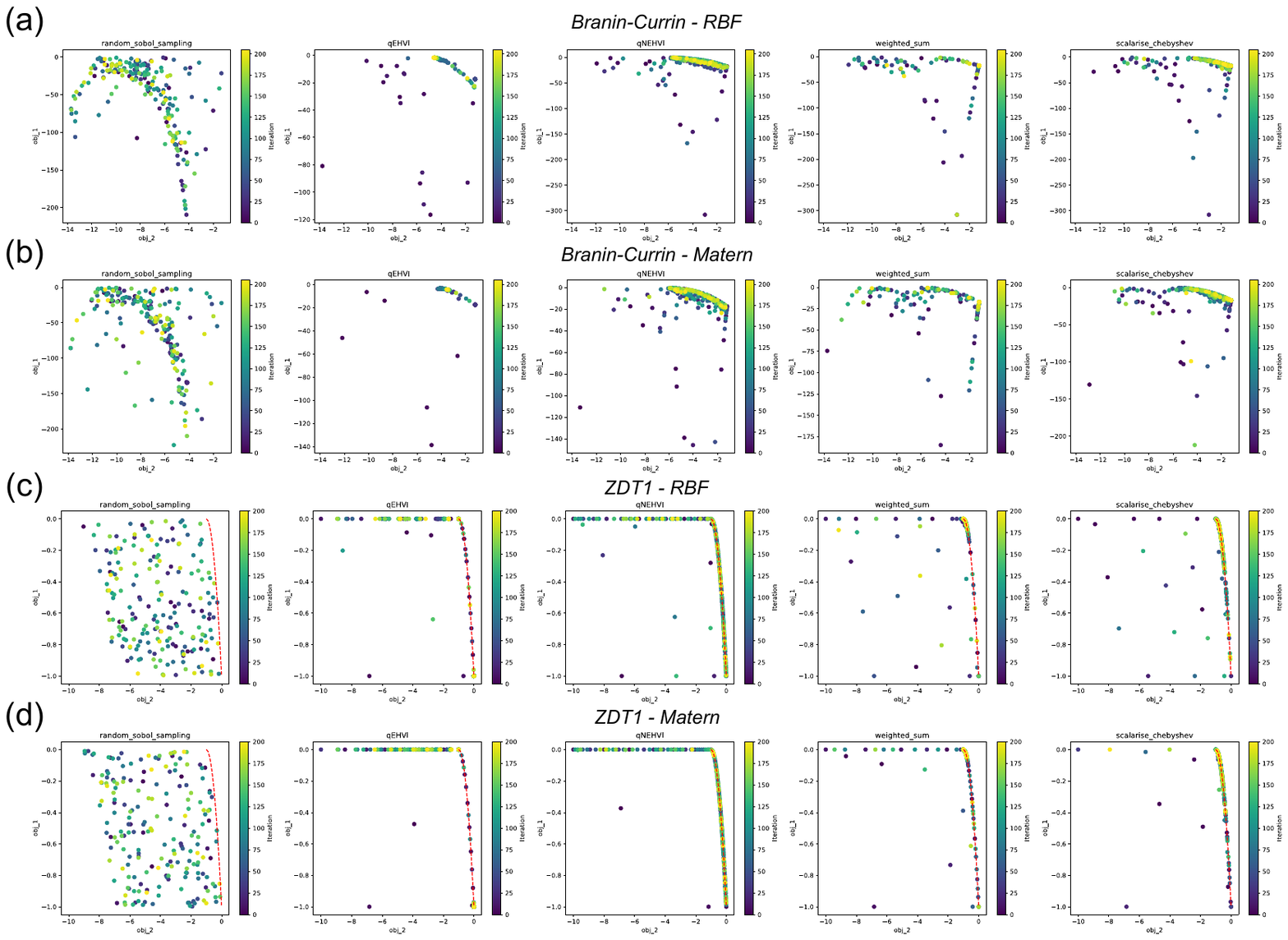}
\caption{\label{fig:benchmarks}Optimisation histories for two benchmark problems that were solved using five methods: Random Sobol sampling, qEHVI, qNEHVI, weighted sum, and Chebyschev scalarisation. For the Branin-Currin problem with artificial observation noise with noise standard deviation of 15.19, and 0.63 for each objective, respectively, the covariance matrix was calculated using (a) the RBF kernel and (b) the Matern kernel. Additionally, subplots (c) and (d) illustrate the solution of the ZDT1 problem with zero additional noise, where the covariance matrix was computed using RBF and Matern kernels, respectively. The red solid line represents the analytical solution for the ZDT1 problem. The colour gradients indicate at which each specific point was obtained.} 
\end{figure}

\subsubsection{Benchmark cumulative hypervolume}
\begin{figure}[!h]
  \centering
   \includegraphics[width=1\textwidth]{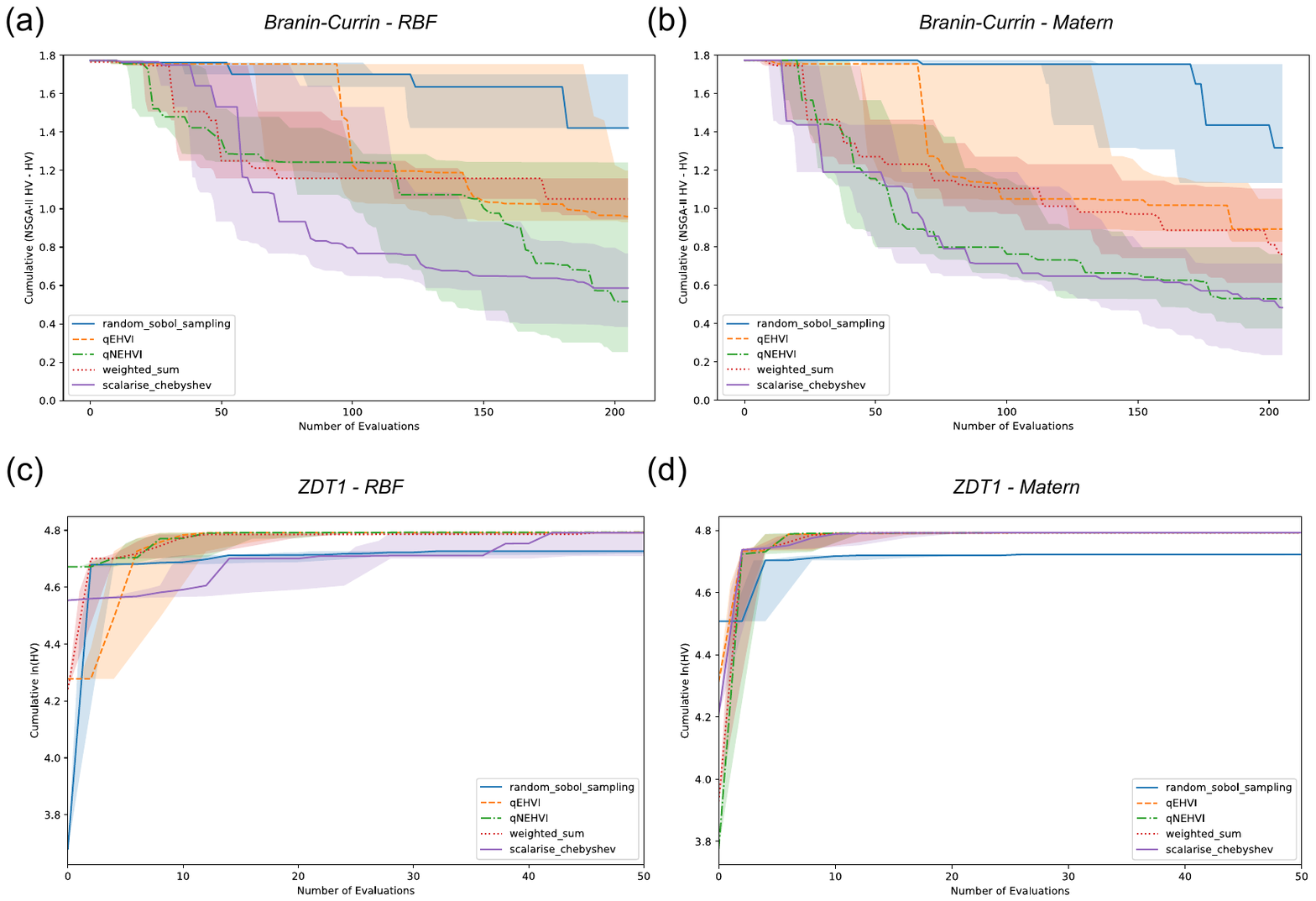}
\caption{\label{fig:benchmarks_median_IQR_plots}The subplots present the hypervolume history over 200 iterations for two benchmark problems: Branin-Currin and ZDT1. The solid lines represent the median hypervolume, while the shaded regions above and below indicate the upper and lower bound of the hypervolumes, respectively. Subplot (a) displays the difference in hypervolumes obtained using MOBO methods compared to the NSGA-II method for the Branin-Currin problem, with the RBF kernel used for the covariance matrix. Subplot (b) shows the hypervolume difference for the Branin-Currin problem, using the Matern kernel for the covariance matrix. (c) An illustration of the hypervolume obtained from various MOBO methods for ZDT1 problem using the RBF kernel function, while (d) shows the corresponding increase in hypervolume for the ZDT1 problem using the Matern kernel. Both (c), and (d) shows the result for the first 50 iterations as beyond this the hypervolume values do not change. The reference point for the Branon-Currin, and the ZDT1 were $(18, 6)$, and $(11, 11)$, respectively.}
\end{figure}

\newpage

\subsection{Pair plot of initial dataset}
\label{appendix:Pair plot of initial dataset}

\begin{figure}[!h]
  \centering
   \includegraphics[width=1\textwidth]{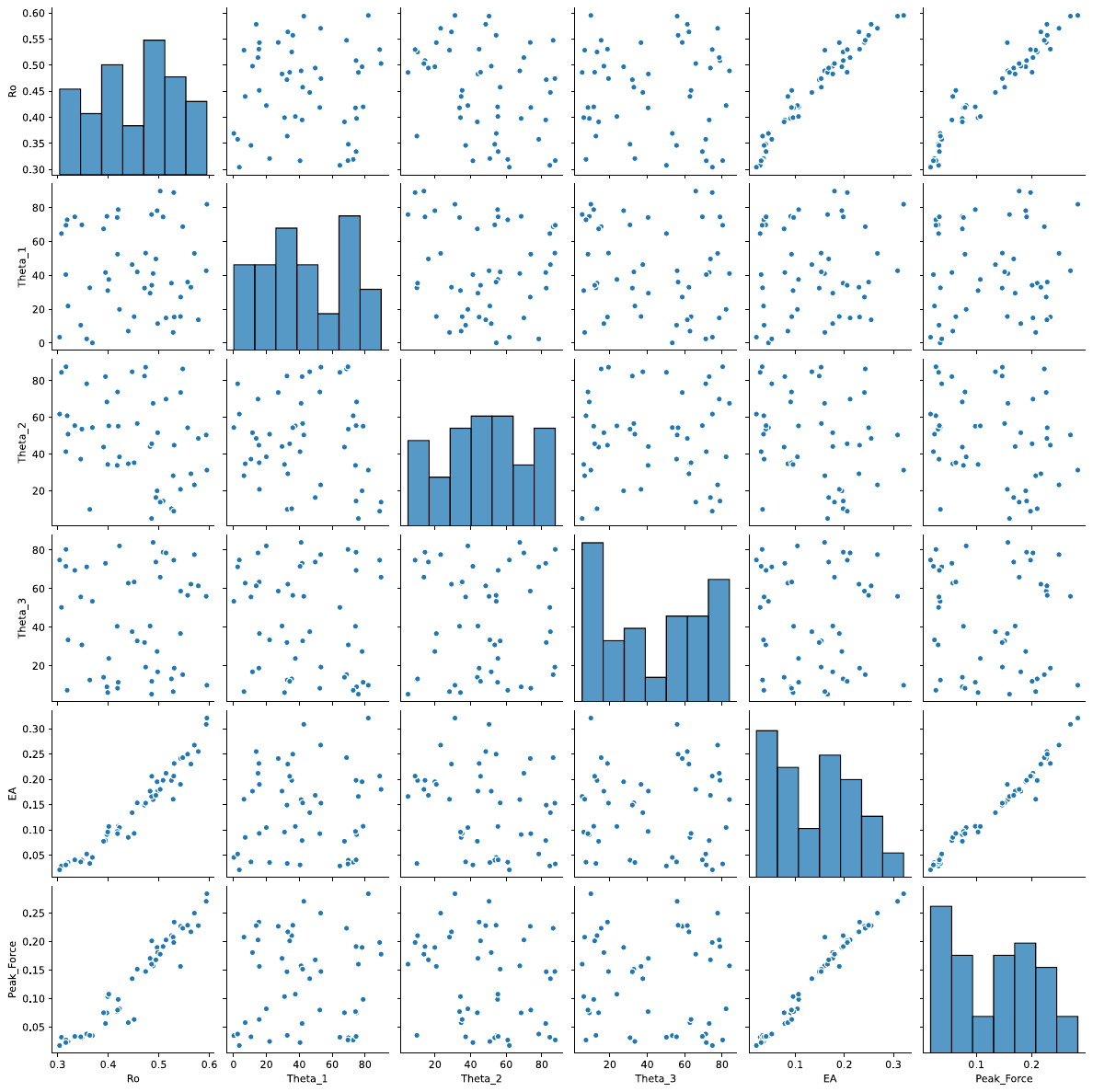}
\caption{\label{fig:pair_plot}Visualisation of the initial dataset containing 50 data points used to initialise the GP using a pair plot.}
\end{figure}

\newpage

\subsection{Optimisation history}
\label{appendix:Optimisation history}
\begin{figure}[!h]
  \centering
   \includegraphics[width=1\textwidth]{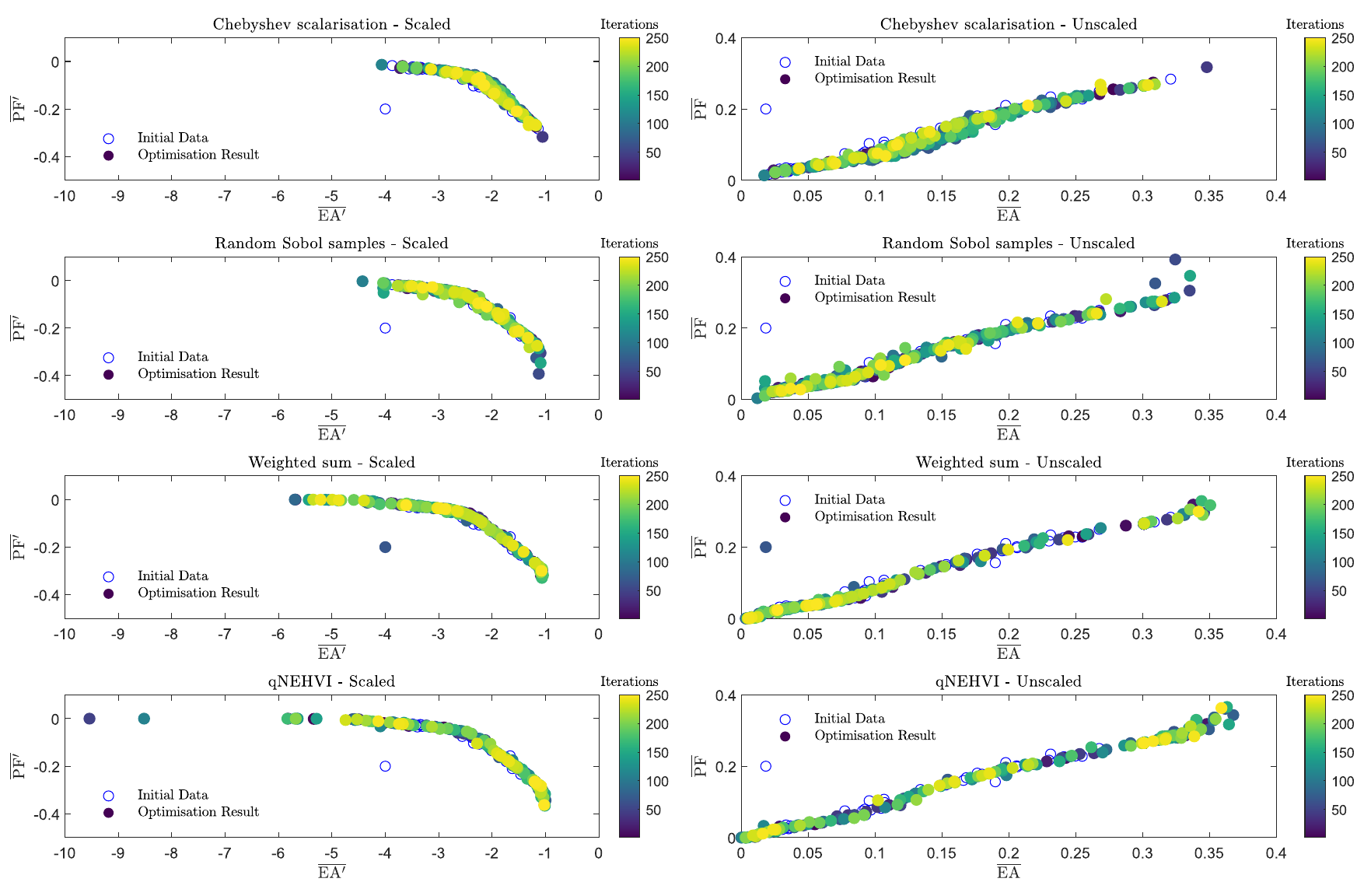}
\caption{\label{fig:optimisation_history}Optimisation history for all methods implemented using the proposed optimisation framework, showing the results of the full 250 iterations and the initial data used to initialise the GP model. The left column shows the scaled outputs, whereas the right column shows the unscaled outputs, with the scaling being done according to Eq.~(\ref{equ:scaling}). The colour bar on the right of each subplot shows the iteration at which the particular output was obtained.}
\end{figure}

\newpage

\subsection{HPC hardware specifications}
\label{appendix:HPC hardware specifications}

Jobs were run on the Apocrita cluster of Queen Mary, University of London, where each node consists of two 24-core Intel Xeon Platinum 8268 processors and 384 GB of RAM. Each job utilised 6 CPUs, with 5 GB of RAM requested per CPU and the time taken to complete a full optimisation task ranged from 10 to 14 days for all five of the methods tested. 

\newpage

\subsection{\textcolor{black}{Optimisation history with three variables}}
\label{appendix:Optimisation history with three variables}

\begin{figure}[!h]
  \centering
   \includegraphics[width=1\textwidth]{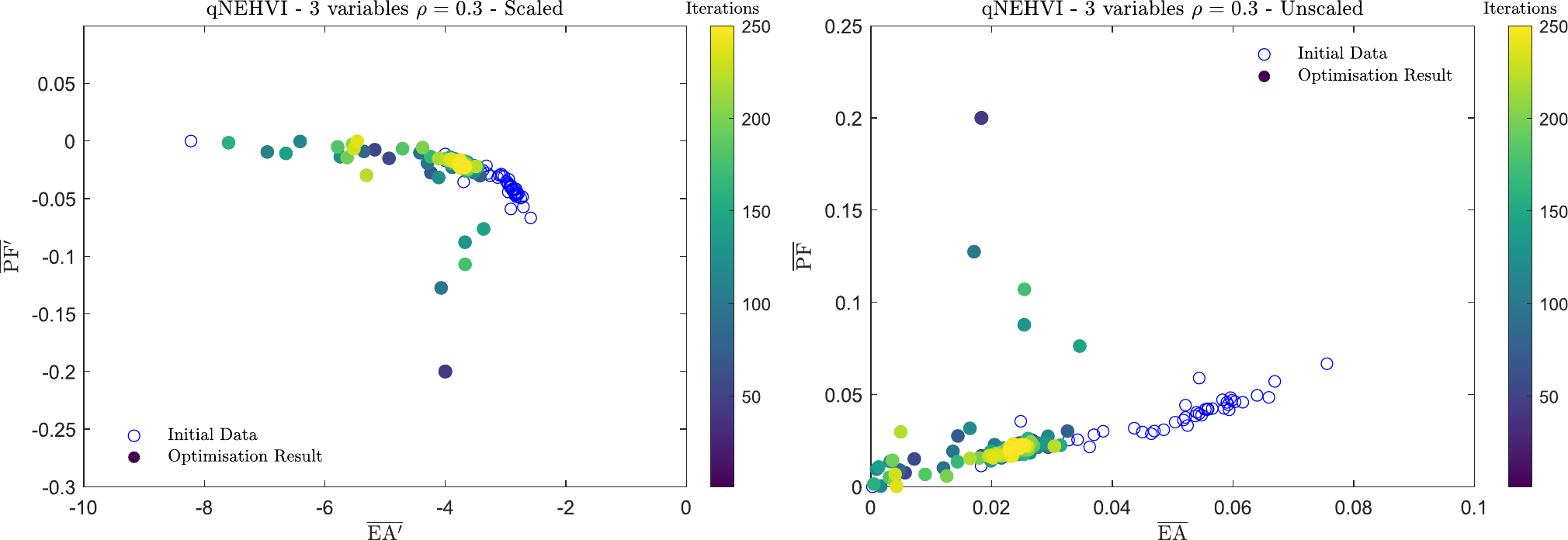}
\caption{\label{fig:optimisation_history_3var}\textcolor{black}{Optimisation history of optimisation task conducted with three variables, $\theta$'s, with fixed relative density at $\rho = 0.3$, showing the results of full 250 iterations with 50 initial data points, conducted using the qNEHVI method. The figure on the left shows the optimisation history with normalised objectives scaled according to Eq.~(\ref{equ:scaling}), while the figure on the right shows the history with the normalised objectives. While the colour bar shows the iteration at which the output was obtained.}}
\end{figure}

\subsection{\textcolor{black}{Evolution of recommended design parameters}}
\label{appendix:Evolution of recommended design parameters}
\begin{figure}[!h]
  \centering
   \includegraphics[width=1\textwidth]{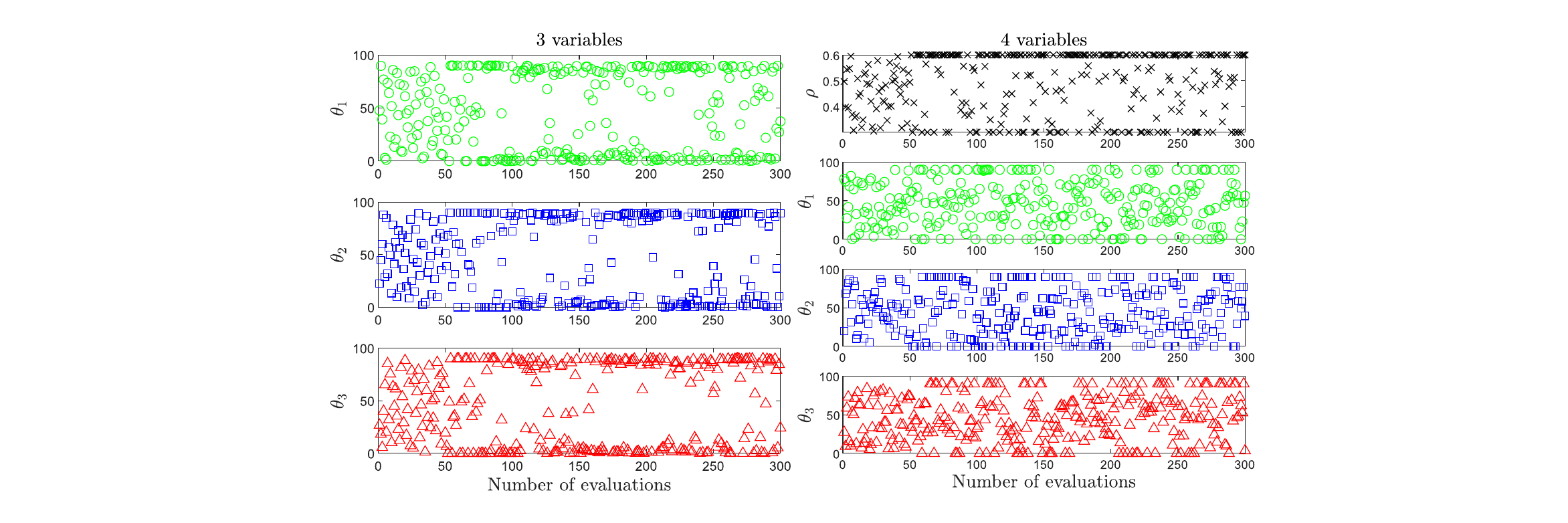}
\caption{\label{fig:variables_vs_iterations}\textcolor{black}{Left column shows the evolution of recommended design parameters for three-variable optimisation with a fixed relative density at $\rho = 0.3$, whereas the right column shows the evolution for four-variable optimisation with $\rho$ varying between 0.3 and 0.6. The first 50 points shown in each row shows the initial data, while the rest show the design parameters required to be evaluated throughout the optimisation.}}
\end{figure}

\end{document}